\PassOptionsToPackage{dvipsnames}{xcolor}
\pdfoutput=1 
\documentclass[acmsmall,screen]{acmart}

\usepackage{booktabs}   %
\usepackage{subcaption} %
\usepackage{xcolor}
\usepackage{relsize} 
\usepackage{pgfplots,pgfplotstable}
\pgfplotsset{compat=1.3}

\usepackage{ottalt}

\newenvironment{ottdefnblock}[3][]{ \framebox{\mbox{#2}} \quad #3 \\[0pt]}{}

\newcommand{\ottnt}[1]{\mathit{#1}}
\newcommand{\ottmv}[1]{\mathit{#1}}

\newcommand{\ottsym}[1]{#1}

\newcommand{\ottafterlastrule}{\\}

  \renewottcommands[ott]

\renewcommand\ottaltinferrule[4]{
  \inferrule*[narrower=1,lab=#1,#2]
  {#3}
  {#4}
}
\usepackage{amsmath}
\usepackage{enumerate}
\newcommand{\defeq}{\overset{\text{\tiny def}}{=}}
\definecolor{grey}{rgb}{0.5,0.5,0.5}
\usepackage{listings}
\newcommand{\code}[1]{\lstinline!#1!}
\lstMakeShortInline[columns=fullflexible]|

\usepackage{enumitem}
\setlist[enumerate]{leftmargin=5.5mm}
\setlist[itemize]{leftmargin=5.5mm}

\usepackage{wrapfig}
\usepackage[utf8]{inputenc}
\usepackage[T1]{fontenc}
\usepackage{microtype}
\usepackage{booktabs}   %
\usepackage{subcaption} %
\usepackage{mathpartir}
\usepackage{stmaryrd}
\usepackage{amsmath}
\usepackage{mathtools}
\usepackage{bbold}
\usepackage{ifthen}
\usepackage{color}
\usepackage{array,multirow}
\usepackage{pgfplots,pgfplotstable}
\usepackage{cleveref}
\usepackage{tikz}
\usepackage{xspace}
\usepackage{lscape}
\usetikzlibrary{shapes,shapes.geometric,backgrounds,patterns,shapes.misc,positioning}
\usetikzlibrary{arrows,automata,calc}
\tikzset{snake it/.style={decorate, decoration=snake}}
\tikzset{elliptic state/.style={draw,ellipse}}
\tikzset{rectangular state/.style={draw,rectangle,rounded corners}}
\pgfdeclarelayer{background}

\pgfplotsset{compat=1.3}
\newcommand{\Implies}{\,?\,}
\newcommand{\apart}{\;\#\;}
\newcommand{\seq}{\circledast}

\newcommand{\fix}[1]{\mu{#1}.\,}

\newcommand{\emptyenv}{\bullet}

\newcommand{\True}{\mathsf{true}}
\newcommand{\False}{\mathsf{false}}

\newcommand{\Follow}{\textsc{Follow}}
\newcommand{\Followlast}{\textsc{FLast}}
\newcommand{\Null}{\textsc{Null}}
\newcommand{\First}{\textsc{First}}
\newcommand{\IfThenElse}[3]{\mathsf{if}\;{#1}\;\mathsf{then}\;{#2}\;\mathsf{else}\;{#3}}

\newcommand{\setof}[1]{\left\{#1\right\}}

\newcommand{\Infer}[3]{\inferrule*[right={\text{\strut#1}}]{{}#2\mathstrut}{{}#3\mathstrut}}

\newcommand{\judgeg}[4]{{#1}; {#2} \vdash {#3} : {#4}}

\newcommand{\mktype}[4][]{\ifthenelse{\equal{#1}{}}
                                     {\left\{ \Null = {#2};\;
                                               \First = {#3};\;
                                               \Followlast = {#4} \right\}}
                                     {\left\{ \begin{array}{lcl}
                                                \Null &=& {#2} \\
                                                \First &=& {#3} \\
                                                \Followlast &=& {#4}
                                              \end{array} \right.}}

\newtheorem{definition}{Definition}

\makeatletter%
\newcommand{\smup}[2]{%
\raise #1\hbox{$\m@th$%
  \csname S@\f@size\endcsname
  \fontsize\sf@size 0%
  \math@fontsfalse\selectfont
#2%
}}%
\DeclareRobustCommand{\FLaP}{F\kern-.00emL\smup{-.2ex}{A}\kern-.00emP}%
\renewcommand{\FLaP}{\textsc{FLaP}}%
\renewcommand{\FLaP}{\texttt{flap}}%

\let\ExtendedVersion

\ifthenelse{\isundefined{\ExtendedVersion}}{
\newcommand{\shortOnly}[1]{#1}
\newcommand{\extendedOnly}[1]{}
}
{
\newcommand{\extendedOnly}[1]{#1}
\newcommand{\shortOnly}[1]{}
}

\gdef\xxDerivationProofCaseColor{N}

\newcommand{\DerivationProofCase}[3]{%
     \smallskip
     \item %
       \parbox[t]{100ex}{%
       \textbf{Case } {#1}: %
       $~$ %
       \if\xxDerivationProofCaseColor N%
           \ensuremath{%
              \Infer{}{#2}{#3}%
            }
       \else%
           \colorbox{\xxDerivationProofCaseColor}{%
              \ensuremath{%
                \Infer{#1}{#2}{#3}%
              }%
           }%
        \fi%
     }%
     \nopagebreak \\[-0.8ex]
  }

\definecolor{sscolor}{HTML}{788ca2}
\newcommand{\sshl}[1]{ {\color{sscolor}{#1}} }
\newcommand{\rulehl}[2][gray!40]{\colorbox{#1}{$\displaystyle#2$}}
\usepackage{xtab}
\newcommand{\tightoverset}[2]{%
  \mathop{#2}\limits^{\vbox to -.7ex{\kern-1ex\hbox{$#1$}\vss}}}
\newcolumntype{C}[1]{>{\centering\let\newline\\\arraybackslash\hspace{0pt}}m{#1}}

\colorlet{nfcolor}{NavyBlue}
\newcommand{\nfhl}[1]{{\color{nfcolor}{#1}}}
\colorlet{lexercolor}{BurntOrange}
\newcommand{\lexerhl}[1]{{\color{lexercolor}{#1}}}
\colorlet{fusedcolor}{Plum}
\newcommand{\fusedhl}[1]{{\color{fusedcolor}{#1}}}

\newcommand{\textnf}[1]{\nfhl{\textsc{#1}}}
\newcommand{\textfused}[1]{\fusedhl{\texttt{#1}}}

\usepackage{thmtools}

\crefformat{section}{\S#2#1#3}
\crefmultiformat{section}{\S\S#2#1#3}{ and~#2#1#3}{, #2#1#3}{, and~#2#1#3}

\Crefformat{figure}{#2Fig.~#1#3}
\Crefformat{section}{\S#2#1#3}
\Crefformat{subsection}{\S#2#1#3}
\Crefformat{subsubsection}{\S#2#1#3}
\usepackage[T1]{fontenc} %
\usepackage[utf8]{inputenc}

\newcommand{\zerodisplayskips}{%
  \setlength{\abovedisplayskip}{2pt}%
  \setlength{\belowdisplayskip}{2pt}%
  \setlength{\abovedisplayshortskip}{2pt}%
  \setlength{\belowdisplayshortskip}{2pt}}
\appto{\normalsize}{\zerodisplayskips}
\appto{\small}{\zerodisplayskips}
\appto{\footnotesize}{\zerodisplayskips}

\makeatletter
\renewcommand\paragraph{\@startsection{paragraph}{4}{\parindent}%
  {-.2\baselineskip \@plus -2\p@ \@minus -.2\p@}%
  {-3.5\p@}%
  {\ACM@NRadjust{\@parfont\@adddotafter}}}
\makeatother

\newcommand*\circled[1]{\tikz[baseline=(char.base)]{
    \node[shape=circle,draw,inner sep=1pt,color=Orchid] (char) {\color{Orchid}\scriptsize#1};}}
 
\extendedOnly{\setcopyright{none}}
\shortOnly{
\setcopyright{rightsretained}
\acmPrice{}
\acmDOI{10.1145/3591269}
\acmYear{2023}
\copyrightyear{2023}
\acmSubmissionID{pldi23main-p312-p}
\acmJournal{PACMPL}
\acmVolume{7}
\acmNumber{PLDI}
\acmArticle{155}
\acmMonth{6}
\received{2022-11-10}
\received[accepted]{2023-03-31}
}

\begin{document}

\title{\FLaP{}: A Deterministic Parser with Fused Lexing}
\extendedOnly{
\subtitle{
  Extended version of a paper published at PLDI 2023
  }
}

\author{Jeremy Yallop}
\affiliation{
  \institution{University of Cambridge}           %
  \country{UK}                   %
}
\email{jeremy.yallop@cl.cam.ac.uk}         %

\author{Ningning Xie}
\affiliation{
  \institution{University of Toronto}           %
  \country{Canada}                   %
}
\email{ningningxie@cs.toronto.edu}         %

\author{Neel Krishnaswami}
\affiliation{
  \institution{University of Cambridge}            %
  \country{UK}                    %
}
\email{Neel.Krishnaswami@cl.cam.ac.uk}          %

\extendedOnly{
\settopmatter{printfolios=false, printacmref=false}
}

\begin{abstract}
  Lexers and parsers are typically defined separately and connected
  by a token stream.
  This separate definition is important for modularity and reduces the
  potential for parsing ambiguity.
  However, materializing tokens as data structures and case-switching
  on tokens comes with a cost.

  We show how to \emph{fuse} separately-defined lexers and
  parsers, drastically improving performance without compromising
  modularity or increasing ambiguity.
  We propose
  a deterministic variant of Greibach Normal Form
  that ensures deterministic parsing with a single token of lookahead
  and makes fusion strikingly simple,
  and
  prove that normalizing context free expressions into the deterministic normal
  form is semantics-preserving.
  Our staged parser combinator library, \FLaP{}, provides a standard
  interface, but generates specialized token-free
  code that runs two to six times faster than \texttt{ocamlyacc} on a
  range of benchmarks.
\end{abstract}




\extendedOnly{
\fancyfoot{}
\thispagestyle{empty}
}

\maketitle

\extendedOnly{
\pagestyle{plain}
}

\section{Introduction}
\label{section:introduction}

Software systems are easiest to understand when their components have clear
interfaces that hide internal details.
For example, a typical compiler uses a separate lexer and parser to
reduce parsing ambiguity~\cite{aho2007compilers}, and connects the two
components via a token stream.

Unfortunately, hiding internal details can also reduce optimization
opportunities.
For parsers, the token stream interface isolates parser definitions from
character syntax details like whitespace, but it also carries overheads that
reduce parsing speed. %
Parsers built for efficiency avoid backtracking and typically need only one
token at any time.
However, even in this case, materializing tokens as data
structures and case-switching on tokens comes with a cost.

In this paper, we present the following contributions:

\begin{itemize}
\item We present a transformation that significantly improves parsing
  performance by fusing together a separately-defined lexer and a parser,
  entirely eliminating tokens.
  \begin{enumerate}
  \item We propose \textit{DGNF}, a \textit{Deterministic} variant of
    \textit{Greibach Normal Form}~\cite{greibach-nf} that ensures deterministic
    parsing with a single token of lookahead, allowing tokens to be discarded
    immediately after inspection (\Cref{sec:overview:dgnf}).
  \item We formalize a \textit{normalization} process that elaborates
    context-free expressions into DGNF, and prove that the elaboration is
    well-defined and preserves semantics (\Cref{section:detgnf}).
  \item We present \textit{lexer-parser fusion}, which transforms a
    separately-defined lexer and \textit{normalized} parser into a
    single piece of code that is specialized for calling contexts,
    avoids materializing tokens, and branches only on individual
    characters, not intermediate structures (\Cref{section:fusion}).
  \end{enumerate}
\item We implement the transformation in a parser combinator library, \FLaP{}
  (\emph{\textbf{f}used \textbf{l}exing \textbf{a}nd
    \textbf{p}arsing})~(\Cref{section:fused-parsing}).
  The lexer and parser are built using standard tools:
  derivative-based lexers by~\citet{DBLP:journals/jfp/OwensRT09}) and
  typed parser combinators by \citet{DBLP:conf/pldi/KrishnaswamiY19}.
\item 
  We demonstrate the effect of our transformation:
  \FLaP{} produces efficient code that runs several times faster than code
  produced by standard tools such as \texttt{ocamllex} and
  \texttt{menhir} (\Cref{section:evaluation}).
\end{itemize}

\noindent
We survey related work in \Cref{section:related}
and set out some directions for further development in
\Cref{section:future}.

\Cref{figure:pipeline} presents the novel code generation architecture of
\FLaP{}. The reader is advised to refer back to this figure while reading the
rest of the article, as what it depicts will gradually come to make sense.
The appendix 
\shortOnly{included in the extended version of the paper~\cite{neel2023fusion}}
includes proofs for the lemmas in the paper.

\begin{figure*}
\begin{tikzpicture}[xscale=2.4,yscale=0.5]
\begin{scope}[draw opacity=0.3]
\node[draw,rounded corners] (parser) at (0,0) { \small parser  };
\node[draw,rounded corners] (first-order) at (0.95,0) { \small first-order };
\node[draw,rounded corners] (typed) at (1.9,0) { \small typed };
\node[draw,rounded corners] (normalized) at (3,0) { \small normalized (\S\ref{section:detgnf}) };
\node[draw,rounded corners,anchor=north west] (lexer) at ($(parser.south west) + (0,-1)$) { \small lexer  };
\node[draw,rounded corners,anchor=north east] (specialized) at ($(typed.south east) + (0,-1)$) { \small specialized (\S\ref{sec:overview:fusion}) };
\node[draw,rounded corners] (fused) at (4,-1) { \small fused (\S\ref{section:fusion})  };
\node[draw,rounded corners] (staged) at (5,-1) { \small staged (\S\ref{section:staged-algorithm}) };
\end{scope}
\draw[out=90,in=-90,->,thick] (parser) -- (first-order);
\draw[out=90,in=-90,->,thick] (first-order) -- (typed);
\draw[out=90,in=-90,->,thick] (typed) -- (normalized);
\draw[->,thick] (normalized) to[out=0,in=180] ($(fused.west) + (0,4pt)$);
\draw[->,thick] (lexer) -- (specialized);
\draw[->,thick] (specialized) to[out=0,in=180] ($(fused.west) + (0,-4pt)$);
\draw[out=90,in=-90,->,thick] (fused) -- (staged);

\begin{scope}{[on background layer}
\draw[dotted,opacity=0.4,fill opacity=0.02,fill=red,%
  rounded corners]
      ($(parser.north west) + (-0.08,0.2)$)
   -- ($(typed.north east) + (0.08,0.2)$)
   -- ($(typed.south east) + (0.08,-0.2)$)
   -- ($(parser.south west) + (-0.08,-0.2)$)
   -- cycle;

\draw[dotted,opacity=0.4,fill opacity=0.02,fill=green,%
  rounded corners]
      ($(lexer.south east) + (0.3,-0.2)$)
   -- ($(lexer.south west) + (-0.08,-0.2)$)
   -- ($(lexer.north west) + (-0.08,0.2)$)
   -- ($(lexer.north east) + (0.3,0.2)$)
   -- cycle;

\draw[dotted,opacity=0.4,fill opacity=0.02,fill=blue,%
  rounded corners]
      ($(normalized.north west) + (-0.16,0.2)$)
   -- ($(staged.north east) + (0.08,1.1)$)
   -- ($(staged.south east) + (0.08,-0.2)$)
   -- ($(fused.south west) + (-0.08,-0.2)$)
   -- ($(specialized.south east) + (0.08,-0.2)$)
   -- ($(specialized.south west) + (-0.52,-0.2)$)
   -- ($(specialized.north west) + (-0.52,0.2)$)
   -- ($(specialized.north east) + (0.15,0.2)$)
   -- cycle;
\end{scope}

\node[anchor=south west] at ($(parser.north west) + (-0.1,0.1)$) { \parbox{6cm}{\itshape\small \color{red!10!black!}
    \citet{DBLP:conf/pldi/KrishnaswamiY19}
  } };
\node[anchor=north west] at ($(lexer.south west) + (-0.1,-0.1)$) { \parbox{4cm}{\itshape\small \color{green!10!black!} 
    \citet{DBLP:journals/jfp/OwensRT09}
  } };
\node (this-paper) at ($(staged.north east) + (0, 0.5) $) {\parbox{1.3cm}{\itshape\small \color{blue!10!black!}
    \FLaP{}} };

\end{tikzpicture}
\caption{\label{figure:pipeline}Architecture of \FLaP}
\end{figure*}

\section{Overview}
\label{sec:overview}

\subsection{Background: Parser Combinators and Typed Context-Free Expressions}
\label{sec:overview-parser}

\emph{Parser combinators}, introduced almost four decades ago
by~\citet{list-of-successes}, provide an elegant way to define parsers
using functions.
A parser combinator library provides functions denoting
token-matching, sequencing, recursion, and so on, allowing the
library user to describe a parser by combining these functions in a
way that reflects the structure of the corresponding grammar.
Here is a partial interface for constructing parsers (type \lstinline!pa!) in this way:

\begingroup
\lstset{
  basicstyle=\footnotesize\ttfamily
}
\begin{tabular}{@{}c@{}c@{}c}
{\begin{lstlisting}
type 'a pa
val tok: 'a tok -> 'a pa
val (>>>): 'a pa -> 'b pa -> ('a * 'b) pa
val fix: ('a pa -> 'a pa) -> 'a pa
\end{lstlisting}}
  &
    {\begin{lstlisting}

        (* token match *)
        (* sequence  *)
        (* recursion *)
\end{lstlisting}}
\end{tabular}
\endgroup

\noindent
The parameterization of \lstinline!pa! allows parsers to construct and return
suitably-typed syntax trees.

The earliest parser combinator libraries represented nondeterministic
parsers, with support for arbitrary backtracking and multiple results.
However, although they enjoyed various pleasant properties (such as a
rich equational theory), they suffered from potentially disastrous
performance.
In a recent departure from the nondeterministic tradition,
\citet{DBLP:conf/pldi/KrishnaswamiY19} define \emph{typed context-free
  expressions}, whose types track properties of languages.
Their design provides the standard set of parser combinators (as defined above),
but adds an additional type-checking step to preclude nondeterminism
and ensure linear-time parsing using a single token of lookahead.

\Cref{figure:cfe-typing} shows the typing rules from
\citet{DBLP:conf/pldi/KrishnaswamiY19}, and we direct the reader to the original
paper for more detailed explanations.
The definition for context-free expressions~(\textit{CFE}) $g$ is standard:
${\bot}$ for the empty language,
${\epsilon}$ for the language containing only the empty string,
${t}$ for the language containing only the single-token string $t$,
variables $\alpha$,
sequences $g_1 \cdot g_2$,
unions $g_1 \vee g_2$, and
the least fixed point operator ${\mu \alpha : \tau.g}$.
A type is a triple recording $\Null$ability, the $\First$ set, and the
$\Followlast$ set. %
Intuitively, a type is an \textit{overapproximation} of the properties of a
language. That is, a language $L$ satisfies a type $\tau$, if the
following is true: %
(1) when the empty string is in $L$, $\tau.\Null$ is $\True$; %
(2) the set of tokens that can start any string in $L$ is a subset of
$\tau.\First$; %
(3) the set of tokens which can follow the last token of a string in $L$
is a subset of $\tau.\Followlast$\footnote{$\Followlast$ sets, originally
  introduced by \citet{Anne1992}, are used as a alternative to the $\Follow$ set
  traditionally used in LL(1) parsing. $\Followlast$ is \textit{compositional},
  so \citet{DBLP:conf/pldi/KrishnaswamiY19} can calculate larger types from
  smaller ones. In contrast, $\Follow$ is the set of tokens following a particular
  nonterminal, and so is a property of a grammar rather than of a language. In
  practice, the typed parser combinators accept languages very close to LL(1);
  there are some differences that only seem to show up in contrived examples.}.

\begin{figure*}
\small
  $\begin{array}{rrcl}
     \mbox{Context-free expression} & g & \Coloneqq
     & \epsilon \mid t \mid \bot \mid \alpha \mid g_1 \cdot g_2 \mid g_1 \vee g_2 \mid \fix{\alpha:\tau}{g}  \\
     \mbox{Types} & \tau & \in &  \left\{ \Null : 2;\; \First : \mathcal{P}(\Sigma);\; \Followlast : \mathcal{P}(\Sigma) \right\} \\
     \mbox{Contexts} & \Gamma, \Delta & \Coloneqq &  \bullet \mid \Gamma, \alpha:\tau  \\
   \end{array}
   $

  \begin{minipage}{0.63\textwidth}
  {\small
  \begin{mathpar}
    \begin{array}{lcl}
    \tau_\epsilon & = & \mktype{\True}{\emptyset}{\emptyset} \\
    \tau_t & = & \mktype{\False}{\setof{t}}{\emptyset} \\
    \tau_\bot & = & \mktype{\False}{\emptyset}{\emptyset} \\[3pt]
    \tau_1 \cdot \tau_2 & = & \mktype[long]{\tau_1.\Null \land \tau_2.\Null}
                                                    {\tau_1.\First \;\cup \tau_1.\Null \Implies \tau_2.\First }
                                                    {\tau_2.\Followlast \;\cup \tau_2.\Null \Implies (\tau_2.\First \cup \tau_1.\Followlast)} \\[15pt]
    \tau_1 \vee  \tau_2 & = & \mktype[long]{\tau_1.\Null \vee \tau_2.\Null}
                                                    {\tau_1.\First \cup \tau_2.\First}
                                       {\tau_1.\Followlast \cup \tau_2.\Followlast}
   \\
   \\
    \tau_1 \seq \tau_2 & \defeq & \tau_1.\Followlast \cap \tau_2.\First = \emptyset \land \lnot \tau_1.\Null \\
    \tau_1 \apart \tau_2 & \defeq & (\tau_1.\First \cap \tau_2.\First = \emptyset) \land \lnot (\tau_1.\Null \land \tau_2.\Null)\\
     b \Implies S & \defeq & \IfThenElse{b}{S}{\emptyset}
          \end{array}
  \end{mathpar}
}
\end{minipage}%
\begin{minipage}{0.34\textwidth}
{\small  
  \begin{mathpar}
    \Infer{}
          {}
          { \judgeg{\Gamma}{\Delta}{\epsilon}{\tau_\epsilon} }
    \quad
    \Infer{}
          {}
          { \judgeg{\Gamma}{\Delta}{t}{\tau_t} }
    \\
    \vspace{4pt}
    \Infer{}
          { }
          { \judgeg{\Gamma}{\Delta}{\bot}{\tau_\bot} }
    \quad
    \Infer{}
          { \alpha:\tau \in \Gamma }
          { \judgeg{\Gamma}{\Delta}{\alpha}{\tau} }
    \\
    \vspace{4pt}
    \Infer{}
          { \judgeg{\Gamma}{\Delta, \alpha:\tau}{g}{\tau}   }
          { \judgeg{\Gamma}{\Delta}{\fix{\alpha:\tau}{g}}{\tau} }
    \\
    \vspace{4pt}
    \Infer{}
          { \judgeg{\Gamma}{\Delta}{g_1}{\tau_1}  \quad
            \judgeg{\Gamma, \Delta}{\emptyenv}{g_2}{\tau_2}  \quad
            \tau_1 \seq \tau_2 \\
          }
          { \judgeg{\Gamma}{\Delta}{g_1 \cdot g_2}{\tau_1 \cdot \tau_2} }
    \\
    \vspace{4pt}
    \Infer{}                                 
          { \judgeg{\Gamma}{\Delta}{g_1}{\tau_1} \quad
            \judgeg{\Gamma}{\Delta}{g_2}{\tau_2}  \quad
            \tau_1 \apart \tau_2 }
          { \judgeg{\Gamma}{\Delta}{g_1 \vee g_2}{\tau_1 \vee \tau_2} }
   \end{mathpar}
}  
\end{minipage}%
\caption{Krishnaswami and Yallop's type system for context-free expressions}
\label{figure:cfe-typing}
\end{figure*}

There is one typing rule for each combinator, whose types are constructed using
corresponding combinators (e.g. $\tau_1 \cdot \tau_2$).
The two contexts $\Gamma$ and $\Delta$ restrict where variables
can occur, disallowing left recursion. Specifically, when typing $ \mu \alpha  :  \tau .~ g $, the variable $\alpha$ is added to $\Delta$. But a variable $\alpha$ is well-typed only if
$\ottsym{(}  \alpha  \ottsym{:}  \tau  \ottsym{)} \, \in \, \Gamma$. The trick is that when typing $ g_{{\mathrm{1}}}  \cdot  g_{{\mathrm{2}}} $, $\Delta$ is
appended to $\Gamma$, where the \textit{separability} side condition $\tau_1 \seq
\tau_2$ ensures that $g_{{\mathrm{1}}}$ cannot be empty,
so that $g_{{\mathrm{2}}}$ can now use $\alpha$.
Additionally, $\tau_1 \seq \tau_2$ also says that $\tau_1.\Followlast$ is disjoint with $\tau_2.\First$,
ensuring strings matched by sequenced parsers have a unique decomposition.
Moreover, the side condition \textit{apartness} $\tau_1 \apart \tau_2$ on the
rule for $ g_{{\mathrm{1}}}  \lor  g_{{\mathrm{2}}} $ ensures that languages matched by alternated parsers do not overlap.

%
%
%
%
  %
  %
  %

\begin{comment}
As an example, consider the following well-typed s-expression (we often omit $\tau$ in $ \fix{\alpha:\tau}{g}$):
%
\[
  \mu \text{ sexp }. (\textsc{lpar} \cdot (\mu \text{ sexps }. \,\epsilon \lor \text{ sexp } \cdot \text{ sexps }) \cdot \textsc{rpar}) \lor \textsc{atom}
\]
The
following code shows how to use Krishnaswami and Yallop's parser combinators to define this
grammar, using a token type with \textsc{lpar}, \textsc{rpar} and \textsc{atom}
constructors, with the explicit fixed point represented using the Kleene star:

\begin{lstlisting}
fix (fun sexp -> (tok $\text{\normalfont\textsc{lpar}}$ >>> star sexp >>> tok $\text{\normalfont\textsc{rpar}}$)
                 $\$$ fun p -> $\ll\;$Sexp (snd (fst $\text{\textasciitilde}$p))$\;\gg$
             <|> tok $\text{\normalfont\textsc{atom}}$ $\$$ fun s -> $\ll\;$Atom $\text{\textasciitilde}$s$\;\gg$)
\end{lstlisting}

The \lstinline!fix!, \lstinline{tok}, \lstinline{>>>} and
\lstinline{star} constructors are from the parser interfaces.
%
%
The example additionally uses alternation \lstinline!<|>! and map
$\$$, which transforms the parsing result via a user-defined function,
and MetaOCaml's quotation and antiquotation constructs $\ll exp \gg$
and~$\text{\textasciitilde}exp$ to build \lstinline!code! values.
%
\end{comment}

\subsection{Overhead of Separate Lexing and Parsing}

These typing rules ensure that well-typed expressions have good
asymptotic performance, supporting linear-time parsing with a single
token of lookahead.
However, even linear-time parsers can be inefficient, using
significant resources at each parsing step.
Some parsing algorithms examine state dispatch tables to determine
what actions to take; similarly, Krishnaswami and Yallop's system
examines the types of context-free expressions to select branches.
To avoid this overhead, Krishnaswami and Yallop apply
\textit{multi-stage programming} to eliminate dispatch on type information,
generating type-specialized parsing code
that has performance competitive with
\texttt{ocamlyacc}.

However, even with these improvements, parsing still caries
considerable overhead.
The main source of inefficiency is the interface between the lexer and
the parser.
In a typical system, a lexer materializes each token it recognizes,
then the parser branches on that token to select an action.
This approach is clearly inefficient: information about which token
has been recognized was available at the point that the token was
created, then discarded and recovered via branching.
Just \emph{how} inefficient it is becomes apparent when we eliminate
the materialization of tokens.
\Cref{section:evaluation} shows that \FLaP{}'s fused lexer and parser,
in which tokens are not materialized, outperforms the unfused
implementation by 2 to 7 times --- that is, the overhead of token
materialization and associated branching accounts for the majority of
parsing time.

\subsection{Our Proposal: A Deterministic Parser with Fused Lexing}
\label{section:overview-proposal}

In this work, we take a systematic approach to fusing a lexer and a
\textit{deterministic} parser. We demonstrate our approach with \FLaP{}, a
parser combinator library that fuses
\begin{enumerate}
  \item
    a lexer based on \textit{derivatives} of regular
    expressions (regexes)~\cite{DBLP:journals/jfp/OwensRT09}, and
  \item
    parser combinators for typed context
  free expressions (\Cref{sec:overview-parser})~\cite{DBLP:conf/pldi/KrishnaswamiY19}.
\end{enumerate}

Lexer-parser fusion is not inherently limited to this particular combination;
it extends to other lexers for regexes and other
deterministic parsers.
In this paper,
\FLaP{} is restricted to LL(1) grammars, and we leave it as
future work to apply \FLaP{} to real programming languages (e.g. Python), and to
adapt the fusion strategy to other grammars such as LR and other practical
programming languages.

For \FLaP{}, the characteristic properties of derivatives and typed CFEs make our implementation
straightforward.
First, derivatives make it straightforward to build
compact deterministic automata that implement regex matchers. Specifically, the
\emph{derivative}~\cite{brzozowski-derivatives} of a regex $r$ with respect to a character $c$ is another
regex $\partial_c\;r$ that matches $s$ exactly when $r$ matches $c \cdot
s$. Therefore, one way to construct an automaton is
to take regexes $r$ as states, with a transition from $r_i$ to
$r_j$ via character $c$ whenever $\partial_c\;r_i = r_j$.
As \citet{DBLP:journals/jfp/OwensRT09} show, lexers based on
derivatives provide a practical basis for real-world lexing tools such
as ml-ulex and the PLT Scheme scanner generator.
We direct the reader to their work for the details about derivative-based lexers that we omit here.

Second, the types in typed context-free expressions correspond to
syntactic constraints in a normal form, DGNF (\Cref{section:detgnf}),
that serves as a basis for lexer-parser fusion.
Since every well-typed context free expression normalizes to
DGNF, we can provide the same parser combinator interface as
Krishnaswami and Yallop, but with a significantly more efficient implementation
(\Cref{section:evaluation}).

\paragraph{The running example}
The following sections illustrate \FLaP{}'s key ideas through a running example
shown in \Cref{figure:overview}. \Cref{figure:overview:syntax} presents the
grammars that will be introduced and used throughout this section, with colors
to help distinguish different grammars for better clarify.

\subsection{Example: The Lexer and Parser for S-Expressions}
\label{sec:overview:lexer}

Consider defining a lexer and a parser for \emph{s-expressions} (sexps)
representing tree-structured data.  Sexps are either (1) 
atoms, or (2) a possibly-empty lists of sexps enclosed in
parentheses $\mathsf{'('~sexps}~')'$.

\paragraph{Lexer.}
We start with the lexer.
\Cref{figure:overview:syntax} defines the syntax for regexes $\lexerhl{r}$ and
lexers $\lexerhl{L}$. %
Regexes $\lexerhl{r}$ include
$\lexerhl{\bot}$ for nothing ,
$\lexerhl{\epsilon}$ for the empty string,
characters $\lexerhl{c}$,
sequencing $\lexerhl{r \cdot s}$,
alternation $\lexerhl{r \mid s}$,
Kleene star $\lexerhl{r*}$,
intersection $\lexerhl{r \mathrel{\&} s}$,
and negation $\lexerhl{\lnot r}$.
A lexer $\lexerhl{L}$ is a mapping\footnote{We canonicalize lexers (\Cref{sec:canonical-lexer}) so there is no overlap between rules.} from regexes to
\emph{actions}, where an action might return a token ($ { \color{lexercolor}  \ottnt{r}  \Rightarrow  \mathbf{Return} ~ t  } $),
invoke the lexer recursively to skip over some input $ { \color{lexercolor}  \ottnt{r}   \Rightarrow   \mathbf{Skip}   } $, or raise
an error otherwise.
Our example sexp lexer (\Cref{figure:sexp-lexer:repeat}) has four
actions: three return tokens \textsc{atom}, \textsc{lpar} and \textsc{rpar},
and one skips whitespace.

\paragraph{Parser.}

\Cref{figure:overview:syntax} repeats the definition of CFE
from \Cref{sec:overview-parser}.
\Cref{fig:eg:grammar} gives a well-typed sexp grammar that
matches sequences of tokens.
For simplicity, we often omit $\tau$ in $\fix{\alpha:\tau}{g}$.

The bottom of \Cref{fig:eg:grammar} shows the BNF form of the grammar to help understanding.
Intuitively, sexp stands
for s-expressions, and sexps stands for lists of s-expressions.
That is,
sexp is either a \textsc{lpar} token followed by a list of s-expressions sexps and a
\textsc{rpar} token, or an \textsc{atom} token; and sexps is either empty ($\epsilon$), or a sexp
followed by another list of s-expressions (sexp sexps).

The rest of this section will show how to fuse the lexer and
parser.  First, however, we need to present DGNF grammars.

\begin{figure}

\begin{subfigure}[b]{\textwidth}
  \centering
  \begin{tabular}{llcl}
    \toprule
    \mbox{regular expression} & $\lexerhl{r}$ & $\Coloneqq$
    & $\lexerhl{\bot} \mid \lexerhl{\epsilon} \mid \lexerhl{c} \mid \lexerhl{r \cdot s} \mid \lexerhl{(r \mid s)} \mid \lexerhl{r*} \mid \lexerhl{(r \mathrel{\&} s)} \mid \lexerhl{\lnot r}$ \\
    lexer         & $\lexerhl{L}$ & $\Coloneqq$ & $\color{lexercolor} \{ \,  { \color{lexercolor}  \ottnt{r}  \Rightarrow  \mathbf{Return} ~ t  }  \,\}  \cup \{\,  { \color{lexercolor}  \ottnt{r}   \Rightarrow   \mathbf{Skip}   }  \,\}$ \\[4pt]
    \mbox{context-free expression} & $g$ & $\Coloneqq$
    & $\bot \mid \epsilon \mid t \mid \alpha \mid g_1 \cdot g_2 \mid g_1 \vee g_2 \mid \fix{\alpha:\tau}{g}$  \\[4pt]
    DGNF grammar            & $\nfhl{D}$ & $\Coloneqq$  & $ \nfhl{\{\,   { \color{nfcolor}   n   \rightarrow    \nfhl{  \nfhl{ t }  ~ \overline{n} }   }   \,\} }  \, \nfhl{\cup} \,  \nfhl{\{\,   { \color{nfcolor}   n   \rightarrow   \epsilon  }   \,\} }   $ \\
    fused grammar  & $\color{fusedcolor}F$ & $\Coloneqq$ & $ \color{fusedcolor} \{\, { \color{fusedcolor}  n   \rightarrow   \ottnt{r} \, \overline{n}  }  \,\} \cup \{\,   { \color{fusedcolor}  n   \rightarrow    ? r   }  \,\}  $ \\
    \bottomrule
  \end{tabular}
  \caption{Syntax of lexers, forms, and grammars in \FLaP{}}
  \label{figure:overview:syntax}
\end{subfigure}

\medskip
\begin{subfigure}[c]{0.12\textwidth}
  \quad
\end{subfigure}
\begin{subfigure}[c]{0.64\textwidth}
\begin{minipage}{0.55\textwidth}
  \[
    \color{lexercolor}
    \begin{array}{lp{1ex}r}
      \begin{array}[t]{rcl}
        \text{id}  &\Rightarrow& \mathbf{Return}\;\textsc{atom}\\
        \text{space}    &\Rightarrow& \mathbf{Skip}\\
        \texttt{(}              &\Rightarrow& \mathbf{Return}\;\textsc{lpar}\\
        \texttt{)}              &\Rightarrow& \mathbf{Return}\;\textsc{rpar}\\
      \end{array}
    \end{array}
  \]
\end{minipage}
\begin{minipage}{0.12\textwidth}
  \[
    \begin{array}{lp{1ex}r}
      \begin{array}[t]{rcl}
        \text{\color{lexercolor}id} &\defeq& \texttt{\color{lexercolor}[a-z]+}\\
        \text{\color{lexercolor}space} &\defeq&    \texttt{\color{lexercolor}\textvisiblespace}
                                                \;\mid\;  \color{lexercolor}\backslash\texttt{n}
      \end{array}
    \end{array}
  \]
\end{minipage}
\caption{\label{figure:sexp-lexer:repeat}A s-expression lexer (\ref{sec:overview:lexer})}
\end{subfigure}
\begin{subfigure}[c]{0.2\textwidth}
\tikzset{
  state/.style={
    rectangle,
    rounded corners,
    minimum height=2em,
    inner sep=2pt,
    text centered,
    align=center
  },
}
\begin{tikzpicture}[->]
  \footnotesize
  \node[state, anchor=center] (b) at (1.2, 0.8) {(b)};
  \node[state, anchor=center] (c) at (0, 0) {(c)};
  \node[state, anchor=center] (d) at (1.2, 0) {(d)};
  \node[state, anchor=center] (e) at (2.4, 0.4) {(e)};
  \draw (b) edge node{} (e);
  \draw (c) edge node[below]{normalizing} (d);
  \draw (d) edge node[above]{fusing} (e);
  \draw (-0.3,-0.5) rectangle (2.6,1.2);
\end{tikzpicture}
\end{subfigure}

\medskip
\begin{subfigure}[b]{\textwidth}
    \[
      \mu \text{ sexp }.\,(\textsc{lpar} \cdot (\mu \text{ sexps }. \,\epsilon \lor (\text{ sexp } \cdot \text{ sexps })) \cdot \textsc{rpar}) \lor \textsc{atom}
    \]
\vspace{-3pt}
    \[
      \begin{array}{lll}
        \begin{array}[t]{rcll}
          \text{ sexp} &::=& \textsc{lpar} \text{ sexps} \textsc{ rpar} &
                                                                          \circled{1} \\
                       &\mid& \textsc{atom} & \circled{2}\\
        \end{array}&
                     \begin{array}[t]{rcll}
                       \text{sexps} &::=& \text{sexp}~\text{sexps} & \circled{3} \\
                                    &\mid& \epsilon & \circled{4} \\
                     \end{array}
      \end{array}
    \]
    \caption{A well-typed s-expression grammar (top), and its BNF form (bottom) (\ref{sec:overview-parser} \& \ref{sec:overview:lexer})}
  \label{fig:eg:grammar}
\end{subfigure}

\medskip
\begin{subfigure}{\textwidth}
\[
  \color{nfcolor}
  \begin{array}{lll}
    \begin{array}[t]{rcl}
      \text{ sexp} &::=& \textsc{lpar} \text{ sexps} \text{ rpar}\\
                   &\mid& \textsc{atom}\\[1ex]
    \end{array}&
                 \begin{array}[t]{rcl}
                   \text{ rpar} &::=& \textsc{rpar}\\[1ex]
                 \end{array}&
                              \begin{array}[t]{rcl}
                                \text{ sexps} &::=& \textsc{lpar} \text{ sexps} \text{ rpar} \text{ sexps }\\
                                              &\mid& \textsc{atom} \text{ sexps}\\
                                              &\mid& \epsilon
                              \end{array}
  \end{array}
\]
\caption{An s-expression DGNF grammar, written in BNF form (\ref{sec:overview:dgnf} \& \ref{sec:overview:normal})}
\label{fig:s-expr:nf}
\end{subfigure}

\medskip
\begin{subfigure}{\textwidth}
\hspace{-13pt}
\begin{minipage}[t]{0.3\textwidth}
\centering
\raisebox{\dimexpr-\height+1.5ex\relax}{
\begin{tikzpicture}
  \node(lexer){
    $
    \color{lexercolor}
    \begin{array}[t]{rcl}
       \text{id}  &\Rightarrow& \mathbf{Return}\;\textsc{atom}\\
       \text{space}    &\Rightarrow& \mathbf{Skip}\\
       \texttt{(}              &\Rightarrow& \mathbf{Return}\;\textsc{lpar}\\
       \texttt{)}              &\Rightarrow& \mathbf{Return}\;\textsc{rpar}\\
       \\
      \color{fusedcolor}
      \text{ sexp} & \fusedhl{::=}&\fusedhl{\texttt{(} \text{ sexps} \text{ rpar}}\\
                  & \fusedhl{\mid}& \fusedhl{\text{id}}\\
                   & \fusedhl{\mid}& \fusedhl{\texttt{space} \text{ sexp}}\\
    \end{array}$

   };
   \draw[red,snake it] ($(lexer.north west) + (10pt,-45pt)$) -- ($(lexer.north east) + (-5pt,-45pt)$);
 \end{tikzpicture}
 }
\end{minipage}
\hfill\vline\hfill
\begin{minipage}[t]{0.3\textwidth}
  \hspace{-12pt}
\raisebox{\dimexpr-\height+1.5ex\relax}{
  \centering
\begin{tikzpicture}
  \node(lexer){
    $
    \color{lexercolor}
    \begin{array}[t]{rcl}
       \text{id}  &\Rightarrow& \mathbf{Return}\;\textsc{atom}\\
       \text{space}    &\Rightarrow& \mathbf{Skip}\\
       \texttt{(}              &\Rightarrow& \mathbf{Return}\;\textsc{lpar}\\
       \texttt{)}              &\Rightarrow& \mathbf{Return}\;\textsc{rpar}\\
       \\
      \color{fusedcolor}
      \text{ rpar} &\fusedhl{::=}& \fusedhl{\texttt{)}}\\
                  &\fusedhl{\mid}& \fusedhl{\texttt{space} \text{ rpar}} \\
    \end{array}$

   };
   \draw[red,snake it] ($(lexer.north west) + (10pt,-10pt)$) -- ($(lexer.north east) + (-5pt,-10pt)$);
   \draw[red,snake it] ($(lexer.north west) + (10pt,-33pt)$) -- ($(lexer.north east) + (-5pt,-33pt)$);
 \end{tikzpicture}
 }
\end{minipage}
\hfill\vline\hfill
\begin{minipage}[t]{0.33\textwidth}
  \hspace{-15pt}
  \raisebox{\dimexpr-\height+1.5ex\relax}{
  \centering
  \begin{tikzpicture}
    \node(lexer){
      $
      \color{lexercolor}
      \begin{array}[t]{rcl}
        \text{id}  &\Rightarrow& \mathbf{Return}\;\textsc{atom}\\
        \text{space}    &\Rightarrow& \mathbf{Skip}\\
        \texttt{(}              &\Rightarrow& \mathbf{Return}\;\textsc{lpar}\\
        \texttt{)}              &\Rightarrow& \mathbf{Return}\;\textsc{rpar}\\
        \\
        \fusedhl{\text{ sexps}}& \fusedhl{::=}& \fusedhl{\texttt{(} \text{ sexps} \text{ rpar} \text{ sexps }}\\
                   &\fusedhl{\mid}& \fusedhl{\text{id} \text{ sexps}}\\
                   &\fusedhl{\mid}& \fusedhl{\texttt{space} \text{ sexps}}\\
                   &\fusedhl{\mid}& \fusedhl{? \lnot \big(\text{id}\;\mid\;\text{space}\;\mid\texttt{(}\big)}
      \end{array}$

    };
    \draw[red,snake it] ($(lexer.north west) + (10pt,-45pt)$) -- ($(lexer.north east) + (-25pt,-45pt)$);
  \end{tikzpicture}
  }
\end{minipage}

\caption{Fusing drops lexing rules that return non-matchable tokens
  (top); the fused s-expr grammar (bottom)  (\ref{sec:overview:fusion})}
\label{fig:fusing}
\end{subfigure}

\vspace{15pt}
 \caption{\FLaP{} running example: s-expression lexing and parsing. Grammars
   are written in BNF form.}
 \label{figure:overview}
\end{figure}
 
\subsection{Deterministic Parsing with DGNF}
\label{sec:overview:dgnf}

To motivate DGNF, we consider how to parse with the s-expression grammar
in \Cref{fig:eg:grammar}.
Linear time, one-token-lookahead, deterministic parsing requires
committing to a particular branch after examining each token.
However the grammar in \Cref{fig:eg:grammar} does not make it clear
how to select branches by examining a single token.

For example, when parsing sexps when the first token matches
$\textsc{lpar}$, it is not immediately clear from the productions for
sexps whether to pursue production \circled{3} or production \circled{4}.
Deterministic parsing systems improve this situation by analysing the
grammar beforehand to calculate the branches that correspond to
particular input tokens.
In Krishnaswami and Yallop's case, the analysis takes the form of type
inference.
Each CFE is annotated with a type whose $\First$
set indicates which tokens can appear at the beginning of the strings
in the corresponding language.
Their parsing algorithm then examines $\First$ sets to select branches.
Using multi-stage programming, they then improve efficiency by
ensuring that $\First$ sets are only examined during analysis, not
during parsing itself.

In this work we take a different approach, transforming the grammar
into a form in which the branch to take at each point is syntactically
manifest.
We call this form \emph{Deterministic Greibach Normal Form} (DGNF),
since it is a deterministic variant of GNF~\cite{greibach-nf}.
\Cref{figure:overview:syntax} shows the syntax for
\textit{DGNF grammars}.
A DGNF grammar $\nfhl{D}$ is a set of productions that map nonterminals to normal
forms,
where all productions are either of form $\color{nfcolor}n \to  \nfhl{  \nfhl{ t }  ~ \overline{n} } $ or $ { \color{nfcolor}   n   \rightarrow   \epsilon  } $, where $\nfhl{n}$ is a nonterminal, $\nfhl{t}$ is a terminal,
and $\nfhl{\overline{n}}$ denotes $\color{nfcolor}n_1 n_2 \ldots n_k$$(k \ge 0)$.
Moreover, a DGNF grammar must also satisfy the following constraints (the formal
definition of DGNF is given later in \Cref{sec:semantics-dgnf}). First, for any
pair of a nonterminal $\color{nfcolor}n$ and a terminal $\color{nfcolor}t$,
there is at most one production beginning $\color{nfcolor}n \to t n_1 n_2 \ldots
n_k$. Second, the $\epsilon$-production may only be used when no terminal symbol
in the non-$\epsilon$ productions matches the input string.

Intuitively speaking, the constraints on the DGNF grammar
are a syntactic analogue of the
constraints enforced by the types in
the typed CFEs.
The constraints
have a simple practical motivation in parsers.
That is, each $\nfhl{n \to t n_1 n_2 \cdots n_k}$ production represents one branch that
matches a distinct terminal $\nfhl{t}$, and $\epsilon$-productions
represent an \text{else} branch that is taken if none of the active productive
branches matches the input.
With those constraints, it is evident that DGNF
ensures deterministic parsing with a single token of lookahead, and branching
(except for $\epsilon$-productions) always consumes tokens immediately.

\paragraph{Examples}
We consider a few examples. For readability, we write the grammar in BNF form,
e.g.~$\color{nfcolor}n ::= \textsc{a} n_1 n_2 \mid \textsc{b}$ corresponds to $\color{nfcolor}n \to \textsc{a} n_1 n_2$
and $ { \color{nfcolor}   n   \rightarrow    \nfhl{  \nfhl{\textsc{b} }  }   } $.
\[
  \color{nfcolor}
  \begin{array}{llll}
    {\color{black}(1)}
    \begin{array}[t]{lcl}
      n &::= &  \nfhl{\textsc{a} }  n_1 n_2 \\
        &\mid &  \nfhl{\textsc{b} }  \\
      n_1 &::=&  \nfhl{\textsc{c} } \\ 
      n_2 &::=&  \nfhl{\textsc{e} } \\ 
    \end{array}&
                 {\color{black}(2)}
                 \begin{array}[t]{lcl}
                   n &::= &  \nfhl{\textsc{a} }   \nfhl{\textsc{b} }  n_1 \\
                   n_1 &::= &  \nfhl{\textsc{c} }  \\
                 \end{array}&
                              {\color{black}(3)}
                              \begin{array}[t]{lcl}
                                n &::= &  \nfhl{\textsc{a} }  n_1 \\
                                 &\mid &  \nfhl{\textsc{a} }  n_2 \\
                                n_1 &::=&  \nfhl{\textsc{c} } \\ 
                                n_2 &::=&  \nfhl{\textsc{e} } \\ 
                              \end{array}&
                                           {\color{black}(4)}
                                           \begin{array}[t]{lcl}
                                             n &::= &  \nfhl{\textsc{a} }  n_1 n_2 \\
                                             n_1 &::= &  \nfhl{\textsc{c} }  \\
                                                 & \mid   & \epsilon \\
                                             n_2 &::= &  \nfhl{\textsc{c} }  \\
                                           \end{array}
  \end{array}
\]
Here (1) is in DGNF, while (2) (3) (4) are not. The reasons why (2) (3) are
not are obvious: In (2), $\nfhl{n}$ starts with two
terminals; in (3), $\nfhl{n}$ has two productions starting with $\nfhl{ \nfhl{\textsc{a} } }$.

(4) is the most subtle case. Consider
matching $\nfhl{n}$ with $\nfhl{ \nfhl{\textsc{a} }  \nfhl{\textsc{c} } }$.
First, $\nfhl{n}$ expands to $\nfhl{ \nfhl{\textsc{a} }  n_1 n_2}$.
But should $\nfhl{n_1}$ then expand to $\nfhl{ \nfhl{\textsc{c} } }$ or $\nfhl\epsilon$?
In a general nondeterministic grammar, it is impossible
to tell simply by looking ahead at the next token $\nfhl{ \nfhl{\textsc{c} } }$:
we may first consider $ { \color{nfcolor}   n_{{\mathrm{1}}}   \rightarrow    \nfhl{  \nfhl{\textsc{c} }  }   } $ and,
finding that $\nfhl{n_{{\mathrm{2}}}}$ fails to match,
backtrack to the other branch
to consider $ { \color{nfcolor}   n_{{\mathrm{1}}}   \rightarrow   \epsilon  } $
and $ { \color{nfcolor}   n_{{\mathrm{2}}}   \rightarrow    \nfhl{  \nfhl{\textsc{c} }  }   } $
and succeed.
However, the second constraint on DGNF grammars
eliminates this choice: only $ { \color{nfcolor}   n_{{\mathrm{1}}}   \rightarrow    \nfhl{  \nfhl{\textsc{c} }  }   } $ applies, and so
the grammar does \textit{not} match $\nfhl{ \nfhl{\textsc{a} }  \nfhl{\textsc{c} } }$.
As we will see,
the formal definition of DGNF (\Cref{sec:semantics-dgnf}) rules
out (4) as a DGNF grammar, ensuring that parsing is deterministic.
As the examples demonstrate,
DGNF ensures that there is never any
ambiguity about whether a production rule applies during parsing.

\vspace{3pt}

\Cref{fig:eg:grammar} is obviously not a DGNF grammar. So next, we discuss a
normalization algorithm that normalizes a context-free expression into a DGNF
grammar.

\subsection{Normalizing Context-Free Expressions to DGNF Grammars}
\label{sec:overview:normal}

We formalize a normalization algorithm (\Cref{section:detgnf}) which takes a
context-free expression, traverses its structure and turns it into a
DGNF grammar. As an example, \Cref{fig:s-expr:nf} presents the DGNF grammar of
normalizing the s-expression grammar in \ref{fig:eg:grammar}. This example
illustrates several points.

First, the normalized DGNF presentation addresses the problem of repeated
branching discussed in the last section.
In particular, parsing $\nfhl{\text{sexps}}$ involves reading the next
token and branching to the first, second or third branch depending on whether
the token is \nfhl{\textsc{lpar}}, \nfhl{\textsc{atom}} or something else.
In the first two cases the token is consumed immediately, and parsing
moves on to the next token in the input.
Only in the last case is the token examined more than
once: after selecting the $\epsilon$ branch that does not consume it,
the token is retained until it selects a non-$\epsilon$ branch that does.

Second, while in this case it seems straightforward to check that the normalized
grammar (\ref{fig:s-expr:nf}) represents exactly the same language as the
original context-free expression (\ref{fig:eg:grammar}), establishing
correctness properties for normalization is generally difficult. A particularly
challenging case is when normalizing a fixed point $ \mu \alpha .~ g $. In such case,
although we do not yet know the normalized grammar for $\alpha$, we must proceed
with normalizing $g$ regardless. Therefore, it is necessary to ``tie the
knot'' when the result of normalizing $g$ becomes available, which requires
us to introduce an intermediate \textit{non-DGNF} grammar form $\nfhl{n \to  {\nfhl{\alpha} }  \, \overline{n} }$, causing extra complication and subtleties during normalization.
We detail the normalization process and its correctness proofs in
\Cref{section:detgnf}.

Lastly, in what cases do we know that normalization will
produce DGNF grammars? For example, it'd be difficult (if not impossible) to
normalize an ambiguous grammar.
Fortunately,
typed context-free expressions give us enough guarantee: we prove that if a
context free expression is well-typed, then the normalization will always
produce a DGNF grammar. This is done by showing that DGNF indeed captures the
constraints enforced by types in the typed context free expressions system.

\subsection{Lexer-Parser Fusion}
\label{sec:overview:fusion}

Now that we have the lexer, and the normalized parser, we
can apply the lexer-parser fusion.
\Cref{figure:overview:syntax} defines the syntax of fused grammars. The fused
grammar $\fusedhl{F}$ is a set of productions, where each production
maps a nonterminal to either a regex followed by a list of nonterminals
$\fusedhl{n \to r~\overline{n}}$, or a single-token lookahead $\fusedhl {n
  \to ? r}$ that matches but not consumes tokens by $\fusedhl{r}$.

Fusion
acts on a lexer and a normalized parser, connected via tokens, and
produces a grammar that is entirely token-free, in which the only
branches involve inspecting individual characters.
\Cref{fig:fusing} fuses the s-expression lexer
(\ref{figure:sexp-lexer:repeat}) and the normalized parser
(\ref{fig:s-expr:nf}), following the steps:
\begin{enumerate}
  \item As the first step, fusion implicitly specializes the lexer to
    each nonterminal $\nfhl{n}$ in the normalized grammar, and
    lexing rules that return tokens not in productions for the
    nonterminal $\nfhl{n}$ are discarded,
    except for skip rules, since skipped characters can precede any token.
    \\
    \textit{Example (\ref{fig:fusing} top)}:
    \nfhl{rpar} has only a single production, which begins with the
    terminal \nfhl{\textsc{rpar}}. We look at the lexing rules, and discard those
    rules that do not return \lexerhl{\textsc{rpar}}, but keep the skip rule.
 \item 
   Then, the algorithm fuses the lexing rules and the parsing rules,
   by substituting the tokens in the parsing rules by regexes in the
   lexing rules that return corresponding tokens.
   Moreover, skip rules generate extra productions that match an arbitrary
   number of skipped characters.
   \\
   \textit{Example (\ref{fig:fusing} bottom)}:
   the fused
   \fusedhl{rpar} has two branches.
   The first branch fuses lexing and parsing, by having the original token
   \textnf{rpar} replaced with the regex \textfused{)}.
   The second branch corresponds to the
   skip rule in the lexer, allowing \fusedhl{rpar} to match an arbitrary
   number of \textfused{space}s.
   Observe how \fusedhl{rpar} now directly matches on characters,
   without referring to any tokens.
 \item For each $\epsilon$-production, fusion generates a
   lookahead rule
   consisting of the complement of the regexes that appear at the
   start of the right hand side of the other productions. \\
   \textit{Example (\ref{fig:fusing} bottom)}:
   when fusing sexps,
   the
   $\epsilon$-production \nfhl{sexps $\to \epsilon$} has been replaced with a
   lookahead rule $\color{fusedcolor}\text{sexp} \to ? \lnot \big(\text{id}\;\mid\;\text{space}\;\mid\texttt{(}\big)$.
\end{enumerate}

\noindent
\Cref{fig:fusing} presents the complete result of fusing the s-expression
lexer and normalized grammar following the fusion steps described above.

Crucially, note how the representation of DGNF grammar allows fusion to be
defined so concisely -- it would be more difficult to
fuse
the original CFE (\ref{fig:eg:grammar}) with
the lexer.
With DGNF grammars,
the constraints on the positions of terminals
make it straightforward to fuse the lexing rules into the grammar
without disrupting its structure.
Additionally, the fused grammar inherits the properties of DGNF: the
productions of a nonterminal start with distinct regexes, and an optional
lookahead rule may only be used when no regexes in other productions match the
input string.

\subsection{Staging}
\label{sec:overview:stagin}

In the last step, \FLaP{} uses MetaOCaml's staging facilities to generate code
for the fused grammar. %
Parsing with staging is very common, and various systems use some form of
staging; for example, ocamlyacc computes parsing tables once in advance, not
repeatedly during parsing.

The staging step in \FLaP{} generates one function for each parser
state (i.e.~for each pair of a nonterminal and a regex vector),
following a parsing algorithm with fused grammars,
but
eliminating information that is statically known, such as the
nullability and derivatives of the regexes associated with each state.
The DGNF representation used in \FLaP{} also makes staging
comparatively straightforward: \FLaP{} does not involve sophisticated
optimization techniques such as the binding-time improvements,
Furthermore, \FLaP{} does not rely on compiler optimizations to
further simplify the code it generates; instead, it directly generates
efficient code, containing no indirect calls, no higher-order
functions and no allocation, except where these elements are inserted
by the user of \FLaP{} in semantic actions.
\Cref{section:fused-parsing} presents the algorithm underlying \FLaP{}'s 
staged parsing implementation in more detail.
\section{Normalizing context-free expressions}
\label{section:detgnf}

This section presents a normalization algorithm that transforms context-free
expressions into DGNF grammars.
The normalization sets the basis for follow-up optimizations of fusion and staging.

\subsection{Normalization to DGNF}
\label{section:normalization}

\newcommand{\rulename}[1]{{\color{nfcolor}(\textit{#1})}\xspace}
\newcommand{\repsilon}{\rulename{epsilon}}
\newcommand{\rchar}{\rulename{token}}
\newcommand{\rbot}{\rulename{bot}}
\newcommand{\rseq}{\rulename{seq}}
\newcommand{\ralt}{\rulename{alt}}
\newcommand{\rfix}{\rulename{fix}}
\newcommand{\rvar}{\rulename{var}}

\begin{figure*}

  \begin{tabular}{llcl}
    \toprule
    normal form        & $\nfhl{N}$ & $\Coloneqq$ & $\nfhl{ \epsilon } \mid  \nfhl{  \nfhl{ t }  ~ \overline{n} }  \mid \rulehl{\nfhl{ {\nfhl{\alpha} }  \, \overline{n}}} $ \\
    normal form grammar            & $\nfhl{G}$ & $\Coloneqq$  & $ \nfhl{\{\,   { \color{nfcolor}   n   \rightarrow   \nfhl{N}  }   \,\} }   $\\[4pt]

    DGNF grammar            & $\nfhl{D}$ & $\Coloneqq$  & $ \nfhl{\{\,   { \color{nfcolor}   n   \rightarrow    \nfhl{  \nfhl{ t }  ~ \overline{n} }   }   \,\} }  \, \cup  \nfhl{\{\,   { \color{nfcolor}   n   \rightarrow   \epsilon  }   \,\} }   $ \\
    \bottomrule
  \end{tabular}
 \vspace{10pt}

{
\setlength{\tabcolsep}{4pt}
\begin{tabular}{llll}
  \multicolumn{4}{c}{$ {\mathcal{N} }  \llbracket \,  g  \, \rrbracket $ returns $ \nfhl{ n }  \Rightarrow  \nfhl{ \nfhl{G} } $, with a grammar $\nfhl{G}$ and the start nonterminal $\nfhl{n}$}
  \\
  \multicolumn{4}{c}{Each rule allocates a fresh nonterminal $\nfhl{n}$,
   except for rule \rfix, which allocates a fresh $\nfhl{ {\nfhl{\alpha} } }$}\\[5pt]
  \repsilon \quad & $ {\mathcal{N} }  \llbracket \,  \epsilon  \, \rrbracket $  & $=$  & $ \nfhl{ n }  \Rightarrow  \nfhl{  \nfhl{\{\,   { \color{nfcolor}   n   \rightarrow   \epsilon  }   \,\} }  } $  \\
  \rchar & $ {\mathcal{N} }  \llbracket \,   t   \, \rrbracket $       & $=$  & $ \nfhl{ n }  \Rightarrow  \nfhl{  \nfhl{\{\,   { \color{nfcolor}   n   \rightarrow    \nfhl{ t }   }   \,\} }  } $ \\
  \rbot & $ {\mathcal{N} }  \llbracket \,  \bot  \, \rrbracket $      & $=$  & $ \nfhl{ n }  \Rightarrow  \nfhl{  \nfhl{ \emptyset }  }  $\\[5pt]
  \rseq & $ {\mathcal{N} }  \llbracket \,   g_{{\mathrm{1}}}  \cdot  g_{{\mathrm{2}}}   \, \rrbracket $& $=$  & $ \nfhl{n}  \Rightarrow  \nfhl{\{}\,  { \color{nfcolor}   n   \rightarrow    \nfhl{ \nfhl{N}_{{\mathrm{1}}} ~ n_{{\mathrm{2}}} }   }  \mid  { \color{nfcolor}   n_{{\mathrm{1}}}   \rightarrow   \nfhl{N}_{{\mathrm{1}}}  }  \in \nfhl{\nfhl{G}_{{\mathrm{1}}}} \,  \nfhl{\}}  \cup   \nfhl{ \nfhl{G}_{{\mathrm{1}}}   \cup   \nfhl{G}_{{\mathrm{2}}} }  $ \\
  &                          &      & where \enskip $ {\mathcal{N} }  \llbracket \,  g_{{\mathrm{1}}}  \, \rrbracket  =  \nfhl{ n_{{\mathrm{1}}} }  \Rightarrow  \nfhl{ \nfhl{G}_{{\mathrm{1}}} }  \land  {\mathcal{N} }  \llbracket \,  g_{{\mathrm{2}}}  \, \rrbracket  =  \nfhl{ n_{{\mathrm{2}}} }  \Rightarrow  \nfhl{ \nfhl{G}_{{\mathrm{2}}} }  $ \\[5pt]
  \ralt & $ {\mathcal{N} }  \llbracket \,   g_{{\mathrm{1}}}  \lor  g_{{\mathrm{2}}}   \, \rrbracket $& $=$  & $\nfhl{n}  \Rightarrow  \nfhl{\{}\,  { \color{nfcolor}   n   \rightarrow   \nfhl{N}_{{\mathrm{1}}}  }  \mid  { \color{nfcolor}   n_{{\mathrm{1}}}   \rightarrow   \nfhl{N}_{{\mathrm{1}}}  }  \in \nfhl{\nfhl{G}_{{\mathrm{1}}}}  \, \nfhl{\}} $\\
&&&$\quad \cup ~\nfhl{\{} \,  { \color{nfcolor}   n   \rightarrow   \nfhl{N}_{{\mathrm{2}}}  }  \mid  { \color{nfcolor}   n_{{\mathrm{2}}}   \rightarrow   \nfhl{N}_{{\mathrm{2}}}  }  \in \nfhl{\nfhl{G}_{{\mathrm{2}}}} \, \nfhl{\}}  \cup   \nfhl{ \nfhl{G}_{{\mathrm{1}}}   \cup   \nfhl{G}_{{\mathrm{2}}} }  $    \\
  &                             &      & where \enskip $ {\mathcal{N} }  \llbracket \,  g_{{\mathrm{1}}}  \, \rrbracket  =  \nfhl{ n_{{\mathrm{1}}} }  \Rightarrow  \nfhl{ \nfhl{G}_{{\mathrm{1}}} }  \land  {\mathcal{N} }  \llbracket \,  g_{{\mathrm{2}}}  \, \rrbracket  =  \nfhl{ n_{{\mathrm{2}}} }  \Rightarrow  \nfhl{ \nfhl{G}_{{\mathrm{2}}} }  $ \\[5pt]
  \rfix& $ {\mathcal{N} }  \llbracket \,   \mu \alpha .~ g   \, \rrbracket $     & $=$  & $\nfhl{ {\nfhl{\alpha} } }  \Rightarrow  { \nfhl{\{} \,  { \color{nfcolor}    {\nfhl{\alpha} }    \rightarrow   \nfhl{N}  }   \mid  { \color{nfcolor}   n   \rightarrow   \nfhl{N}  }  \in \nfhl{G} \, \nfhl{\}} }    \hfill{\circled{1}\qquad\enskip} $ \\
                  & & & $ \qquad  \cup ~{ \nfhl{\{} \,  { \color{nfcolor}   n'   \rightarrow    \nfhl{ \nfhl{N} ~ \overline{n}' }   }  \mid  { \color{nfcolor}   n'   \rightarrow    \nfhl{  \nfhl{  {\nfhl{\alpha} }  }  ~ \overline{n}' }   }  \in \nfhl{G} \land   { \color{nfcolor}   n   \rightarrow   \nfhl{N}  }  \in \nfhl{G} \, \nfhl{\}} }  \hfill{\circled{2}\qquad\enskip}$ \\
                  & & & $ \qquad  \cup ~{{\nfhl{G}}\backslash_{ { \color{nfcolor}   n'   \rightarrow    \nfhl{  \nfhl{  {\nfhl{\alpha} }  }  ~ \overline{n}' }   } }} \hfill{\circled{3}\qquad\enskip} $\\
                   &                       &      & where \enskip  $ {\mathcal{N} }  \llbracket \,  g  \, \rrbracket  =  \nfhl{ n }  \Rightarrow  \nfhl{ \nfhl{G} } $ \\
                   &                       & &  \qquad\quad\, $ \nfhl{G}\backslash_{ { \color{nfcolor}   n'   \rightarrow    \nfhl{  \nfhl{  {\nfhl{\alpha} }  }  ~ \overline{n}' }   } }$ is $\nfhl{G}$ with all $ { \color{nfcolor}   n'   \rightarrow    \nfhl{  \nfhl{  {\nfhl{\alpha} }  }  ~ \overline{n}' }   } $ removed for any $n'$ and $\overline{n}'$
  \\[4pt]
  \rvar & $ {\mathcal{N} }  \llbracket \,  \alpha  \, \rrbracket $           & $=$  & $\nfhl{n}  \Rightarrow   \nfhl{\{\,   { \color{nfcolor}   n   \rightarrow    \nfhl{  {\nfhl{\alpha} }  }   }   \,\} } $ \\
\end{tabular}
}

\caption{\label{figure:normalization}Normalization of well-typed context-free
  expressions.}
\end{figure*}

\Cref{figure:normalization} presents the syntax for normal forms and the
normalization algorithm.

The normal form grammar $\nfhl{G}$ maps nonterminals to
the normal forms. Note the difference from the DGNF grammar $\nfhl{D}$:
$\nfhl{G}$ includes an extra \textit{non-DGNF} form $ { \color{nfcolor}   n   \rightarrow    \nfhl{  \nfhl{  {\nfhl{\alpha} }  }  ~ \overline{n} }   } $,
which makes $\nfhl{G}$ a superset of $\nfhl{D}$.
As discussed in \Cref{sec:overview:normal},
this non-DGNF form is necessary for normalizing fixpoints, where $\nfhl{ {\nfhl{\alpha} } }$ is interpreted as a special
kind of nonterminal.
As a nonterminal,
$\nfhl{ {\nfhl{\alpha} } }$ itself may appear as part of a $\nfhl{ \nfhl{  \nfhl{ t }  ~ \overline{n} } }$ (e.g.~$\nfhl{ \nfhl{  \nfhl{ t }  ~  {\nfhl{\alpha} }  } }$).
We show later that $\nfhl{ {\nfhl{\alpha} }  \, \overline{n}}$ is an intermediate form
that is entirely eliminated in the final result,
turning the grammar into DGNF. We will mostly use $\nfhl{G}$ in
this section.

The key to normalization is the normalization function $ {\mathcal{N} }  \llbracket \,  g  \, \rrbracket $ that
normalizes $g$ and yields $ \nfhl{ n }  \Rightarrow  \nfhl{ \nfhl{G} } $, with a distinguished start
nonterminal $\nfhl{n}$ and a normalized grammar $\nfhl{G}$. There are seven cases
for the seven context-free expression constructors, and each case involves
allocating a fresh nonterminal ($\nfhl{n}$ or $\nfhl{ {\nfhl{\alpha} } }$) to use as the
start symbol.
The cases with sub-expressions ($ g_{{\mathrm{1}}}  \cdot  g_{{\mathrm{2}}} $, $ g_{{\mathrm{1}}}  \lor  g_{{\mathrm{2}}} $ and
$ \mu \alpha .~ g $) are defined compositionally in terms of the
normalization of those sub-expressions.
Since normalization simply merges together all the production sets
resulting from sub-expressions, the situation frequently arises where
some productions are not reachable from the start symbol; the definition
here ignores this issue, since it is easy to trim unreachable
productions in the implementation.

Rules \repsilon, \rchar, and \rbot are 
straightforward.
For each of $ \epsilon $ and $t$, normalization produces a
grammar with a single production whose right-hand side is
$ \epsilon $ or $t$ respectively.
For $ \bot $, normalization produces an empty grammar, with a start
symbol and no productions.

Normalization of $ g_{{\mathrm{1}}}  \cdot  g_{{\mathrm{2}}} $ (rule \rseq)
is defined compositionally in terms of the
normalization of $g_{{\mathrm{1}}}$ and $g_{{\mathrm{2}}}$,
which produces start symbols $\nfhl{n_{{\mathrm{1}}}}$ and $\nfhl{n_{{\mathrm{2}}}}$ respectively.
We then want $ { \color{nfcolor}   n   \rightarrow    \nfhl{  \nfhl{ n_{{\mathrm{1}}} }  ~ n_{{\mathrm{2}}} }   } $, i.e.:

\newcommand{\rseqone}{\rulename{seq1}}
\begin{center}
  \begin{tabular}{llll}
    \rseqone \qquad  & $ {\mathcal{N} }  \llbracket \,   g_{{\mathrm{1}}}  \cdot  g_{{\mathrm{2}}}   \, \rrbracket $     & $=$  & $n  \Rightarrow   \{  { \color{nfcolor}   n   \rightarrow    \nfhl{  \nfhl{ n_{{\mathrm{1}}} }  ~ n_{{\mathrm{2}}} }   }   \}  \cup   \nfhl{ \nfhl{G}_{{\mathrm{1}}}   \cup   \nfhl{G}_{{\mathrm{2}}} }  $
                                                        \qquad where \enskip  $ {\mathcal{N} }  \llbracket \,  g_{\ottmv{i}}  \, \rrbracket  =  \nfhl{ n_{\ottmv{i}} }  \Rightarrow  \nfhl{ \nfhl{G}_{\ottmv{i}} } , i = 1, 2$ \\
  \end{tabular}
\end{center}
However, while this is semantically correct, $ { \color{nfcolor}   n   \rightarrow    \nfhl{  \nfhl{ n_{{\mathrm{1}}} }  ~ n_{{\mathrm{2}}} }   } $ is not
in normal form.
Therefore, rule \rseq instead copies each production
$\nfhl{\nfhl{N}_{{\mathrm{1}}}}$ of $\nfhl{n_{{\mathrm{1}}}}$, appending to each the start
symbol $\nfhl{n_{{\mathrm{2}}}}$, producing $ \nfhl{ \nfhl{N}_{{\mathrm{1}}} ~ n_{{\mathrm{2}}} } $.
Rule \ralt is similar, with
normalization merging the productions for the start symbols $\nfhl{n_{{\mathrm{1}}}}$ of $g_{{\mathrm{1}}}$
and $\nfhl{n_{{\mathrm{2}}}}$ of $g_{{\mathrm{2}}}$ into the productions for the new
start symbol $\nfhl{n}$.

Finally, rules \rfix and \rvar deal with the binding fixed point
operator $ \mu \alpha .~ g $ and with bound variables $\alpha$.
In rule \rfix,
we assume we can always rename bound variables to avoid clashes.
Normalizing $ \mu \alpha .~ g $ takes place
in two stages.
First, the body $g$ is normalized, yielding a start symbol
$\nfhl{n}$.
Then, according to the semantics of fixed point, we should
proceed to tie the knot by producing $ { \color{nfcolor}    {\nfhl{\alpha} }    \rightarrow    \nfhl{ n }   } $ and return $ {\nfhl{\alpha} } $
as the start symbol. That is:

\newcommand{\rfixone}{\rulename{fix1}}
\begin{center}
\begin{tabular}{llll}
  \rfixone \qquad  & $ {\mathcal{N} }  \llbracket \,   \mu \alpha .~ g   \, \rrbracket $     & $=$  & $ {\nfhl{\alpha} }   \Rightarrow  \{  { \color{nfcolor}    {\nfhl{\alpha} }    \rightarrow    \nfhl{ n }   }  \}  \cup  \nfhl{G} $
                                            \qquad where \enskip  $ {\mathcal{N} }  \llbracket \,  g  \, \rrbracket  =  \nfhl{ n }  \Rightarrow  \nfhl{ \nfhl{G} } $ \\
\end{tabular}
\end{center}
However, $ { \color{nfcolor}    {\nfhl{\alpha} }    \rightarrow    \nfhl{ n }   } $ is
not in normal form so, as with \rseq,
we instead copy the productions for $n$ into the rules for
$ {\nfhl{\alpha} } $:

\newcommand{\rfixtwo}{\rulename{fix2}}
\begin{center}
  \begin{tabular}{llll}
    \rfixtwo \qquad  & $ {\mathcal{N} }  \llbracket \,   \mu \alpha .~ g   \, \rrbracket $     & $=$  & $ {\nfhl{\alpha} }   \Rightarrow  \{  { \color{nfcolor}    {\nfhl{\alpha} }    \rightarrow   \nfhl{N}  }  \mid  { \color{nfcolor}   n   \rightarrow   \nfhl{N}  }  \in \nfhl{G} \}  \cup  \nfhl{G} $
                                                             \qquad where \enskip  $ {\mathcal{N} }  \llbracket \,  g  \, \rrbracket  =  \nfhl{ n }  \Rightarrow  \nfhl{ \nfhl{G} } $ \\
  \end{tabular}
\end{center}
But there is some extra work. In
particular, productions in $\nfhl{G}$ might start with $ {\nfhl{\alpha} } $ (e.g.~$ { \color{nfcolor}   n'   \rightarrow    \nfhl{  \nfhl{  {\nfhl{\alpha} }  }  ~ \overline{n}' }   } $).
While such form is allowed by the syntax of
$\nfhl{N}$, our ultimate goal is to get rid of it and turn the productions into DGNF.
Now that we learn the rules of $ {\nfhl{\alpha} } $, we can look up and substitute
in $\nfhl{G}$ all productions that start with $ {\nfhl{\alpha} } $. For example, if
$ { \color{nfcolor}    {\nfhl{\alpha} }    \rightarrow    \nfhl{  \nfhl{\textsc{b} }  }   } $ and $ { \color{nfcolor}   n'   \rightarrow    \nfhl{  \nfhl{  {\nfhl{\alpha} }  }  ~ \overline{n} }   } $, then after substitution we have
$ { \color{nfcolor}   n'   \rightarrow    \nfhl{  \nfhl{  \nfhl{\textsc{b} }  }  ~ \overline{n} }   } $. Note that $ {\nfhl{\alpha} } $, as a special kind
of nonterminal, may still appear in the middle of a
production; for example, $ { \color{nfcolor}   n'   \rightarrow    \nfhl{  \nfhl{ t }  ~  {\nfhl{\alpha} }  }   } $ won't get
substituted. Performing the substitution would not be correct:
if $ { \color{nfcolor}    {\nfhl{\alpha} }    \rightarrow    \nfhl{  \nfhl{\textsc{b} }  }   } $, then after substitution $ { \color{nfcolor}   n'   \rightarrow    \nfhl{  \nfhl{ t }     \nfhl{  \nfhl{\textsc{b} }  }  }   } $ is not in
DGNF.

Rule \rfix in \Cref{figure:normalization} presents the final form of
normalizing a fixed point. \circled{1} first copies the
productions for $n$ into the rules for $ {\nfhl{\alpha} } $,
then \circled{2} substitutes in $\nfhl{G}$ all productions that start with
$ {\nfhl{\alpha} } $,
and \circled{3} finally adds back all productions in $\nfhl{G}$ that
do not start with $ {\nfhl{\alpha} } $. As we will see, rule \rfix
effectively guarantees that normalizing closed context-free
expressions produces DGNF.

Lastly, rule \rvar creates the singleton production $ { \color{nfcolor}   n   \rightarrow    \nfhl{  {\nfhl{\alpha} }  }   } $.
Combining \rfix with \rvar, normalization treats $ {\nfhl{\alpha} } $
as a placeholder for the productions denoted by the fixed point. Once
$ {\nfhl{\alpha} } $ is known, it is replaced with its productions if
necessary (as in rule \rfix). It is tempting to return
$ \nfhl{  {\nfhl{\alpha} }  }  \Rightarrow  \nfhl{  \nfhl{ \emptyset }  } $ with $ {\nfhl{\alpha} } $ as a start symbol and no productions, but that
is incorrect: $ \nfhl{  {\nfhl{\alpha} }  }  \Rightarrow  \nfhl{  \nfhl{ \emptyset }  } $ means an empty grammar.
\begin{figure*}
\renewcommand{\rulehl}[2][gray!15]{\colorbox{#1}{$\displaystyle#2$}}
\scalebox{0.96}{
\begin{mathpar}
  \hspace{-10pt}
  \footnotesize
  \inferrule{
    \inferrule*{
      \inferrule*{}{
        \inferrule*{
          \inferrule*{
            \inferrule*{}
            {
               {\mathcal{N} }  \llbracket \textsc{lpar} \rrbracket = \cdots
            }
            \and
            \inferrule*{}
            {
              \inferrule{
                \inferrule{}
                {
                  \cdots
                }
              }
              {
               {\mathcal{N} }  \llbracket \mu \text{
                ss }. \,\epsilon \lor \text{ s } \cdot \text{ ss }
               \rrbracket
               = \nfhl{
                 \text{ss}  \Rightarrow  \{ \,  \text{ss}  \rightarrow   \epsilon , \text{ss}
                \rightarrow  \text{s}\, \text{ss}
                \,\}
                }
              }
            }
          }{
             {\mathcal{N} }  \llbracket \textsc{lpar} \cdot (\mu \text{
              ss }. \,\epsilon \lor \text{ s } \cdot \text{ ss }) \rrbracket
            = \nfhl{
            n_{{\mathrm{1}}}  \Rightarrow  \{ \, \rulehl{n_{{\mathrm{1}}}  \rightarrow  \textsc{lpar}\, \text{ss}},
            \text{ss}  \rightarrow   \epsilon ,
            \text{ss}  \rightarrow  \text{s}\, \text{ss}
            \,\}
            }
          }
          \and
          \inferrule*{}
          {
             {\mathcal{N} }  \llbracket \textsc{rpar} \rrbracket = \cdots
          }
        }
        {
         {\mathcal{N} }  \llbracket \textsc{lpar} \cdot (\mu \text{
          ss }. \,\epsilon \lor \text{ s } \cdot \text{ ss }) \cdot
        \textsc{rpar} \rrbracket
        = \nfhl{
          n_{{\mathrm{2}}}  \Rightarrow  \{ \, \rulehl{n_{{\mathrm{2}}}  \rightarrow  \textsc{lpar}\, \text{ss}\, \text{rpar}},
          \text{ss}  \rightarrow   \epsilon ,
          \text{ss}  \rightarrow  \text{s}\, \text{ss},
          \rulehl{\text{rpar}   \rightarrow  \textsc{rpar}}
        \,\}
        }
        }
      }
      \and
      \inferrule{}{
         {\mathcal{N} }  \llbracket \textsc{atom} \rrbracket = \cdots
      }
    }
    {
     {\mathcal{N} }  \llbracket (\textsc{lpar} \cdot (\mu \text{
      ss }. \,\epsilon \lor \text{ s } \cdot \text{ ss }) \cdot
    \textsc{rpar}) \lor \textsc{atom} \rrbracket
    = \nfhl{
      n_{{\mathrm{3}}}  \Rightarrow  \{ \,
      \rulehl{n_{{\mathrm{3}}}  \rightarrow  \textsc{lpar}\, \text{ss}\, \text{rpar}},
    \rulehl{n_{{\mathrm{3}}}  \rightarrow  \textsc{atom}},
    \text{ss}  \rightarrow   \epsilon , \text{ss}
     \rightarrow  \text{s}\, \text{ss},
    \text{rpar}  \rightarrow  \textsc{rpar}
    \,\}
    }
    }
  }
  {  {\mathcal{N} }  \llbracket
    g
    \rrbracket
    = \nfhl{
      \text{s}  \Rightarrow  \{ \, \rulehl{\text{s}  \rightarrow  \textsc{lpar}\, \text{ss}\, \text{rpar}},
    \rulehl{\text{s}  \rightarrow  \textsc{atom}},
    \text{ss}  \rightarrow   \epsilon ,
    \rulehl{\text{ss}  \rightarrow  \textsc{lpar}\, \text{ss} \, \text{rpar} \, \text{ss}},
    \rulehl{\text{ss}  \rightarrow  \textsc{atom}\, \text{ss}},
    \text{rpar}  \rightarrow  \textsc{rpar}
    \,\}
    }
  }
\end{mathpar}
}
\smallskip
\caption{Normalizing s-expression $g= \mu \text{ s }. (\textsc{lpar} \cdot (\mu \text{ ss }. \,\epsilon \lor \text{ s } \cdot \text{ ss }) \cdot \textsc{rpar}) \lor \textsc{atom}$}
\label{figure:norm:example}
\end{figure*}
 
\paragraph{Example.}
\Cref{figure:norm:example}
presents a simplified normalization derivation for the grammar in \Cref{fig:eg:grammar}:
\[ g =
\mu \text{ s }. (\textsc{lpar} \cdot (\mu \text{ ss }. \,\epsilon \lor \text{ s } \cdot \text{ ss }) \cdot \textsc{rpar}) \lor \textsc{atom}
\]
For space reasons, we write s for sexp, and ss for sexps.
We highlight new productions generated during derivation in light gray,
and omit some details via $\cdots$ for space reasons and also since normalizing
tokens is straightforward;
the complete derivation tree is given in the appendix\shortOnly{ in the extended version of the paper}.
Of particular interest is the last step, which normalizes a
fixed point. In this case, \nfhl{s} is used as the variable bound by the fixed
point, and we have a non-DGNF production $\nfhl{\text{ss}  \rightarrow  \text{s}\,
\text{ss}}$.
First, \nfhl{s} copies all productions from $\nfhl{n_{{\mathrm{3}}}}$. Then, since
$\nfhl{\text{ss}  \rightarrow  \text{s}\, \text{ss}}$ starts with \nfhl{s},
it expands to
two productions where \nfhl{s} is replaced by its two normal forms
respectively.

\subsection{Semantics of DGNF}
\label{sec:semantics-dgnf}

Recall that \Cref{sec:overview:dgnf} gave a high-level description of DGNF.
This section defines the formal semantics of DGNF.
We start with the expansion relation:

\begin{definition}[Expansion ($\nfhl{G}  \vdash   \leadsto $)]
  \label{def:expansion}
  Given a grammar $\nfhl{G}$, we define the expansion relation by
  (1) (Base) $ \nfhl{G}   \vdash  \nfhl{ n }  \leadsto    \nfhl{ n }  $;
  (2) (Step) if $ \nfhl{G}   \vdash  \nfhl{ n }  \leadsto    \nfhl{ \overline{t} }  \,  \nfhl{ n_{{\mathrm{1}}} }  \,  \nfhl{ \overline{n} }   $ and $( { \color{nfcolor}   n_{{\mathrm{1}}}   \rightarrow   \nfhl{N}  }  \, \in \, \nfhl{G})$,
    then $\nfhl{G}  \vdash  \nfhl{n}  \leadsto   \nfhl{ \overline{t} }  \,  \nfhl{ \nfhl{N} }  \,  \nfhl{ \overline{n} } $.
    We write $ \nfhl{G}   \vdash  \nfhl{ n }  \leadsto   w $ when $\nfhl{n}$ expands to a complete word $w$.
\end{definition}

The expansion relation essentially captures what a nonterminal can expand to.
For example, if $ { \color{nfcolor}   n   \rightarrow    \nfhl{  \nfhl{  \nfhl{\textsc{b} }  }  ~ n_{{\mathrm{1}}} }   }  \, \in \, \nfhl{G}$ and $ { \color{nfcolor}   n_{{\mathrm{1}}}   \rightarrow    \nfhl{  \nfhl{\textsc{c} }  }   }  \, \in \, \nfhl{G}$, then $ \nfhl{G}   \vdash  \nfhl{ n }  \leadsto    \nfhl{  \nfhl{  \nfhl{\textsc{b} }  }     \nfhl{  \nfhl{\textsc{c} }  }  }  $. We enforce a left-to-right expansion order for clarity and to
stay close to the parsing behavior, but
that is not necessary: it is easy to imagine an arbitrary order expansion, but
any order leads to the same set of words.

With the notion of expansion, we define what it means for a grammar to be in
DGNF precisely.

\begin{definition}[Deterministic Greibach Normal Form]
  \label{def:dgnf:new}
  A grammar $\nfhl{G}$ is in DGNF (i.e.~it is a $\nfhl{D}$ grammar), if
  all productions are either of form $\nfhl{ { \color{nfcolor}   n   \rightarrow    \nfhl{  \nfhl{ t }  ~ \overline{n} }   } }$ or $ { \color{nfcolor}   n   \rightarrow   \epsilon  } $, and moreover,
\begin{itemize}
  \item (Determinism)
    for any pair of a nonterminal $\nfhl{n}$ and a terminal $\nfhl{t}$,
    if there are two distinct productions $\ottsym{(}   { \color{nfcolor}   n   \rightarrow    \nfhl{  \nfhl{ t_{{\mathrm{1}}} }  ~ \overline{n}_{{\mathrm{1}}} }   }   \ottsym{)} \, \in \, \nfhl{G}$
    and $\ottsym{(}   { \color{nfcolor}   n   \rightarrow    \nfhl{  \nfhl{ t_{{\mathrm{2}}} }  ~ \overline{n}_{{\mathrm{2}}} }   }   \ottsym{)} \, \in \, \nfhl{G}$,
    we have $\nfhl{t_{{\mathrm{1}}}} \neq \nfhl{t_{{\mathrm{2}}}}$;
  \item (Guarded $\epsilon$-productions)
    if $ \nfhl{G}   \vdash  \nfhl{ n }  \leadsto    \nfhl{ \overline{t} }  \,  \nfhl{ n_{{\mathrm{1}}} \, n_{{\mathrm{2}}} }  \,  \nfhl{ \overline{n} }  $
    and $\ottsym{(}   { \color{nfcolor}   n_{{\mathrm{1}}}   \rightarrow   \epsilon  }   \ottsym{)} \, \in \, \nfhl{G}$,
    then for any $\nfhl{t}$, either
    $\ottsym{(}   { \color{nfcolor}   n_{{\mathrm{1}}}   \rightarrow    \nfhl{  \nfhl{ t }  ~ \overline{n}_{{\mathrm{1}}} }   }   \ottsym{)} \, \notin \, \nfhl{G}$
    or $\ottsym{(}   { \color{nfcolor}   n_{{\mathrm{2}}}   \rightarrow    \nfhl{  \nfhl{ t }  ~ \overline{n}_{{\mathrm{2}}} }   }   \ottsym{)} \, \notin \, \nfhl{G}$ for any $\nfhl{\overline{n}_{{\mathrm{1}}}}$, $\nfhl{\overline{n}_{{\mathrm{2}}}}$.
\end{itemize}
\end{definition}

The Determinism condition is straightforward, while
the Guarded $\epsilon$-productions condition needs more explanation. In
\Cref{sec:overview:dgnf}, we mentioned that the $\epsilon$-production
may only be used when no terminal symbol in other productions matches the input
string. Consider that the next token to match is $ \nfhl{\textsc{c} } $. The case when both the
$\epsilon$-production $ { \color{nfcolor}   n_{{\mathrm{1}}}   \rightarrow   \epsilon  } $ and a production $ { \color{nfcolor}   n_{{\mathrm{1}}}   \rightarrow    \nfhl{  \nfhl{\textsc{c} }  }   } $ can
match raises when $\nfhl{n_{{\mathrm{1}}}}$'s follow-up nonterminal $\nfhl{n_{{\mathrm{2}}}}$ can also match
$ \nfhl{\textsc{c} } $, making it possible to use the $ \epsilon $-production while
$ { \color{nfcolor}   n_{{\mathrm{1}}}   \rightarrow    \nfhl{  \nfhl{\textsc{c} }  }   } $ also matches. \Cref{def:dgnf:new} captures such cases,
requiring that $\nfhl{n_{{\mathrm{1}}}}$ and $\nfhl{n_{{\mathrm{2}}}}$ cannot match the same terminal if $\nfhl{n_{{\mathrm{1}}}}$
has an $ \epsilon $-production, and thus rules out example (4) in \Cref{sec:overview:dgnf}.

Now we can formally define the important property of DGNF that makes it
practically useful.

\begin{restatable}[Deterministic Parsing]{theorem}{theoremdetparsing}
  If $\nfhl{G}$ is a DGNF grammar,
  then for any expansion $ \nfhl{G}   \vdash  \nfhl{ n }  \leadsto   w $,
  there is a unique derivation for this expansion.
\end{restatable}

\subsection{Well-definedness and Correctness}
\label{section:normalization-properties}

Since normalization serves as the basis for the parsing
algorithm, establishing its correctness is crucial for
\FLaP{}.
In this section, we prove three key properties of
normalization:
normalization always succeeds for well-typed expressions
(\Cref{section:totality});
the normalization result does not include the internal form $\nfhl{ {\nfhl{\alpha} }  \, \overline{n}}$ (\Cref{sec:norm:no:internal});
and the result of normalization is a DGNF grammar (\Cref{sec:norm:norm-form}).

\subsubsection{Normalization is well-defined}
\label{section:totality}

To understand what well-definedness means,
consider
normalizing $ g_{{\mathrm{1}}}  \cdot  g_{{\mathrm{2}}} $. Rule \rseq returns $ \nfhl{ \nfhl{N}_{{\mathrm{1}}} ~ n_{{\mathrm{2}}} } $
with $ { \color{nfcolor}   n_{{\mathrm{1}}}   \rightarrow   \nfhl{N}_{{\mathrm{1}}}  } $ from $g_{{\mathrm{1}}}$, and $\nfhl{n_{{\mathrm{2}}}}$ from $g_{{\mathrm{2}}}$.
However,
in order for $ \nfhl{ \nfhl{N}_{{\mathrm{1}}} ~ n_{{\mathrm{2}}} } $ to be well-formed, we
must ensure that $\nfhl{\nfhl{N}_{{\mathrm{1}}}}$ is not $\nfhl{ \epsilon }$, or otherwise
$\nfhl{ \epsilon ~n_{{\mathrm{2}}}}$ is ill-formed.
The case for sequencing is one of
several places the typing information is useful.
In particular, if $ g_{{\mathrm{1}}}  \cdot  g_{{\mathrm{2}}} $ is well-typed,
then the typing rule for sequencing
(\Cref{figure:cfe-typing}) guarantees
$\tau_1 \seq \tau_2$, which says $\lnot \tau_1.\Null$.
We then prove below that if an
expression is not nullable, its normalization cannot have
an $ \epsilon $-production. Thus
$\nfhl{\nfhl{N}_{{\mathrm{1}}}}$ cannot be
$\nfhl{ \epsilon }$, ensuring that the normalization result 
is in normal form.

\begin{restatable}[Productions of $\Null$]{lemma}{lemmatotalnullfalse}
  \label{lemma:total:null:false}
  Given $\Gamma  \ottsym{;}  \Delta  \vdash  g  \ottsym{:}  \tau$
  and $ {\mathcal{N} }  \llbracket \,  g  \, \rrbracket $ returns $ \nfhl{ n }  \Rightarrow  \nfhl{ \nfhl{G} } $,
  we have
  $\tau.{\Null} = \True$
  if and only if
  (1) $ { \color{nfcolor}   n   \rightarrow   \epsilon  }  \in \nfhl{G}$;
  or (2) $ { \color{nfcolor}   n   \rightarrow    \nfhl{  {\nfhl{\alpha} }  }   }  \in \nfhl{G}$
  where $\ottsym{(}  \alpha  \ottsym{:}  \tau'  \ottsym{)} \, \in \, \Gamma$ and $\tau'.{\Null} = \True$.
  In other words, if $\tau.{\Null} = \False$,
  then $ { \color{nfcolor}   n   \rightarrow   \epsilon  }  \notin \nfhl{G}$.
\end{restatable}

With \Cref{lemma:total:null:false} and similar reasoning about typing for other
constructs (such as alternations), we prove that normalization is well-defined for well-typed
expressions.

\begin{restatable}[Well-definedness]{theorem}{lemmatotality}
  \label{lemma:well-definedness}
  If $\Gamma  \ottsym{;}  \Delta  \vdash  g  \ottsym{:}  \tau$,
  then $ {\mathcal{N} }  \llbracket \,  g  \, \rrbracket $ returns $ \nfhl{ n }  \Rightarrow  \nfhl{ \nfhl{G} } $
  for some $\nfhl{G}$ and $n$.
\end{restatable}

\subsubsection{Normalizing closed expressions produces no $\nfhl{ {\nfhl{\alpha} }  \, \overline{n}}$ form}
\label{sec:norm:no:internal}

\Cref{lemma:well-definedness} says that if an expression is well-typed then
normalizing it returns a grammar $\nfhl{G}$. However, $\nfhl{G}$ may include
$ { \color{nfcolor}   n   \rightarrow    \nfhl{  \nfhl{  {\nfhl{\alpha} }  }  ~ \overline{n} }   } $, which is not valid DGNF.
In this part, we prove that normalizing \textit{closed} well-typed
expressions will not generate $\nfhl{ {\nfhl{\alpha} }  \, \overline{n}}$ productions.
To do so, we need to reason about the occurrences of $ {\nfhl{\alpha} } $. The
following lemma says that every $ {\nfhl{\alpha} } $ returned as the head of a
production must be in the typing context.

\begin{restatable}[Internal normal form]{lemma}{lemmanafv}
  \label{lemma:na:fv}
  Given $\Gamma  \ottsym{;}  \Delta  \vdash  g  \ottsym{:}  \tau$
  and $ {\mathcal{N} }  \llbracket \,  g  \, \rrbracket $ returns $ \nfhl{ n }  \Rightarrow  \nfhl{ \nfhl{G} } $,
  \begin{itemize}
    \item if $\ottsym{(}   { \color{nfcolor}   n   \rightarrow    \nfhl{  \nfhl{  {\nfhl{\alpha} }  }  ~ \overline{n} }   }   \ottsym{)} \, \in \, \nfhl{G}$, then $\alpha \in \mathsf{dom} \, \ottsym{(}  \Gamma  \ottsym{)}$;
    \item if $\ottsym{(}   { \color{nfcolor}   n'   \rightarrow    \nfhl{  \nfhl{  {\nfhl{\alpha} }  }  ~ \overline{n} }   }   \ottsym{)} \, \in \, \nfhl{G}$ for any $n'$, then
      $\alpha \in \mathsf{fv} \, \ottsym{(}  g  \ottsym{)} $, and thus $\alpha \in \mathsf{dom} \, \ottsym{(}  \Gamma  \ottsym{,}  \Delta  \ottsym{)}$.
  \end{itemize}
\end{restatable}
\noindent
Note that the first result applies only to the start symbol $\nfhl{n}$,
and its proof relies on the typing rule where $\alpha$ is well-typed
only if $\alpha \in \Gamma$ (\Cref{figure:cfe-typing}). 
The second
result applies to any $\nfhl{n'}$, and the most tricky case in the proof is
when normalizing $ \mu \alpha .~ g $, where we need to prove that the productions of the start symbol
of $g$ cannot start with $ {\nfhl{\alpha} } $, or otherwise normalizing $ \mu \alpha .~ g $ would copy all productions from $g$ for $ {\nfhl{\alpha} } $ which would result in
(e.g.) $ { \color{nfcolor}    {\nfhl{\alpha} }    \rightarrow    \nfhl{  {\nfhl{\alpha} }  }   } $ that fails the lemma as we are getting out of the scope of $ {\nfhl{\alpha} } $.
Fortunately, that is exactly what the first result tells us: when typing $ \mu \alpha .~ g $, we add $\alpha$ to $\Delta$ (\Cref{figure:cfe-typing}), and
thus normalizing $g$ cannot have $ {\nfhl{\alpha} } $ at the head of a
production for its start symbol.

Our goal then follows as a corollary of \Cref{lemma:na:fv}, which says
that normalizing any closed well-typed expression produces only the
desired normal forms.

\begin{restatable}[Normalizing without internal normal form]{corollary}{lemmaformcorrect}
  \label{lemma:form:correct}
  Given $\bullet  \ottsym{;}  \bullet  \vdash  g  \ottsym{:}  \tau$,
  if $ {\mathcal{N} }  \llbracket \,  g  \, \rrbracket $ returns $ \nfhl{ n' }  \Rightarrow  \nfhl{ \nfhl{G} } $,
  then any production in $\nfhl{G}$ is either $ { \color{nfcolor}   n   \rightarrow   \epsilon  } $
  or $ { \color{nfcolor}   n   \rightarrow    \nfhl{  \nfhl{ t }  ~ \overline{n} }   } $ for some $\nfhl{n}$, $\nfhl{t}$ and $\nfhl{\overline{n}}$.
\end{restatable}

\subsubsection{Normalization returns  DGNF grammars}
\label{sec:norm:norm-form}

Finally, we prove that normalization returns DGNF grammars. That
 requires productions to satisfy the conditions given in
 \Cref{def:dgnf:new}.

\textit{(1) Determinism}: 
We prove that
all $ { \color{nfcolor}   n   \rightarrow    \nfhl{  \nfhl{ t }  ~ \overline{n} }   } $ for the same $\nfhl{n}$ start
with different $\nfhl{t}$.
To this end,
we establish the relation between starting terminals in productions and the $\textsc{First}$ set
of types.
\begin{restatable}[Terminals in $\textsc{First}$]{lemma}{lemmaterminalfirst}
  \label{lemma:terminal:first}
  Given $\Gamma  \ottsym{;}  \Delta  \vdash  g  \ottsym{:}  \tau$
  and $ {\mathcal{N} }  \llbracket \,  g  \, \rrbracket $ returns $ \nfhl{ n }  \Rightarrow  \nfhl{ \nfhl{G} } $,
  we have
  $\nfhl{t} \in \tau.{\textsc{First}}$
  if and only if
  (1) $\ottsym{(}   { \color{nfcolor}   n   \rightarrow    \nfhl{  \nfhl{ t }  ~ \overline{n} }   }   \ottsym{)} \, \in \, \nfhl{G}$;
  or (2) $\ottsym{(}   { \color{nfcolor}   n   \rightarrow    \nfhl{  \nfhl{  {\nfhl{\alpha} }  }  ~ \overline{n} }   }   \ottsym{)} \, \in \, \nfhl{G}$
   where $\ottsym{(}  \alpha  \ottsym{:}  \tau'  \ottsym{)} \, \in \, \Gamma$ and $\nfhl{t}  \in  \tau'.{\textsc{First}}$.
\end{restatable}
\noindent
This lemma is particularly important when proving the case for normalizing
$ g_{{\mathrm{1}}}  \lor  g_{{\mathrm{2}}} $, where the typing condition $\tau_{{\mathrm{1}}} \apart \tau_{{\mathrm{2}}}$ ensures that
$g_{{\mathrm{1}}}$ and $g_{{\mathrm{2}}}$ have disjoint $\textsc{First}$, which in turn ensures that
rule \ralt only copies distinct head terminals from $g_{{\mathrm{1}}}$ and $g_{{\mathrm{2}}}$.

\textit{(2) Guarded  $\epsilon$-productions:} The proof is more involved, as it essentially
requires us to show that during expansion $ \nfhl{G}   \vdash  \nfhl{ n }  \leadsto    \nfhl{ \overline{t} }  \,  \nfhl{ n_{{\mathrm{1}}} \, n_{{\mathrm{2}}} }  \,  \nfhl{ \overline{n} }  $,
the $\First$ set of $\nfhl{n_{{\mathrm{1}}}}$ is disjoint with the $\First$ set of $\nfhl{n_{{\mathrm{2}}}}$, if
$\nfhl{n_{{\mathrm{1}}}}$ is nullable.
The proof relies on showing that ``expansion preserves typing''.
More concretely, think
from the well-typed context free expressions' point of view:
if $ \ottsym{(}   g_{{\mathrm{1}}}  \lor  g_{{\mathrm{2}}}   \ottsym{)}  \cdot  g_{{\mathrm{3}}} $
is well-typed, then $ g_{{\mathrm{1}}}  \cdot  g_{{\mathrm{3}}} $ (and $ g_{{\mathrm{2}}}  \cdot  g_{{\mathrm{3}}} $) must also be well-typed,
and going from $ \ottsym{(}   g_{{\mathrm{1}}}  \lor  g_{{\mathrm{2}}}   \ottsym{)}  \cdot  g_{{\mathrm{3}}} $ to $ g_{{\mathrm{1}}}  \cdot  g_{{\mathrm{3}}} $ is one step of branching,
similar to one step of expansion. 
We refer the
reader to the appendix\shortOnly{ of the extended version of the paper} for more details.

With all the conditions proved, we conclude our goal.

\begin{restatable}[$ {\mathcal{N} }  \llbracket \,  g  \, \rrbracket $ produces DGNF]{theorem}{theoremgnf}
  \label{theorem:dgnf}
  If $\bullet  \ottsym{;}  \bullet  \vdash  g  \ottsym{:}  \tau$,
  then $ {\mathcal{N} }  \llbracket \,  g  \, \rrbracket $ returns $  \nfhl{ n }  \Rightarrow  \nfhl{ \nfhl{D} } $ for some $\nfhl{n}$, $\nfhl{\nfhl{D}}$.
\end{restatable}

\subsection{Normalization Soundness}
\label{sec:norm:correctness}

Our final piece of normalization metatheory establishes that
normalization is sound with respect to the
denotational semantics of typed context-free expressions.
The denotational semantics $ { {\llbracket  g  \rrbracket}_{ \gamma } } $
interprets $g$ as a language (i.e.~a set of strings matched
by $g$), where $\gamma$ interprets free variables in $g$:

\begin{center}
\begin{tabular}{llllll} \toprule
  \enskip $ { {\llbracket  \epsilon  \rrbracket}_{ \gamma } } $  & = &  $\{  \epsilon  \}$ &
  \enskip $ { {\llbracket   g_{{\mathrm{1}}}  \cdot  g_{{\mathrm{2}}}   \rrbracket}_{ \gamma } } $  & = &  $\{ w \cdot w' \mid  w \in  { {\llbracket  g_{{\mathrm{1}}}  \rrbracket}_{ \gamma } }  \land w' \in  { {\llbracket  g_{{\mathrm{2}}}  \rrbracket}_{ \gamma } }   \}$\\

  \enskip $ { {\llbracket   t   \rrbracket}_{ \gamma } } $  & = &  $\{ t\}$ &
  \enskip $ { {\llbracket  \alpha  \rrbracket}_{ \gamma } } $  & = &  $ \gamma(a) $ \\

  \enskip $ { {\llbracket  \bot  \rrbracket}_{ \gamma } } $  & = &  $\emptyset$ &
  \enskip $ { {\llbracket   \mu \alpha .~ g   \rrbracket}_{ \gamma } } $  & =  & $\mathsf{fix}(\lambda \mathtt{L}.  { {\llbracket  g  \rrbracket}_{ \ottsym{(}  \gamma  \ottsym{,}   \mathtt{L}  /  \alpha   \ottsym{)} } } )$ \\

  \enskip $ { {\llbracket   g_{{\mathrm{1}}}  \lor  g_{{\mathrm{2}}}   \rrbracket}_{ \gamma } } $  & = &  $ { {\llbracket  g_{{\mathrm{1}}}  \rrbracket}_{ \gamma } }  \cup  { {\llbracket  g_{{\mathrm{2}}}  \rrbracket}_{ \gamma } }  $ \qquad \quad &

                                                                                 \multicolumn{3}{l}{ \quad
                                                                                 \multirow{2}{*}{
  \enskip $\mathsf{fix}(f)  = \bigcup\limits_{i \in \mathbb{N}} \mathtt{L}_i$ where $\begin{array}{ll}
                                                                          \mathtt{L}_0 &  = \emptyset\\
                                                                          \mathtt{L}_{i+1} &  = f(\mathtt{L}_i) \\
                                                                  \end{array}$
                                                                                               }}
  \\[12pt] \bottomrule \\
\end{tabular}
\end{center}

\noindent
Most cases are straightforward:
$ \epsilon $ denotes the singleton set containing the empty
string, $t$ the singleton containing the one-token string $t$,
$ \bot $ the empty language, and $ g_{{\mathrm{1}}}  \lor  g_{{\mathrm{2}}} $ a union of sets. The
interpretation of $ g_{{\mathrm{1}}}  \cdot  g_{{\mathrm{2}}} $ appends a string from $g_{{\mathrm{1}}}$ to a string from
$g_{{\mathrm{2}}}$. Variables $\alpha$ draw interpretations from the environment
$\gamma$, and $ \mu \alpha .~ g $ denotes the least fixed point of $g$
with respect to $\alpha$.

To prove that our normalization is sound, we show that the normalized DGNF
denotes exactly the same language as the denotation semantics of an expression.
Recall that we have defined the expansion relation in \Cref{def:expansion}, where
$ \nfhl{G}   \vdash  \nfhl{ n }  \leadsto   w $ denotes that $n$ expands to a
complete string $w$, where all non-terminals have been expanded.
We prove the normalized grammar can expand to a string \textit{if and only if}
the string is included in the denotational semantics of the expression.
The proof is done by induction first on the length of $w$
and then on the structure of $g$.

\begin{restatable}[Soundness]{theorem}{theoremcorrectness}
  Given $\bullet  \ottsym{;}  \bullet  \vdash  g  \ottsym{:}  \tau$
  and $ {\mathcal{N} }  \llbracket \,  g  \, \rrbracket  $ returns $  \nfhl{ n }  \Rightarrow  \nfhl{ \nfhl{G} } $,
  we have $w \, \in \,  { {\llbracket  g  \rrbracket}_{ \bullet } } $ if and only if
  $ \nfhl{G}   \vdash  \nfhl{ n }  \leadsto   w $
  for any $w$.
\end{restatable}

\subsection{Implementation}
\label{section:implementing-normalization}

The compositionality of the normalization algorithm simplifies the
implementation of normalization in \FLaP{}.
For example, if $g$ and $g'$ are \FLaP{} parsers in normal form, then
$g >>> g'$ is also a parser in normal form, built from $g$ and $g'$
using the rules in \Cref{figure:normalization}.

Unsurprisingly, the most intricate part of the algorithm --- dealing
with fixed points --- is also the subtlest part of the implementation.
The implementation follows the formal algorithm closely, inserting
placeholders ($ {\nfhl{\alpha} } $s) that are tracked using an environment and resolved later.
This kind of ``backpatching'' mirrors the way in which recursion is
commonly implemented in eager functional languages such as
OCaml~\cite{DBLP:journals/pacmpl/ReynaudSY21}; if \FLaP{} were instead
implemented in a lazy language then it would be possible to implement
fixed point normalization with less fuss.

\section{Fusion}
\label{section:fusion}

This section shows how \FLaP{} fuses a separately-defined
lexer and normalized parser,
eliminating tokens from generated code altogether.

\paragraph{Canonicalizing lexer}
\label{sec:canonical-lexer}

We use \textit{canonicalized} lexers:
we assume
that rules are disjoint on the left
(i.e.~there is no string that is matched by more than one regular
expression in a set of rules),
and on the right (i.e.~there is exactly one Skip
rule, and no token appears in more than one Return rule).
Negation and intersection make it easy to transform a lexer that
does not obey these constraints into an equivalent lexer that does, so
there is no need to restrict the interface exposed to the user.

\paragraph{The fusion algorithm}
\label{section:fusion-algorithm}

\newcommand{\Fuse}[1]{{\mathcal{F} \llbracket\, \lexerhl{L}, \nfhl{#1}\, \rrbracket }}

\begin{figure}
\begin{tabular}{llll}
 fused grammar  & $\color{fusedcolor}F$ & $\Coloneqq$ & $ \color{fusedcolor} \{\, { \color{fusedcolor}  n   \rightarrow   \ottnt{r} \, \overline{n}  }  \,\} \cup \{\,   { \color{fusedcolor}  n   \rightarrow    ? r   }  \,\}  $ \\
\end{tabular}

\smallskip

\setlength{\tabcolsep}{2pt}
\begin{tabular}{rllllll}
    $\mathcal{F} \llbracket\, \lexerhl{L}, \nfhl{G} \,\rrbracket$ &  $=$ & \multicolumn{3}{l}{$\mathcal{F}_1 \cup \mathcal{F}_2 \cup \mathcal{F}_3$} \\
     where & $ \mathcal{F}_{1}$  & $=$  & $ \{ \,   { \color{fusedcolor}  n   \rightarrow   \ottnt{r} \, \overline{n}  }   $ & $\mid$ & $  { \color{lexercolor}  \ottnt{r}  \Rightarrow  \mathbf{Return} ~ t  }  \in \lexerhl{L} \land  { \color{nfcolor}   n   \rightarrow    \nfhl{  \nfhl{ t }  ~ \overline{n} }   }  \in {\nfhl{G}}      \,\} $ \qquad\qquad & \sshl{\textit{(inline the lexer)}}  \\
                & $ \mathcal{F}_{2}$  & $=$  & $\{ \,    { \color{fusedcolor}  n   \rightarrow   \ottnt{r} \, n  }  $ & $ \mid $ & $   { \color{lexercolor}  \ottnt{r}   \Rightarrow   \mathbf{Skip}   }  \in \lexerhl{L} \land \nfhl{n} \in {\nfhl{G}} \, \} $ & \sshl{\textit{(whitespace)}} \\
                & $ \mathcal{F}_{3}$  & $=$  & $ \{ \,  \fusedhl{n  \rightarrow   ? \lnot {\ottnt{r}} } $ & $\mid$ & $  { \color{nfcolor}   n   \rightarrow   \epsilon  }  \, \in \, \nfhl{G} \land \fusedhl{{\ottnt{r}}} = \bigvee \{ \, \fusedhl{\ottnt{r}} \mid  { \color{fusedcolor}  n   \rightarrow   \ottnt{r} \, \overline{n}  }  \in \mathcal{F}_1 \cup \mathcal{F}_2 \,\}  \} $ & \sshl{\textit{(epsilon productions)}}
\end{tabular}
\caption{\label{figure:fusion-algorithm}Lexer-parser fusion}
\end{figure}

\Cref{figure:fusion-algorithm} formally defines the fusion
algorithm.
$\Fuse{\nfhl{G}}$, which operates on a
canonicalized
lexer $\lexerhl{L}$ and a normalized grammar $\nfhl{G}$,
yielding a fused grammar $\color{fusedcolor}F$.

The fused result consists of three parts.
First, we
replace each production $ { \color{nfcolor}   n   \rightarrow    \nfhl{  \nfhl{ t }  ~ \overline{n} }   } $ with a new production
$ { \color{fusedcolor}  n   \rightarrow   \ottnt{r} \, \overline{n}  } $, retrieving the regex $\lexerhl{r}$ that is associated
with the token $\lexerhl{t}$ in the lexer $\lexerhl{L}$ ($\mathcal{F}_1$).
This is where
the fusion function implicitly specializes the lexer to each
nonterminal in the normalized grammar, and discards
lexing rules that
return tokens not in productions for the nonterminal.
Canonicalizing the lexer to enforce disjointness simplifies this
discarding of rules.

Then, we
add an additional production $ { \color{fusedcolor}  n   \rightarrow   \ottnt{r} \, n  } $ for the \textbf{skip}
regex $\lexerhl{r}$ (which may be $\lexerhl{\bot}$) for each nonterminal,
allowing each nonterminal to match an arbitrary number of the skip regex ($\mathcal{F}_2$).

Finally,
for nonterminals with an $ \epsilon $-production,
the discarded regexes, along with the skip regex, are incorporated into a lookahead
regex ($\mathcal{F}_3$).
That is,
we add a lookahead
production $\fusedhl{n  \rightarrow  ? \lnot {\ottnt{r}}}$ for the regex
that is the complement of the regexes that appear in other productions for $\fusedhl{n}$.

Fusion with normalized grammars is strikingly simple;
it would be much more involved to directly fuse context-free
expressions with the lexing rules.
As with normalized grammars, an expansion relation for
fused grammars would guarantee that every expansion has a
unique derivation.

\section{Implementation of parsing}
\label{section:fused-parsing}

\newcommand{\dyn}[2][red!20]{\colorbox{#1}{$\displaystyle#2$}}

\begin{figure}
\begin{minipage}[t]{0.57\textwidth}\small
$\arraycolsep=3pt
\begin{array}{lll}
\multicolumn{3}{l}{\mathcal{L}ex(L,s) = \mathcal{L}(L,\textsc{no},[], s)} \\[1.3ex]
\mathcal{L}(L',k,rs,[]) & = & \mathcal{M}(k,rs) \\
\mathcal{L}(L',k,rs,c{::}cs) & = & \text{if } L'_c \overset{\tiny ?}{=} \emptyset \text{ then } \mathcal{M}(k,rs)\\
& &  \text{else } \text{case } K \text{ of } \emptyset \mapsto \mathcal{L}(L'_c, k, rs, cs) \\
                       && \qquad\qquad\;\; \{k'\} \mapsto \mathcal{L}(L'_c, k', cs, cs)\\
\multicolumn{3}{l}{
\begin{array}{rll}
\text{\textit{where} } \;L'_c &=& \{ \partial_c(r) \Rightarrow k \mid r \Rightarrow k \in L' \wedge \partial_c(r) \neq \bot \} \\
\multicolumn{1}{r}{K} &=&  \{k \mid r \Rightarrow k \in L'_c \wedge \nu(r)\} %
\end{array}
} 
\end{array}
$
\end{minipage}%
\begin{minipage}[t]{0.43\textwidth}\small
$\arraycolsep=3pt
\begin{array}{lll}
  \\
\mathcal{M}(\textsc{no},rs) &=& \textsc{fail} \\
\mathcal{M}(\text{Skip}, []) &=& []\\ 
\mathcal{M}(\text{Skip}, c{::}cs) &=& \mathcal{L}(L,\textsc{no},[],c{::}cs)\\
\mathcal{M}(\text{Return } t, []) &=& [t]\\
\mathcal{M}(\text{Return } t, c{::}cs) &=& t \,::\, \mathcal{L}(L,\textsc{no}, [], c::cs)\\
\end{array}
$

\end{minipage}%
\caption{\label{figure:lexing-algorithm}Lexing algorithm}

\begin{minipage}[t]{0.57\textwidth}\small
$\arraycolsep=2pt
\begin{array}{lll}
  \multicolumn{3}{l}{\mathcal{P}arse(n \Rightarrow G,s) = \mathcal{P}(n,s)}\\[1.3ex]
\mathcal{P}(n,[])     &=& \text{if } n \to \epsilon \in  G \text{ then } [] \text{ else } \textsc{fail} \\[0.3ex]
\mathcal{P}(n,t{::}ts)  &=& \text{if } n \to t \overline{n} \in  G \text{ then } \mathcal{Q}(\overline{n},ts) \\[0.3ex]
                           &   & \text{else if } n \to \epsilon \in  G \text{ then } t{::}ts \text{ else } \textsc{fail} \\
\end{array}
$
\end{minipage}%
\begin{minipage}[t]{0.43\textwidth}\small
$\arraycolsep=2pt
\begin{array}{lll}
  \\
\mathcal{Q}([],ts)    &=& ts \\
\mathcal{Q}(n{::}ns,ts) &=& \mathcal{Q}(ns,\mathcal{P}(n,ts))\\[1ex]
\end{array}
$
\end{minipage}

\caption{\label{figure:parsing-algorithm}Parsing algorithm for DGNF grammars}

\begin{minipage}[t]{0.57\textwidth}\small
$\arraycolsep=2pt
  \begin{array}{llll}
  \multicolumn{3}{l}{\mathcal{FP}arse(n \Rightarrow F,s) = \mathcal{G}([n], s)}\\[1.3ex]
     \multicolumn{3}{l}{\mathcal{F}(F_n,k, rs, s) =}\\
                              & \quad \text{case } s \text{ of } [] \mapsto Step(k,rs) \\[0.2ex]
                              & \qquad\qquad\;\;\; c{::}cs \mapsto \text{if } F_n' \overset{\tiny ?}{=} \emptyset \text{ then } Step(k,rs) \\[0.2ex]
                              & \qquad\qquad\qquad\quad\;\;\;\text{ else }\text{case } K \text{ of } \emptyset \mapsto  \mathcal{F}(F_n', k, rs, cs) \\[0.2ex]
                              & \qquad\qquad\qquad\qquad\qquad\qquad\;\;\, \{ns\} \mapsto  \mathcal{F}(F_n', \textsc{on}\;ns, cs, cs)\\[0.4ex]
     \multicolumn{3}{l}{\arraycolsep=3pt
     \begin{array}{rll}
       \text{\textit{where} }\; F_n' &=& \{ \langle\partial_c(r), k\rangle  \mid \langle r, k \rangle \in F_n \wedge \partial_c(r) \neq \bot \} \\[0.2ex]
       \multicolumn{1}{r}{K} &=&  \{k \mid \langle r, k\rangle \in F_n' \wedge \nu(r) \}
     \end{array}
                                 }
   \end{array}
   $
\end{minipage}%
\begin{minipage}[t]{0.43\textwidth}\small
$\arraycolsep=2pt
\begin{array}{l}
\begin{array}{lll}
    \\[19pt]
\mathcal{G}([], s) &=& s\\
\mathcal{G}(n{::}ns, s) &=& \mathcal{G}(ns, \mathcal{F}(F_n, k, s, s))\\[0.2ex]
\multicolumn{3}{r}{
  \begin{array}{rll}
\text{\textit{where} }\; F_n &=& \{ \langle r, \overline{n} \rangle \mid n \to r \overline{n} \in F \} \\[0.2ex]
k &=& \text{if } n \to ?r \in F \text{ then } \textsc{back} \text{ else } \textsc{no}
  \end{array}
}
\end{array}
      \\
        \begin{array}{lll}
          \\
  Step(\textsc{back},s) &=& s\\
  Step(\textsc{on}\;ns,s) &=& \mathcal{G}(ns, s)\\
  Step(\textsc{no},s) &=& \textsc{fail}\\[1.5ex]
\end{array}
\end{array}
$
\end{minipage}

\caption{\label{figure:fused-parsing-algorithm}Parsing algorithm for fused grammars}

\begin{minipage}[t]{0.57\textwidth}\small
$\arraycolsep=0pt
\begin{array}{llll}
    \multicolumn{3}{l}{\dyn{\mathcal{SP}arse_{n \Rightarrow F}(s)} = \mathcal{T}([n],\dyn{s})} \\[1.4ex]
  \multicolumn{3}{l}{\dyn{\strut\mathcal{S}_{F_n,k}(rs, s)} \dyn{\strut=}}\\
                                        & \quad \dyn{\strut\text{case } s \text{ of } [] \mapsto} Step(k,\dyn{rs}) \\[-0.2ex]
                            & \qquad\qquad\;\;\dyn{\strut c_i{::}cs \mapsto}\text{if } F_{n,i}' \overset{\tiny ?}{=} \emptyset \text{ then } Step(k,\dyn{rs}) \\
                            & \qquad\qquad\qquad\qquad\; \text{ else }\text{case } K_i \text{ of } \emptyset \mapsto  \dyn{\strut\mathcal{S}_{F_{n,i}', k}(rs, cs)} \\
                            & \qquad\qquad\qquad\qquad\qquad\qquad\quad\; \{ns\} \mapsto  \dyn{\strut\mathcal{S}_{F_{n,i}', \textsc{on}\;ns}(cs, cs)}\\[-1.5ex]
                            & \qquad\qquad\;\;\dyn{c_j{::}cs \mapsto} \ldots\\[0.2ex]
\multicolumn{3}{l}{\arraycolsep=3pt
  \begin{array}{rll}
    \text{\textit{where} }\; F_{n,i}' &=& \{ \langle\partial_{c_i}(r), k\rangle  \mid \langle r, k \rangle \in F_n \wedge \partial_{c_i}(r) \neq \bot \} \\[0.2ex]
    \multicolumn{1}{r}{K_i} &=&  \{k \mid \langle r, k\rangle \in F_{n,i}' \wedge \nu(r) \}  \\[0.2ex]
\end{array}
}
\end{array}
$
\end{minipage}%
\begin{minipage}[t]{0.43\textwidth}\small
  $\arraycolsep=2pt
  \begin{array}{l}
  \begin{array}{lll}
  \\
    \mathcal{T}([],\dyn{s}) &=& \dyn{s}\\
    \mathcal{T}(n{::}ns,\dyn{s}) &=& \mathcal{T}(ns,\dyn{\mathcal{S}_{F_n, k}(s, s)})\\[0.4ex]
    \multicolumn{3}{r}{
    \begin{array}{rll}
      \text{\textit{where} }\; F_n &=& \{ \langle r, \overline{n} \rangle \mid n \to r \overline{n} \in F \} \\[0.2ex]
      k &=& \text{if } n \to ?r \in F \text{ then } \textsc{back} \text{ else } \textsc{no}
    \end{array}
            }
  \end{array}
    \\
    \begin{array}{lll}
      \\
    Step(\textsc{back},\dyn{s}) &=& \dyn{s}\\
    Step(\textsc{on}\;ns,\dyn{s}) &=& \mathcal{T}(ns,\dyn{s})\\
    Step(\textsc{no},\dyn{s}) &=& \dyn{\textsc{fail}}\\[1.5ex]
  \end{array}
  \end{array}
  $
\end{minipage}

\caption{\label{figure:staged-parsing-algorithm}Staged parsing algorithm}
\end{figure}

This section describes the lexing and parsing algorithms, shows how to
stage the parsing algorithm to improve performance, and
explains details of the implementation of the algorithms in \FLaP{}.

\subsection{The lexing algorithm}
\label{section:lexing-algorithm}

\Cref{figure:lexing-algorithm} presents the lexing algorithm.
The algorithm has conventional
  \emph{longest-match} semantics:
each token returned
 corresponds to
    the rule
      matching
         the longest possible prefix of the input.
This behaviour is implemented by
repeatedly updating
  the best match
    seen \emph{so far}
  until
    no rule
      matches.

The top-level function $\mathcal{L}ex$ takes lexing rules $L$
and input string $s$.
For simplicity, we assume utility functions $\mathcal{L}$ and $\mathcal{M}$ can freely access $L$.
At a high level,
$\mathcal{L}$
  reads
    a single token
    from a prefix of a string,
  pairs the token action
   with the remainder of the string,
and passes it to
$\mathcal{M}$.
$\mathcal{M}$
  constructs a sequence of tokens,
  updating the sequence
    according to
      the action
        passed from $\mathcal{L}$.

$\mathcal{L}$ has four arguments:
  the lexing rules $L'$;
  a token action $k$
   representing
     the best match so far;
  the remainder string $rs$
     for the best match;
  the input string $s$.
For empty input
 the best match information $k$ and $rs$ is passed to $\mathcal{M}$.
For non-empty input $c{::}cs$,
  the result depends on
   $L'_c$,
     the lexing rules
     updated to use
       the non-empty \emph{derivatives with respect to $c$} of the string.
If $L'$ is empty,
  lexing cannot advance,
  so $\mathcal{L}$ transfers control to $\mathcal{M}$.
Otherwise,
  the result depends on
   the rule $r \Rightarrow a$
     that matches
      the string up to this point
        including $c$
      (i.e.~the rule that accepts $\epsilon$ after consuming $c$). 
If there is no such rule,
  then lexing continues with $k$.
If there is such a rule,
  it is unique
  (since lexing rules are disjoint (\Cref{sec:canonical-lexer})),
  and lexing continues with the new longest match $k'$. 

The $\mathcal{M}$ function has two arguments:
  an action $k$, and
  a remainder string $rs$.
The sentinel $\textsc{no}$
  indicates that lexing has failed.
For $\text{Skip}$,
  lexing continues
    if the remainder $rs$ is non-empty.
For $\text{Return }t$,
  $t$ is added to the output sequence,
  and lexing continues
    if the remainder $rs$ is non-empty.
In the cases where lexing continues,
it commences
  by supplying $\textsc{no}$
  for the best-match-so-far,
so that
  reading the next token
    only succeeds if
      $\mathcal{L}$ matches
        a non-empty prefix
          of the remaining input. 

\subsection{The DGNF parsing algorithm}
\label{section:parsing-algorithm}

\Cref{figure:parsing-algorithm} presents the parsing algorithm for DGNF grammars.
Deterministic parsing makes the algorithm simple, since there is no need for backtracking.

$\mathcal{P}arse$ is the top-level parsing algorithm which takes the parsing
grammar $n \Rightarrow G$ and a sequence of tokens $s$.
There are two key functions:
$\mathcal{P}$ parses using a single nonterminal $n$,
and $\mathcal{Q}$ parses using a sequence of nonterminals $ns$.
Again, we assume $\mathcal{P}$ and $\mathcal{Q}$ can freely access $G$.

$\mathcal{P}$ takes the nonterminal $n$ and a sequence of tokens and returns the remainder of the sequence after parsing.
For empty sequences parsing succeeds only if the grammar has a rule $n \to \epsilon$.
For non-empty sequences $t{::}ts$, if the grammar has a rule $n \to t\overline{n}$, $\mathcal{P}$ consumes $t$ and parses $ts$ with $\mathcal{Q}$.
Otherwise, parsing succeeds (consuming nothing) only if the grammar has a rule $n \to \epsilon$.

$\mathcal{Q}$ takes a sequence of nonterminals $ns$ and a sequence of tokens $ts$ and parses successive prefixes of $s$ with each nonterminal in $ns$.

\subsection{The parsing algorithm for fused grammars}
\label{section:fused-parsing-algorithm}

In practice, \FLaP{} does not use separately-defined lexing and DGNF parsing algorithms,
since it fuses lexing and parsing.
We presented those algorithms to allow a direct comparison with
the algorithm for fused grammars.

\Cref{figure:fused-parsing-algorithm} shows an algorithm for
parsing with fused grammars.
The algorithm
  combines the features of
    the lexing algorithm (\Cref{figure:lexing-algorithm})
    and
    the parsing algorithm (\Cref{figure:parsing-algorithm}):
  like the lexing algorithm
    it maintains
      a set of derivatives
      and
      an action and remainder string for the current \emph{best match};
  like the parsing algorithm,
     it keeps track of
       the current non-terminal.

$\mathcal{FP}arse$ takes the fused grammar $n \Rightarrow F$ and an input string
$s$,
with two key functions:
$\mathcal{F}$ parses using a single nonterminal $n$,
and $\mathcal{G}$ parses using a sequence of nonterminals $ns$ using $F$.

$\mathcal{F}$ takes four arguments:
$F_n$, a set of pairs representing non-epsilon productions for $n$;
$k$, an action;
$rs$, a remainder string; and
$s$, an input string.
For empty input strings the best match information $k$ and $rs$ is
passed to $\mathcal{G}$ (via the auxiliary function $Step$).
For non-empty input strings $c{::}cs$,
  the result depends on
   $F'_n$,
     the production pairs for $n$
     updated to use
       the non-empty \emph{derivatives with respect to $c$} (\Cref{section:overview-proposal}) of the string.
If $F'_n$ is empty,
  parsing cannot proceed any further, and
  so $\mathcal{F}$ transfers control to $\mathcal{G}$ (via $Step$),
  passing the best match information.
  Otherwise,
  the result depends on
   the production pair $\langle r, \overline{n}\rangle$
     for which $r$ matches
      the string up to this point
        including $c$
      (i.e.~the rule that accepts $\epsilon$ after consuming $c$). 
If there is no such rule,
  then parsing continues with $k$.
If there is such a rule,
  it is unique
   (since the regexes for a particular nonterminal are disjoint), and
  it represents a new longest-match $\overline{ns}$, and
  parsing continues, updating the best match information to $\textsc{on}\;\overline{ns}$.
Here $\textsc{on}\;\overline{ns}$ represents one of three continuation
types, and indicates that parsing should continue using the
nonterminal sequence $\overline{ns}$;
the others are
$\textsc{back}$,
  indicating that
    parsing with $n$ should succeed,
      consuming no input, and
$\textsc{no}$,
  indicating that
    parsing with $n$ should fail.
The $Step$ function matches these three cases, and takes an action
appropriate to each continuation.

The $\mathcal{G}$ function takes a sequence of nonterminals $ns$ and a sequence of characters $s$
and parses successive prefixes of $s$ with each nonterminal in $ns$ by calling $\mathcal{F}$.
The value of $\mathcal{F}$'s $k$ argument depends on whether there is
an epsilon rule for $n$ in the fused grammar:
if so, then a parsing failure with $n$ should backtrack, consuming no input;
if not, then parsing returns $\textsc{fail}$.

We draw attention to two salient features of the fused parsing algorithm:
first,
 it consists of elements from the lexing and parsing algorithms of Sections~\ref{section:lexing-algorithm} and~\ref{section:parsing-algorithm};
second,
 it does not materialize the tokens produced by the lexing algorithm,
  instead operating directly on the character string.   
The final algorithm in the next section makes this even more apparent.

\subsection{The staged parsing algorithm}
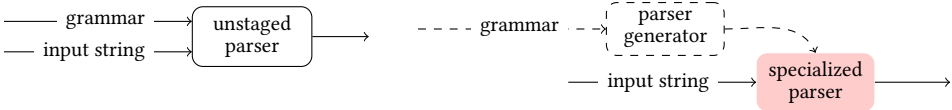
\label{section:staged-algorithm}

\begin{tikzpicture}
\node[draw,rectangle,rounded corners] (parser) at (0,0.6) { \parbox{9ex}{\centering\footnotesize unstaged\\[-0.5ex] parser} };
\draw[->] ($(parser.west) + (-2.5,0.2)$) -- ($(parser.west) + (0,0.2)$);
\draw[->] ($(parser.west) + (-2.5,-0.2)$) -- ($(parser.west) + (0,-0.2)$);
\draw[->] ($(parser.east)$) -- ($(parser.east) + (0.75,0)$);
\node[anchor=east,fill=white] at ($(parser.west) + (-0.5,0.2)$) { \footnotesize grammar  };
\node[anchor=east,fill=white] at ($(parser.west) + (-0.5,-0.2)$) {  \footnotesize input string };

\node[draw,rectangle,dashed,rounded corners,inner sep=3pt] (parser-generator) at (5.5,0.7) { \parbox{9ex}{\centering\footnotesize parser\\[-0.5ex] generator } };
\node[fill=red!20,rectangle,rounded corners,inner sep=3pt] (specialized-parser) at (7.5,0) { \parbox{9ex}{\centering\footnotesize specialized\\[-0.5ex] parser} };
\draw[->,dashed] ($(parser-generator.west) + (-2.5,0)$) -- ($(parser-generator.west)$);
\draw[->,dashed] ($(parser-generator.east)$) to[out=0,in=115] ($(specialized-parser.north)$);
\draw[->] ($(specialized-parser.west) + (-2.5,0)$) -- ($(specialized-parser.west)$);
\draw[->] ($(specialized-parser.east)$) -- ($(specialized-parser.east) + (1,0)$);

\node[anchor=east,fill=white] at ($(parser-generator.west) + (-0.5,0)$) { \footnotesize grammar };
\node[anchor=east,fill=white] at ($(specialized-parser.west) + (-0.5,0)$) { \footnotesize input string };

\end{tikzpicture}

The parsing algorithm for fused grammars described in
\Cref{section:fused-parsing-algorithm} is practically inefficient.
For each character of the input, the algorithm computes derivatives
and checks emptiness and nullability for sets of regexes.
However, since the regexes and other information about the
grammar are known in advance of parsing, the inefficient algorithm can
be \emph{staged}~\cite{Walid:1999:MPT:890806} to produce an efficient
algorithm.
The idea of staging is to identify those parts of the algorithm that
do depend only on static information --- i.e.~on the grammar --- and
execute them first, leaving only the parts that depend on dynamic
information --- i.e.~on the input string --- for later.
The result of staging, as illustrated in \Cref{figure:staged-parsing-algorithm}, is to transform the
unstaged parser into a parser generator that produces as output a
parser specialized to the input grammar.

\Cref{figure:staged-parsing-algorithm} shows a staged version of
the fused parsing algorithm.
The structure of the algorithm is very close to the fused grammar parsing
algorithm of \Cref{section:fused-parsing-algorithm}:
$\mathcal{S}$ corresponds to $\mathcal{F}$
and
$\mathcal{T}$ corresponds to $\mathcal{G}$.
However, there are three key differences.

First, those parts of the algorithm that depend on the input string
are marked as \emph{dynamic}, indicated with \dyn{\text{red highlighting}}.
These dynamic elements are not executed immediately; instead they
become part of the generated specialized parser produced by the first
stage of execution.

Second, in the function $\mathcal{S}$, $F_n$ and $k$ have become
indexes rather than arguments.
Consequently, rather than being passed to the function at run-time,
those arguments serve to distinguish generated functions: each
instantiation of $F_n$ and $k$ generate a distinct function
$\mathcal{S}$ in the specialized parser.

Finally, the $\text{case}$ match in $\mathcal{S}$ is expanded to
include a distinct case for each character $c_i$, $c_j$, etc.
This expansion resolves a tension in the distinction between static
and dynamic data: the static computation of derivatives
$\partial_c(r)$ in the first stage depends on the value of $c$, which
is only available dynamically.
In the expanded $\text{case}$ match the value of $c_i$ is known on the
right-hand side of the corresponding case, making it possible to
compute derivatives valid within that program context.
This scrutiny of a statically-unknown expression using a $\text{case}$
match over its statically-known set of possible values is known 
as ``The Trick'' in the partial evaluation literature~\cite{DBLP:journals/toplas/DanvyP96}.

The evaluation of the staged parsing algorithm is largely standard:
the unhighlighted (static) expressions are executed first, producing
the highlighted (dynamic) expressions as output.
Each call to a dynamic indexed function $\mathcal{S}_{F_n,k}$ triggers
the generation of a dynamic function whose body consists of the result
of executing the right-hand side of $\mathcal{S}_{F_n,k}$ in
\Cref{figure:staged-parsing-algorithm}.
To ensure that the generation process terminates, the generation of
these indexed functions is memoized: there is at most one
generated function $\mathcal{S}_{F_n,k}$ for any particular $F_n$ and $k$.
The result of the algorithm is a set of mutually recursive functions
that operate only on strings, not on components of the grammar:
\[
\begin{array}{llll}
S_{n \to r \overline{n},\ldots,\textsc{back}}(r,s) &=& \text{case } s \text{ of} 
                                        & \text{\lstinline!'a'!}::cs \mapsto S_{n \to r_a \overline{n},\textsc{back}}(r,cs)\\
                                    & & & \text{\lstinline!'b'!}::cs \mapsto S_{n \to r_a \overline{n},\textsc{on}\;\overline{ns}}(cs,cs)\\[-0.5ex]
                                    & & & \ldots\\[-0.7ex]
S_{n \to r \overline{n},\ldots,\textsc{on}\;\overline{ns}}(r,s) &=& \ldots
\end{array}
\]

\subsection{Implementing the staged parsing algorithm}
\label{section:staging-practicalities}

\FLaP{} generates code for fused grammars using
MetaOCaml's staging facilities
together with
\citeauthor{generating-mutually-recursive-definitions}'s \citeyearpar{generating-mutually-recursive-definitions} \emph{letrec insertion}
library for creating
the indexed mutually-recursive functions produced by the staged
parsing algorithm (\Cref{section:staged-algorithm}).

There are
 three key differences between
   the pseudocode algorithm in \Cref{figure:staged-parsing-algorithm}
   and
   \FLaP{}'s implementation.
First,
 while the pseudocode presents a recognizer that either consumes
 input or fails, \FLaP{} supports \emph{semantic actions} ---
 i.e.~constructing and returning ASTs or other values when parsing
 succeeds --- as described in \Cref{sec:overview-parser}.

Second,
 while the input to the pseudocode is a character linked list,
 \FLaP{} operates on OCaml's flat array
 representation of strings, using indexes to keep track of string
 positions as parsing proceeds.
 Relatedly, \FLaP{} also optimizes the end of input test by
 using the fact that OCaml's strings are
 null-terminated, like C's.
 This representation allows the end of input check to be
 incorporated into the per-character branch in the generated
 code: a null character \lstinline!'\000'! indicates a \emph{possible}
 end of input, which can subsequently be confirmed by checking the
 string length.

Third,
 while the pseudocode generates a case in each branch for each possible 
 character in the input, \FLaP{} generates a smaller number of cases
 by grouping characters with equivalent behaviour into classes,
 as described in detail by~\citet{DBLP:journals/jfp/OwensRT09}.
 Branching on these character classes rather than treating characters
 individually leads to a substantial reduction in code size.

Here is an excerpt of the code generated by \FLaP{} for the s-expression parser:

\begin{lstlisting}[basicstyle=\ttfamily\small]
and parse$_5$ r i len s = match s.[i] with
   | ' '|'\n' -> parse$_6$ r (i + 1) len s
   | '('      -> parse$_9$ r (i + 1) len s
   | 'a'..'z' -> parse$_3$ r (i + 1) len s
   | '\000'   -> if i = len then [] else failwith "unexpected"
   | _        -> []
\end{lstlisting}

\noindent
This excerpt shows the code generated for a single indexed function
$\dyn{\mathcal{S}_{F_n,k}}$.
There are four arguments, representing
  the beginning of the current token \lstinline!r!
    (to support backtracking in the lookahead transition),
  the current index \lstinline!i!,
  the input length \lstinline!len!,
  and the input string \lstinline!s!.

The subscripts~$5$,~$6$, etc.~attached to the \lstinline!parse!
functions correspond to the indexes $F_n,k$ in the pseudocode
algorithm; the letrec insertion library assigns a fresh
subscript to each distinct index.

The character range pattern \lstinline!'a'..'z'! illustrates the character class
optimization described above,
without which each of the characters from \lstinline!'a'!
to \lstinline!'z'! would have a separate case in the \lstinline!match!.

The check \lstinline!i = len! determines whether
\lstinline!'\000'! indicates end of input or a null in
the input string.

The value \lstinline![]! corresponds to a semantic action: it is the
empty list returned when an empty sequence of s-expressions is parsed.
It appears twice in the generated code, since (as
\Cref{figure:staged-parsing-algorithm} shows), parsing for a
particular nonterminal can end in two ways: when it encounters the end
of input, and when it encounters a non-matching character.

OCaml compiles tail calls to known functions such as
\lstinline!parse$_6$! to unconditional jumps.  As
\Cref{section:evaluation} shows, the resulting code is extremely fast.

\section{Evaluation}
\label{section:evaluation}

This section evaluates the performance of
\FLaP{}, and shows that lexer-parser fusion drastically improves
performance.
Many parser combinator libraries suffer from poor performance,
but the experiments described here show that combinator parsing does
not need to be slow.

In part, \FLaP{}'s speed is a consequence of the linear-time guarantee
provided by the type system of \Cref{sec:overview-parser} and by the
application of staging to eliminate the overhead arising from parsing
abstractions.
This section shows that lexer-parser fusion provides a substantial
further performance improvement by eliminating the overhead that
arises from defining lexers and parsers separately, which accounts for
most of the remaining running time.

\pgfplotsset{
   every axis/.append style={
    font=\small,
  },
}
\pgfplotstableread[col sep = comma]{generated-code/sizes.csv}\sizesdata
\pgfplotstableread[col sep = comma]{csv/sexp.csv}\sexpdata
\pgfplotstableread[col sep = comma]{csv/intexp.csv}\intexpdata
\pgfplotstableread[col sep = comma]{csv/ppm.csv}\ppmdata
\pgfplotstableread[col sep = comma]{csv/pgn.csv}\pgndata
\pgfplotstableread[col sep = comma]{csv/csv.csv}\csvdata
\pgfplotstableread[col sep = comma]{csv/json.csv}\jsondata

\pgfplotstableread[col sep = comma]{csv/throughput.csv}\throughputininedata

\begin{figure}[!t]
\begin{tikzpicture}
  \centering
  \begin{axis}[
        ybar, axis on top,
        height=5.2cm, width=0.95\textwidth,
        bar width=0.2cm,
        tick align=inside,
        major grid style={draw=white},
        enlarge y limits={value=.1,upper},
        ymin=0, ymax=1350,
        axis x line*=bottom,
        axis y line*=right,
        y axis line style={opacity=0},
        tickwidth=0pt,
        enlarge x limits=true,
        legend columns=3,
        legend style={
            at={(0.0,0.55)},
            anchor=south west,
            legend columns=1,
            /tikz/every even column/.append style={column sep=0.2cm}
        },
        ylabel={Throughput (MB/s)},
        symbolic x coords={pgn,ppm,sexp,csv,json,arith},
        xtick=data,
        nodes near coords={
        \pgfmathprintnumber[precision=0]{\pgfplotspointmeta}
       },
       every node near coord/.append style={rotate=90, anchor=west},
    ]
   \addplot [draw=none, text=black,fill=red!60!black,error bars/.cd,x dir=both,y dir=both,y explicit] 
       table [x index = {0}, y expr = (\thisrow{ocamlyacc}) ]{\throughputininedata};

    \addplot+[draw=none,text=black,fill=orange,error bars/.cd,x dir=both,y dir=both,y explicit] 
       table [x index = {0}, y expr = (\thisrow{menhir_table}) ]{\throughputininedata};

    \addplot+[draw=none,text=black,fill=blue!20!black,error bars/.cd,x dir=both,y dir=both,y explicit] 
       table [x index = {0}, y expr = (\thisrow{menhir_code}) ]{\throughputininedata};

    \addplot+[draw=none,text=black,fill=green!60!black,error bars/.cd,x dir=both,y dir=both,y explicit] 
       table [x index = {0}, y expr = (\thisrow{fused}) ]{\throughputininedata};

    \addplot+[draw=none,text=black,fill=green!20!white,error bars/.cd,x dir=both,y dir=both,y explicit] 
       table [x index = {0}, y expr = (\thisrow{normalized}) ]{\throughputininedata};

   \addplot [draw=none,y filter/.expression={y==0 ? nan : y},text=black,fill=blue!20!white,error bars/.cd,x dir=both,y dir=both,y explicit] 
       table [x index = {0}, y expr = (\thisrow{staged}) ]{\throughputininedata};

   \addplot [draw=none,y filter/.expression={y==0 ? nan : y},text=black,fill=red!20!white,error bars/.cd,x dir=both,y dir=both,y explicit] 
       table [x index = {0}, y expr = (\thisrow{parts}) ]{\throughputininedata};

    \legend{ocamlyacc,menhir+table,menhir+code,flap,normalized,asp,ParTS}
  \end{axis}
\end{tikzpicture}

\caption{\label{figure:throughput-graph}
  Parser throughput: ocamlyacc, menhir, \FLaP{}, asp and ParTS} 
\vspace{-2ex}
\end{figure}

\paragraph{Benchmarks}
\label{section:benchmarks}

We compare seven implementations.  All
seven guarantee
deterministic, linear-time parsing,
and use staging,
generating code
specialized to a given grammar.
Our aim is to evaluate whether \FLaP{} is faster than other
asymptotically-efficient systems, so it is not possible to make
meaningful comparisons with systems with different complexity
(e.g.~GLR or backtracking recursive-descent):

\smallskip
The parser implementations are:

\begin{tabular}{lp{0.4\textwidth}lp{0.4\textwidth}}
(a) & \texttt{ocamlyacc} &
(b) & \texttt{menhir} in table-generation mode \\
(c) & \texttt{menhir} in code-generation mode &
(d) & \FLaP{} \\
(e) & \texttt{asp}~\cite{DBLP:conf/pldi/KrishnaswamiY19}  &
(f) & ParTS~\cite{casinghino-roux}\\
(g) & \multicolumn{2}{l}{Parsing with normalized but unfused grammars }
\end{tabular}

\noindent
Implementations (a)--(c) are widely-used parser-generation tools.
Implementation (d) is described in this paper.
Implementations (e) and (f) are existing parser combinator libraries that
guaranteee deterministic, linear-time parsing.
Implementation (g) is a variant of (d) in which the grammars used for
parsing are normalized by \FLaP{} and lexers are implemented using
\FLaP{}, but parsers and lexers are connected via OCaml's
\texttt{Stream} type (as in \texttt{asp}) rather than fused together (as in
\texttt{flap}).

\begin{figure}[t]
  \begin{tikzpicture}[scale=0.9]
  \begin{axis}[xshift=0\textwidth,width=0.25\textwidth,height=3.5cm,
    xlabel={\small input size~(MB)},
    ylabel={run time (ms)},
    axis x line*=left,
    axis y line*=left,
    scaled ticks=false,
    tick label style={/pgf/number format/fixed},
    title=sexp,every axis title/.style={below right,at={(0,1)}}
    ]
    \addplot[red!60!black,mark=square,mark size=1pt] table[x index = {0}, x expr=(\thisrow{index}/1048576), y expr=(\thisrow{ocamlyacc})]{\sexpdata};
    \addplot[orange,mark=*,mark size=1pt] table[x index = {0}, x expr=(\thisrow{index}/1048576), y expr=(\thisrow{menhir_table})]{\sexpdata};
    \addplot[blue!20!black,mark=*,mark size=1pt] table[x index = {0}, x expr=(\thisrow{index}/1048576), y expr=(\thisrow{menhir_code})]{\sexpdata};
    \addplot[green!60!black,mark=triangle,mark size=1pt] table[x index = {0}, x expr=(\thisrow{index}/1048576), y expr=(\thisrow{fused})]{\sexpdata};
    \addplot[green!20!white,mark=triangle,mark size=1pt] table[x index = {0}, x expr=(\thisrow{index}/1048576), y expr=(\thisrow{normalized})]{\sexpdata};
    \addplot[blue!20!white,mark=*,mark size=1pt] table[x index = {0}, x expr=(\thisrow{index}/1048576), y expr=(\thisrow{staged})]{\sexpdata};
    \addplot[red!20!white,mark=*,mark size=1pt] table[x index = {0}, x expr=(\thisrow{index}/1048576), y expr=(\thisrow{parts})]{\sexpdata};
  \end{axis}

  \begin{axis}[xshift=0.18\textwidth,width=0.25\textwidth,height=3.5cm,
    xlabel={\small input size~(MB)},
    axis x line*=left,
    axis y line*=left,
    scaled ticks=false,
    tick label style={/pgf/number format/fixed},
    title=arith,every axis title/.style={below right,at={(0,1)}}
    ]

    \addplot[red!60!black,mark=square,mark size=1pt] table[x index = {0}, x expr=(\thisrow{index}/1048576), y expr=(\thisrow{ocamlyacc}]{\intexpdata};
    \addplot[orange,mark=*,mark size=1pt] table[x index = {0}, x expr=(\thisrow{index}/1048576), y expr=(\thisrow{menhir_table})]{\intexpdata};
    \addplot[blue!20!black,mark=*,mark size=1pt] table[x index = {0}, x expr=(\thisrow{index}/1048576), y expr=(\thisrow{menhir_code})]{\intexpdata};
    \addplot[green!60!black,mark=triangle,mark size=1pt] table[x index = {0}, x expr=(\thisrow{index}/1048576), y expr=(\thisrow{fused})]{\intexpdata};
    \addplot[green!20!white,mark=triangle,mark size=1pt] table[x index = {0}, x expr=(\thisrow{index}/1048576), y expr=(\thisrow{normalized})]{\intexpdata};
    \addplot[blue!20!white,mark=*,mark size=1pt] table[x index = {0}, x expr=(\thisrow{index}/1048576), y expr=(\thisrow{staged})]{\intexpdata};
  \end{axis}

  \begin{axis}[xshift=0.35\textwidth,width=0.25\textwidth,height=3.5cm,
    xlabel={\small input size~(kpixels)},
    axis x line*=left,
    axis y line*=left,
    scaled ticks=false,
    tick label style={/pgf/number format/fixed},
    title=ppm,every axis title/.style={below right,at={(0,1)}}
    ]
    \addplot[red!60!black,mark=square,mark size=1pt] table[x index = {0}, x expr=(\thisrow{index}/1000), y expr=(\thisrow{ocamlyacc}/1000)]{\ppmdata};
    \addplot[orange,mark=*,mark size=1pt] table[x index = {0}, x expr=(\thisrow{index}/1000), y expr=(\thisrow{menhir_table}/1000)]{\ppmdata};
    \addplot[blue!20!black,mark=*,mark size=1pt] table[x index = {0}, x expr=(\thisrow{index}/1000), y expr=(\thisrow{menhir_code}/1000)]{\ppmdata};
    \addplot[green!60!black,mark=triangle,mark size=1pt] table[x index = {0}, x expr=(\thisrow{index}/1000), y expr=(\thisrow{fused}/1000)]{\ppmdata};
    \addplot[green!20!black,mark=triangle,mark size=1pt] table[x index = {0}, x expr=(\thisrow{index}/1000), y expr=(\thisrow{normalized}/1000)]{\ppmdata};
    \addplot[blue!20!white,mark=*,mark size=1pt] table[x index = {0}, x expr=(\thisrow{index}/1000), y expr=(\thisrow{staged}/1000)]{\ppmdata};
  \end{axis}

  \begin{axis}[xshift=0.53\textwidth,width=0.25\textwidth,height=3.5cm,
    xlabel={\small input size~(games)},
    axis x line*=left,
    axis y line*=left,
    scaled ticks=false,
    tick label style={/pgf/number format/fixed},
    title=pgn,every axis title/.style={below right,at={(0,1)}}
    ]
    \addplot[red!60!black,mark=square,mark size=1pt] table[x index = {0}, y expr=(\thisrow{ocamlyacc}/1000)]{\pgndata};
    \addplot[orange,mark=*,mark size=1pt] table[x index = {0}, y expr=(\thisrow{menhir_table}/1000)]{\pgndata};
    \addplot[blue!20!black,mark=*,mark size=1pt] table[x index = {0}, y expr=(\thisrow{menhir_code}/1000)]{\pgndata};
    \addplot[green!60!black,mark=triangle,mark size=1pt] table[x index = {0}, y expr=(\thisrow{fused}/1000)]{\pgndata};
    \addplot[green!20!white,mark=triangle,mark size=1pt] table[x index = {0}, y expr=(\thisrow{normalized}/1000)]{\pgndata};
    \addplot[blue!20!white,mark=*,mark size=1pt] table[x index = {0}, y expr=(\thisrow{staged}/1000)]{\pgndata};
  \end{axis}

  \begin{axis}[xshift=0.71\textwidth,width=0.25\textwidth,height=3.5cm,
    xlabel={\small input size~(MB)},
    axis x line*=left,
    axis y line*=left,
    scaled ticks=false,
    tick label style={/pgf/number format/fixed},
    title=csv,every axis title/.style={below right,at={(0,1)}}
    ]
    \addplot[red!60!black,mark=square,mark size=1pt] table[x index = {0}, x expr=(\thisrow{size}/1048576),y expr=(\thisrow{ocamlyacc}/1000)]{\csvdata};
    \addplot[orange,mark=*,mark size=1pt] table[x index = {0}, x expr=(\thisrow{size}/1048576),y expr=(\thisrow{menhir_table}/1000)]{\csvdata};
    \addplot[blue!20!black,mark=*,mark size=1pt] table[x index = {0}, x expr=(\thisrow{size}/1048576),y expr=(\thisrow{menhir_code}/1000)]{\csvdata};
    \addplot[green!60!black,mark=triangle,mark size=1pt] table[x index = {0}, x expr=(\thisrow{size}/1048576),y expr=(\thisrow{fused}/1000)]{\csvdata};
    \addplot[green!20!white,mark=triangle,mark size=1pt] table[x index = {0}, x expr=(\thisrow{size}/1048576),y expr=(\thisrow{normalized}/1000)]{\csvdata};
  \end{axis}

  \begin{axis}[xshift=0.89\textwidth,width=0.25\textwidth,height=3.5cm,
    xlabel={\small input size~(msgs)},
    axis x line*=left,
    axis y line*=left,
    scaled ticks=false,
    tick label style={/pgf/number format/fixed},
    title=json,every axis title/.style={below right,at={(0,1)}}
    ]
    \addplot[red!60!black,mark=square,mark size=1pt] table[x index = {0}, y expr=(\thisrow{ocamlyacc}/1000)]{\jsondata};
    \addplot[orange,mark=*,mark size=1pt] table[x index = {0}, y expr=(\thisrow{menhir_table}/1000)]{\jsondata};
    \addplot[blue!20!black,mark=*,mark size=1pt] table[x index = {0}, y expr=(\thisrow{menhir_code}/1000)]{\jsondata};
    \addplot[green!60!black,mark=triangle,mark size=1pt] table[x index = {0}, y expr=(\thisrow{fused}/1000)]{\jsondata};
    \addplot[green!20!white,mark=triangle,mark size=1pt] table[x index = {0}, y expr=(\thisrow{normalized}/1000)]{\jsondata};
    \addplot[blue!20!white,mark=*,mark size=1pt] table[x index = {0}, y expr=(\thisrow{staged}/1000)]{\jsondata};
    \addplot[red!20!white,mark=*,mark size=1pt] table[x index = {0}, y expr=(\thisrow{parts}/1000)]{\jsondata};
  \end{axis}
\end{tikzpicture}

\caption{\label{figure:linear-parsing}Linear-time parsing (colors as \Cref{figure:throughput-graph})} %
\end{figure}

For lexing we use \texttt{ocamllex} for (a)--(c), and the
combinators supplied by each library for (d)--(g).
Implementations (a)--(c) use identically structured grammars
(since \texttt{menhir}~\cite{menhir} accepts \texttt{ocamlyacc} files as input)
and lexers based on \texttt{ocamllex}.
Implementations (d)--(g) also use identically structured grammars
based on the standard parser combinator
interface (\Cref{sec:overview-parser}).
However, (d)--(g) use differently-structured lexers: (e) and (f) reuse
the deterministic parser combinators for lexing, while \FLaP{} and the
normalized but unfused parser use the more conventional lexing
interface from \Cref{figure:overview:syntax}.

The benchmarks are largely taken from
\citet{DBLP:conf/pldi/KrishnaswamiY19} (using the same test corpora),
except for the CSV benchmark (which uses a set of files of various
sizes and dimensions, using a random variety of textual and numeric
data).  They are:

\begin{enumerate}[label={(\arabic*)}]
\item (pgn) Parse 6759 \href{https://en.wikipedia.org/wiki/Portable_Game_Notation}{Portable
    Game Notation} chess game descriptions, and extract game results.

\item (ppm) 
  Parse and check semantic properties (e.g.~pixel count, color range) of
  \href{https://en.wikipedia.org/wiki/Netpbm_format}{Netpbm} files.

\item (sexp) Parse S-expressions with alphanumeric atoms, returning
  the atom count.

\item (csv) Parse CSV files (\citet{rfc4180}, with mandatory
 terminating CRLF), checking row lengths. 
  This benchmark has no
  \texttt{asp} implementation, because distinguishing escaped double-quotes
  \verb!""! from unescaped quotes \verb!"! in the lexer needs
  multiple characters of lookahead.

\item (json) Parse JSON using the grammar by \citet{staged-parser-combinators},
 returning the object count.

\item (arith) Parse and evaluate terms in a mini language (arithmetic/comparison/binding/branching).

\end{enumerate}

The benchmarks were compiled with BER MetaOCaml N111 with flambda
optimizations enabled and run on a single Intel i9-12900K core with
1GB memory running Debian Linux, using the \verb!Core_bench!
micro-benchmarking library~\cite{core-bench}.

\paragraph{Running time}
\label{section:running-time}

\Cref{figure:throughput-graph} shows the throughput of the seven
implementations using the benchmark grammars.
\Cref{figure:linear-parsing} illustrates that all seven
produce parsers
with running time linear in input length.

As \Cref{figure:throughput-graph} shows, our experiments confirm the results reported by
\citeauthor{DBLP:conf/pldi/KrishnaswamiY19}: the staged implementation
of typed CFEs in \texttt{asp} generally
outperforms \texttt{ocamlyacc}.
The addition of lexer-parser fusion makes \FLaP{} considerably faster
than both \texttt{asp} and \texttt{ocamlyacc}, reaching around 1.4GB/s (a
little over 2.3 cycles per byte) on the \texttt{json} benchmark.
The throughput ratios of \texttt{flap} to \texttt{asp}
($\frac{286}{81} = 3.5\times$, $\frac{104}{27} = 3.9\times$,
$\frac{213}{92} = 2.3\times$, $\frac{1359}{169}=8.0\times$,
$\frac{57}{29}=2.0\times$) indicate the additional performance benefit
provided by the combination of fusion and staging over staging alone.
The throughput ratios of \texttt{flap} to the normalized but unfused
implementation ($\frac{286}{48} = 6.0\times$, $\frac{104}{14} =
7.4\times$, $\frac{213}{125} = 1.7\times$,
$\frac{1359}{344}=4.0\times$, $\frac{57}{29}=2.0\times$) show that the
normalization step in \FLaP{} is not sufficient to account for \texttt{flap}'s
superior performance: performing lexer-parser fusion after grammar
normalization provides a substantial additional speedup.

\paragraph{Code size}
\label{section:code-size}

\begin{table*}
\begin{minipage}[t]{0.73\textwidth}
  \small
  \begin{tabular}{|r|cc|cc|c|c|}
& \multicolumn{2}{|c|}{{\textbf{Input}\normalfont}} & \multicolumn{2}{|c|}{{\textbf{Normalized}}} & {{\textbf{Fused}}} &   \textbf{Output}\\
\hline
{\textbf{Grammar}} & {Lex rules} &  {CFEs} & {NTs} & {Prods} & {Prods} & {Functions} \\
\hline
pgn    & 13 & 95  & 38 & 53 & 91 & 203 \\
ppm    & 6  & 10  & 5  & 6  & 16 & 55  \\
sexp   & 4  & 11  & 3  & 6  & 9  & 11  \\
csv    & 3  & 14  & 5  & 7  & 7  & 17  \\
json   & 12 & 42  & 9  & 33 & 42 & 93  \\
arith  & 14 & 143 & 28 & 55 & 83 & 209 \\
\end{tabular}
\caption{\label{table:sizes} Sizes of inputs, intermediate forms, and generated code}
\end{minipage}%
\begin{minipage}[t]{0.27\textwidth}
\small
  \begin{tabular}{|c|}%
    \parbox{0.9\textwidth}{\centering\textbf{Compilation time \\ (ms)}}\\
    \hline
    212\\
    3.60\\
    0.331\\
    0.499\\
    28.5\\
    460\\
  \end{tabular}
  \caption{\label{table:compilation-time}Compilation time (type-checking, normalization, fusion, code generation)}
\vspace{-4ex}
\end{minipage}
\end{table*}

A second important measure of usefulness for parsing:
if parsing tools are to be usable
in practice, it is essential that they do not generate unreasonably
large code.

There are several reasons to be apprehensive about the size of code
generated by \FLaP{}.
First, conversion to Greibach Normal Form is known to
substantially increase grammar size; for example, in the
procedure given by~\citet{BLUM1999112} the result of converting a
grammar $G$ has size $O(\vert G \vert^3)$.
Second, fusion is inherently duplicative, repeatedly
copying lexer rules into grammar productions.
Finally, experience in the multi-stage programming community shows
that it is easy to inadvertently generate large programs,
since antiquotation makes it easy to duplicate terms.

However, measurements largely dispel these concerns.
Table~\ref{table:sizes} lists parser representation sizes
at various stages in \FLaP{}'s pipeline.
The leftmost columns show the size of the input
parsers, measured as the number of lexer rules (both
\textbf{Return} and \textbf{Skip}) and the number of
CFE nodes, as described in
\Cref{figure:overview:syntax}.
The central columns show the number of nonterminals and
productions after conversion to DGNF using
the procedure in \Cref{section:detgnf};
they show that normalization for typed
CFEs does not produce the drastic increases in
size that occur in the more general conversion to GNF.
The next column to the right shows the grammar size after
fusion (\Cref{section:fusion}).
Fusion does not alter the number of nonterminals, but can add
productions;
for example, the \textbf{Skip} rules in the sexp lexer add
additional productions to each nonterminal.
Finally, the rightmost column shows the number of function bindings in
the code generated by \FLaP{}.
Comparing this generated function count with the number of
CFEs in the input reveals an unalarming
relationship: with one exception (ppm), their ratio
barely exceeds $2$.

\paragraph{Sharing}
The entries for \emph{pgn} and \emph{arith} hint at opportunities for
further improvement.
In both cases, the number of CFEs in the
grammar (95 and 143) is surprisingly high, since both languages are
fairly simple.
Inspecting the grammar implementations reveals the cause: in
several places, the combinators that construct the grammar duplicate
subexpressions.
For example, here is the implementation of a Kleene plus operator used
in \emph{pgn}:
\begin{lstlisting}
let oneormore e = (e >>> star e) $\ldots$
\end{lstlisting}
Normalization turns these two occurrences of \lstinline!e! into
multiple entries in the normalized form, and ultimately to multiple
functions in the generated code.

The core problem is that the parser combinator interface
(\Cref{sec:overview-parser}) provides no way to express
sharing of subgrammars.
Since duplication of this sort is common, it is likely that extending
\texttt{flap} with facilities to express and maintain sharing could
substantially reduce generated code size.

\paragraph{Compilation time}
\label{section:compilation-time}

A final measure of practicality is the time taken to perform the fusion transformation.
Slow compilation times can have a significant effect on usability; as
\citet{nielsen:usability} notes, software that takes more
than ten seconds to respond can cause a user to lose focus.

Table~\ref{table:compilation-time} shows the compilation time for the
benchmark grammars.
For each, the total time taken to type-check and normalize the
grammar, fuse the grammar and lexer and generate code
is below half a second.
Measurements indicate that the compilation time of the OCaml code
generated by flap is also fairly low, at approximately 20ms/function,
and linear in the size of the generated code.

\section{Related work}
\label{section:related}

\paragraph{Deterministic Greibach Normal Form}

There are several longstanding results related to deterministic
variants of Greibach Normal Form.
For example, \citet{DBLP:journals/dm/GellerHH76} show that every
strict deterministic language can be given a strict deterministic
grammar in Greibach Normal Form,
and \citet{DBLP:journals/eik/Nijholt79} gives a translation into
Greibach Normal Form that preserves strict deterministicness.
The distinctive contributions of this paper are the new normal form
that is well suited to fusion, and the compositional normalization
procedure from typed context-free expressions, allowing deterministic
GNF to be used in the implementation of parser combinators.

\paragraph{Combining lexers and parsers}

The work most closely related to ours, by~\citet{casinghino-roux}
investigates the application of traditional stream fusion techniques
to parser combinators in the ParTS system.
We have included their two published benchmarks in the evaluation of
\Cref{section:evaluation} and found that, as they report, 
when the flambda compiler optimizations are applied to their code,
its performance is similar to the results achieved
by~\citet{DBLP:conf/pldi/KrishnaswamiY19}.
A major difference between their work and ours is that they
approach fusion as a traditional optimization problem, in which
transformations are applied to code that satisfies certain heuristics,
and are not applied in more complex cases.
In contrast, we treat fusion as a sequence of total
transformations guaranteed to convert every parser into
a form with good performance.
More concretely, in \Cref{figure:throughput-graph}, \FLaP{} achieves
two and ten times the throughputs of ParTS on the
\texttt{sexp} and \texttt{json} benchmarks.

Another line of work, on \emph{Scannerless GLR
parsing}~\cite{DBLP:conf/cc/EconomopoulosKV09,DBLP:conf/cc/BrandSVV02},
also aims to eliminate the boundary between lexers and parsers,
but in the interface (not just in the implementation, as in \FLaP{}).
The principal aim is a principled way to handle lexical ambiguity.
Scannerless parsing carries considerable cost, often
running orders of magnitude slower than \FLaP{} according to the
figures given by
\citet{DBLP:conf/cc/EconomopoulosKV09}.

Similarly, ANTLR 4~\cite{adaptive-ll-star} supports scannerless
parsing based on a top-down algorithm, ALL(*), that performs grammar
analysis dynamically, during parsing.
Like Scannerless GLR, it has superlinear (here $O(n^4)$) complexity in
theory, but often enjoys linear performance in practice.

The \emph{packrat} algorithm~\cite{packrat} also supports a form of
scannerless parsing; in contrast to Scannerless GLR and ALL(*), it is
restricted to deterministic grammars.
Packrat parsers are structured like backtracking recursive-descent
parsers, but use lazy evaluation to construct and memoize intermediate
results during parsing, reducing needless recomputation and
guaranteeing linear time complexity.
However, packrat has some sigificant performance limitations.  Since
it retains all intermediate structures, it uses space linear in the
input size; further, its reported throughput (around 25 kb/second) is
orders of magnitude slower than flap.

Unlike scannerless systems, \FLaP{} does not provide a more
powerful parsing interface to eliminate the need for a separate lexer.
In \FLaP{} parsers are defined using a traditional parser combinator
interface and lexers are defined separately: it is only in the code
generated by \FLaP{}, not in the interface, that tokens are statically
eliminated.

\emph{Context-aware scanning}, introduced
by~\citet{VanWyk:2007:CSP:1289971.1289983} is another variant on the
parser-lexer interface focused on disambiguation;
it passes contextual information from parser to lexer about
the set of valid tokens at a particular point, in a similar way to the
lexer specialization in \Cref{sec:overview:fusion} of
this paper.
However, \citeauthor{VanWyk:2007:CSP:1289971.1289983}'s framework goes
further, and allows the automatic selection of a lexer (not just a subset of
lexing rules) based on parsing context.

\paragraph{Fusion}

The notion of fusion, in the sense of merging computations to
eliminate intermediate structures, has been applied in several
domains, including query
engines~\cite{DBLP:journals/jfp/ShaikhhaD018}, GPU
kernels~\cite{DBLP:journals/tjs/FilipovicMFM15} and tree
traversals~\cite{DBLP:conf/pldi/SakkaSN019}.

Perhaps the most widespread is stream fusion, which 
originated with Wadler's deforestation~\cite{deforestation}, and has
been applied as both a traditional compiler
optimization~\cite{DBLP:conf/icfp/CouttsLS07} and a staged
library~\cite{DBLP:conf/popl/KiselyovBPS17} with guarantees
similar to \FLaP{}'s.

\paragraph{Parser optimization}

Finally, in contrast to the constant-time speedups resulting from
lexer-parser fusion, we note an intriguing piece of work
by~\citet{two-level-supercompilation} that applies
two-level-supercompilation to parser optimization, leading to
asymptotic improvements.

\section{Future work}
\label{section:future}

There are a number of promising avenues for future work.
First, extending \FLaP's rather minimal lexer and parser interfaces to
support common needs such as left-recursive grammars, lexers and
parsers with multiple entry points, mechanisms for maintaining state
during parsing, and more expressive lexer semantic action could make
the library substantially more usable in practice.

Second, applying the fusion techniques to more powerful parsing
algorithms (e.g.~LR(1)) in
a traditional parser generator
could make lexer-parser fusion available to many more programmers.

Finally, it may be that fusion can be extended to longer pipelines
than the lexer-parser interface that we investigate here.  Might it be
possible to fuse together (e.g.) decompression, unicode
decoding, lexing and parsing into a single computation that does not
materialize intermediate values?

\begin{acks}
We thank Paul Gazzillo for shepherding the paper,
the anonymous reviewers for their helpful comments,
and members of IFIP WG 2.11 for feedback on a presentation of this work.

This work was supported in part by a European Research Council (ERC)
Consolidator Grant for the project “TypeFoundry”, funded under the
European Union’s Horizon 2020 Framework Programme (grant agreement
no.~101002277), and in part by a grant from the Isaac Newton Trust
(grant no.~G101121).
\end{acks}

\section*{Artifact}

We have made available an artifact and accompanying instructions that
allow the interested reader to reproduce the claims in this
paper~\cite{flap-artifact}.

\bibliography{fusion}

\extendedOnly{
\cleardoublepage
\appendix

\pagebreak

\begin{landscape}

\section{Complete Derivation}

This section presents the complete derivation for normalizing
\[ g =
  \mu \text{ sexp }. (\textsc{lpar} \cdot (\mu \text{ sexps }. \,\epsilon \lor \text{ sexp } \cdot \text{ sexps }) \cdot \textsc{rpar}) \lor \textsc{atom}
\]
We automatically remove unreachable productions in the result.

\begin{mathpar}
  \footnotesize
  \inferrule*{
    \inferrule*{
      \inferrule*{}{
        \inferrule*{
          \inferrule*[leftskip=5em,rightskip=8em,vdots=3em]{
            \inferrule*{}
            {
               {\mathcal{N} }  \llbracket \textsc{lpar} \rrbracket = \nfhl{n_{{\mathrm{6}}}  \Rightarrow  \{\,
              n_{{\mathrm{6}}}  \rightarrow  \textsc{lpar}\,\} }
            }
            \and
            \inferrule*{}
            {
              \inferrule*[leftskip=8em,rightskip=8em,vdots=3em]{
                \inferrule*{
                  \inferrule*{}{
                   {\mathcal{N} }  \llbracket \epsilon
                  \rrbracket
                  = \nfhl{n_{{\mathrm{1}}}  \Rightarrow  \{\,  { \color{nfcolor}   n_{{\mathrm{1}}}   \rightarrow   \epsilon  }  \,\} }
                  }
                  \and
                  \inferrule*{
                    \inferrule*{}{
                     {\mathcal{N} }  \llbracket \text{ sexp }
                    \rrbracket
                    = \nfhl{n_{{\mathrm{2}}}  \Rightarrow  \{ \,  n_{{\mathrm{2}}}  \rightarrow  \text{sexp} \,\} }
                  }
                  \and
                  \inferrule*{}
                  {
                     {\mathcal{N} }  \llbracket \text{ sexps }
                    \rrbracket
                    = \nfhl{n_{{\mathrm{3}}}  \Rightarrow  \{ \, n_{{\mathrm{3}}}  \rightarrow  \text{sexps}  \,\} }
                  }
                  }{
                     {\mathcal{N} }  \llbracket \text{ sexp } \cdot \text{ sexps }
                    \rrbracket
                    = \nfhl{n_{{\mathrm{4}}}  \Rightarrow  \{ \,  n_{{\mathrm{4}}}  \rightarrow  \text{sexp}\, n_{{\mathrm{3}}}, n_{{\mathrm{3}}}  \rightarrow  \text{sexps}  \,\} }}
                }
                {
                   {\mathcal{N} }  \llbracket \epsilon \lor \text{ sexp } \cdot \text{ sexps }
                  \rrbracket
                  =
                  \nfhl { n_{{\mathrm{5}}}  \Rightarrow  \{ \,  { \color{nfcolor}   n_{{\mathrm{5}}}   \rightarrow   \epsilon  } , n_{{\mathrm{5}}}  \rightarrow  \text{sexp}\, n_{{\mathrm{3}}}, n_{{\mathrm{3}}}  \rightarrow  \text{sexps}  \,\} }
                }
              }
              {
               {\mathcal{N} }  \llbracket \mu \text{
                sexps }. \,\epsilon \lor \text{ sexp } \cdot \text{ sexps }
               \rrbracket
               = \nfhl{ \text{sexps}  \Rightarrow  \{ \,  \text{sexps}  \rightarrow   \epsilon , \text{sexps}
                \rightarrow  \text{sexp}\, n_{{\mathrm{3}}}, n_{{\mathrm{3}}}  \rightarrow   \epsilon , n_{{\mathrm{3}}}
                \rightarrow  \text{sexp}\, n_{{\mathrm{3}}}  \,\} }
              }
            }
          }{
             {\mathcal{N} }  \llbracket \textsc{lpar} \cdot (\mu \text{
              sexps }. \,\epsilon \lor \text{ sexp } \cdot \text{ sexps }) \rrbracket
            = \nfhl{ n_{{\mathrm{7}}}  \Rightarrow  \{ \, n_{{\mathrm{7}}}  \rightarrow  \textsc{lpar}\, \text{sexps},  \text{sexps}  \rightarrow   \epsilon , \text{sexps}
             \rightarrow  \text{sexp}\, n_{{\mathrm{3}}}, n_{{\mathrm{3}}}  \rightarrow   \epsilon , n_{{\mathrm{3}}}
             \rightarrow  \text{sexp}\, n_{{\mathrm{3}}}  \,\} }
          }
          \and
          \inferrule*{}
          {
             {\mathcal{N} }  \llbracket \textsc{rpar} \rrbracket =
            \nfhl{ \text{rpar}  \Rightarrow  \{ \,  \text{rpar}  \rightarrow  \textsc{rpar} \, \} }
          }
        }
        {
         {\mathcal{N} }  \llbracket \textsc{lpar} \cdot (\mu \text{
          sexps }. \,\epsilon \lor \text{ sexp } \cdot \text{ sexps }) \cdot
        \textsc{rpar} \rrbracket
        = \nfhl{ n_{{\mathrm{8}}}  \Rightarrow  \{ \, n_{{\mathrm{8}}}  \rightarrow  \textsc{lpar}\, \text{sexps}\, \text{rpar},  \text{sexps}  \rightarrow   \epsilon , \text{sexps}
         \rightarrow  \text{sexp}\, n_{{\mathrm{3}}}, n_{{\mathrm{3}}}  \rightarrow   \epsilon , n_{{\mathrm{3}}}
         \rightarrow  \text{sexp}\, n_{{\mathrm{3}}},
        \text{rpar}  \rightarrow  \textsc{rpar}
        \,\} }
        }
      }
      \and
      \inferrule*{}{
         {\mathcal{N} }  \llbracket \textsc{atom} \rrbracket = \nfhl{n_{{\mathrm{9}}}  \Rightarrow  \{ \,
        n_{{\mathrm{9}}}  \rightarrow  \textsc{atom} \, \} }
      }
    }
    {
     {\mathcal{N} }  \llbracket (\textsc{lpar} \cdot (\mu \text{
      sexps }. \,\epsilon \lor \text{ sexp } \cdot \text{ sexps }) \cdot
    \textsc{rpar}) \lor \textsc{atom} \rrbracket
    = \nfhl{ n_{{\mathrm{10}}}  \Rightarrow  \{ \, n_{{\mathrm{10}}}  \rightarrow  \textsc{lpar}\, \text{sexps}\, \text{rpar},
    n_{{\mathrm{10}}}  \rightarrow  \textsc{atom},
    \text{sexps}  \rightarrow   \epsilon , \text{sexps}
     \rightarrow  \text{sexp}\, n_{{\mathrm{3}}}, n_{{\mathrm{3}}}  \rightarrow   \epsilon , n_{{\mathrm{3}}}
     \rightarrow  \text{sexp}\, n_{{\mathrm{3}}},
    \text{rpar}  \rightarrow  \textsc{rpar}
    \,\} }
    }
  }
  {  {\mathcal{N} }  \llbracket
    g
    \rrbracket
    =
    \nfhl{ \text{sexp}  \Rightarrow  \{ \, \text{sexp}  \rightarrow  \textsc{lpar}\, \text{sexps}\, \text{rpar},
    \text{sexp}  \rightarrow  \textsc{atom},
    \text{sexps}  \rightarrow   \epsilon ,
    \text{sexps}  \rightarrow  \textsc{lpar}\, \text{sexps} \, \text{rpar} \, n_{{\mathrm{3}}},
    \text{sexps}  \rightarrow  \textsc{atom}\, n_{{\mathrm{3}}},
    n_{{\mathrm{3}}}  \rightarrow   \epsilon ,
    n_{{\mathrm{3}}}  \rightarrow  \textsc{lpar}\, \text{sexps} \, \text{rpar} \, n_{{\mathrm{3}}},
    n_{{\mathrm{3}}}  \rightarrow  \textsc{atom}\, n_{{\mathrm{3}}},
    \text{rpar}  \rightarrow  \textsc{rpar}
    \,\} }
  }
\end{mathpar}

Comparing the simplified derivation in \Cref{section:normalization} with the
complete derivation, we note the following simplification: first, we omit
the derivation of tokens; second, when normalizing $\text{sexps}$ we produce a
nonterminal $n_{{\mathrm{3}}}$ with a production $n_{{\mathrm{3}}}  \rightarrow  \text{sexps}$. That means
$n_{{\mathrm{3}}}$ is equivalent to $\text{sexps}$. However, this $n_{{\mathrm{3}}}$ is retained in
the final result, making the final grammar a big larger. It's easy to check that
the grammar is equivalent to the one given in the paper.

It is easy to consider an optimization
process that gets rid of $n_{{\mathrm{3}}}$ in the middle of the derivation. For example,
for the result for $n_{{\mathrm{5}}}$, instead of
\[
  \nfhl{n_{{\mathrm{5}}}  \Rightarrow   { \color{nfcolor}   n_{{\mathrm{5}}}   \rightarrow   \epsilon  } , n_{{\mathrm{5}}}  \rightarrow  \text{sexp}\, n_{{\mathrm{3}}}, n_{{\mathrm{3}}}  \rightarrow  \text{sexps}}
\]
We can have
\[
  \nfhl{n_{{\mathrm{5}}}  \Rightarrow   { \color{nfcolor}   n_{{\mathrm{5}}}   \rightarrow   \epsilon  } , n_{{\mathrm{5}}}  \rightarrow  \text{sexp}\, \text{sexps}}
\]
Then the normalization result would be exactly the same as the one in the paper.

\end{landscape}

\pagebreak

\section{Deterministic Parsing}

\theoremdetparsing*

\proof

By straightforward induction on $ \nfhl{G}   \vdash  \nfhl{ n }  \leadsto   w $.

\qed

\section{Well-typed Normalization}

This section presents well-typed normalization, which shows how normalization
captures the type information, and then proves its properties that are important
for later proofs.

First, we note that during normalization (\Cref{figure:normalization}), we
create one fresh nonterminal exactly for one context-free expression.
Therefore, we can attach to each nonterminal its type information. That is,
instead of $n$, we have $ \nfhl{ { n }_{ \tau } } $, where $\tau$ indicates the
type of $n$. We also write $ { \alpha }_{ \tau } $ where $\tau$ is the type
of $\alpha$ as in $ \mu \alpha  :  \tau .~ g $.

Refining the normalization, we have:

\noindent
\begin{small}
\begin{tabular}{llll}
  \multicolumn{4}{l}{\qquad\qquad $ {\mathcal{N} }  \llbracket \,  g  \, \rrbracket $ returns $ \nfhl{  \nfhl{ { n }_{ \tau } }  }  \Rightarrow  \nfhl{ \nfhl{G} } $, with a grammar $\nfhl{G}$ and the start nonterminal $n$ of type $\tau$ (with $ n~\mathsf{fresh} $)}
  \\[5pt]
  \repsilon \qquad & $ {\mathcal{N} }  \llbracket \,  \epsilon  \, \rrbracket $  & $=$  & $ \nfhl{  \nfhl{ { n }_{  \tau_{\epsilon}  } }  }  \Rightarrow  \nfhl{  \nfhl{\{\,   { \color{nfcolor}    \nfhl{ { n }_{  \tau_{\epsilon}  } }    \rightarrow   \epsilon  }   \,\} }  } $  \\
  \rchar & $ {\mathcal{N} }  \llbracket \,   t   \, \rrbracket $       & $=$  & $ \nfhl{  \nfhl{ { n }_{  \tau_{c}  } }  }  \Rightarrow  \nfhl{  \nfhl{\{\,   { \color{nfcolor}    \nfhl{ { n }_{  \tau_{c}  } }    \rightarrow    \nfhl{ t }   }   \,\} }  } $ \\
  \rbot & $ {\mathcal{N} }  \llbracket \,  \bot  \, \rrbracket $      & $=$  & $ \nfhl{  \nfhl{ { n }_{  \tau_{\bot}  } }  }  \Rightarrow  \nfhl{  \nfhl{ \emptyset }  }  $\\
  \\
  \rseq & $ {\mathcal{N} }  \llbracket \,   g_{{\mathrm{1}}}  \cdot  g_{{\mathrm{2}}}   \, \rrbracket $& $=$  & $  \nfhl{ { n }_{  \tau_{{\mathrm{1}}}  \cdot  \tau_{{\mathrm{2}}}  } }   \Rightarrow  \{  { \color{nfcolor}    \nfhl{ { n }_{  \tau_{{\mathrm{1}}}  \cdot  \tau_{{\mathrm{2}}}  } }    \rightarrow    \nfhl{ \nfhl{N}_{{\mathrm{1}}} ~  \nfhl{ { n_{{\mathrm{2}}} }_{ \tau_{{\mathrm{2}}} } }  }   }  \mid  { \color{nfcolor}    \nfhl{ { n_{{\mathrm{1}}} }_{ \tau_{{\mathrm{1}}} } }    \rightarrow   \nfhl{N}_{{\mathrm{1}}}  }  \in \nfhl{G}_{{\mathrm{1}}}  \}  \cup   \nfhl{ \nfhl{G}_{{\mathrm{1}}}   \cup   \nfhl{G}_{{\mathrm{2}}} }  $ \\
  &                          &      & where \enskip $ {\mathcal{N} }  \llbracket \,  g_{{\mathrm{1}}}  \, \rrbracket  =  \nfhl{  \nfhl{ { n_{{\mathrm{1}}} }_{ \tau_{{\mathrm{1}}} } }  }  \Rightarrow  \nfhl{ \nfhl{G}_{{\mathrm{1}}} }  \land  {\mathcal{N} }  \llbracket \,  g_{{\mathrm{2}}}  \, \rrbracket  =  \nfhl{  \nfhl{ { n_{{\mathrm{2}}} }_{ \tau_{{\mathrm{2}}} } }  }  \Rightarrow  \nfhl{ \nfhl{G}_{{\mathrm{2}}} }  $ \\
  \\[2pt]
  \ralt & $ {\mathcal{N} }  \llbracket \,   g_{{\mathrm{1}}}  \lor  g_{{\mathrm{2}}}   \, \rrbracket $& $=$  & $ \nfhl{ { n }_{ \tau_{{\mathrm{1}}}  \vee  \tau_{{\mathrm{2}}} } }   \Rightarrow  \{  { \color{nfcolor}    \nfhl{ { n }_{ \tau_{{\mathrm{1}}}  \vee  \tau_{{\mathrm{2}}} } }    \rightarrow   \nfhl{N}_{{\mathrm{1}}}  }  \mid  { \color{nfcolor}    \nfhl{ { n_{{\mathrm{1}}} }_{ \tau_{{\mathrm{1}}} } }    \rightarrow   \nfhl{N}_{{\mathrm{1}}}  }  \in \nfhl{G}_{{\mathrm{1}}} \}  \cup  \{  { \color{nfcolor}    \nfhl{ { n }_{ \tau_{{\mathrm{1}}}  \vee  \tau_{{\mathrm{2}}} } }    \rightarrow   \nfhl{N}_{{\mathrm{2}}}  }  \mid  { \color{nfcolor}    \nfhl{ { n_{{\mathrm{2}}} }_{ \tau_{{\mathrm{2}}} } }    \rightarrow   \nfhl{N}_{{\mathrm{2}}}  }  \in \nfhl{G}_{{\mathrm{2}}} \} $    \\
                   & &       & $ \cup  \,\,  \nfhl{ \nfhl{G}_{{\mathrm{1}}}   \cup   \nfhl{G}_{{\mathrm{2}}} }  $ \\
  &                             &      & where \enskip $ {\mathcal{N} }  \llbracket \,  g_{{\mathrm{1}}}  \, \rrbracket  =  \nfhl{  \nfhl{ { n_{{\mathrm{1}}} }_{ \tau_{{\mathrm{1}}} } }  }  \Rightarrow  \nfhl{ \nfhl{G}_{{\mathrm{1}}} }  \land  {\mathcal{N} }  \llbracket \,  g_{{\mathrm{2}}}  \, \rrbracket  =  \nfhl{  \nfhl{ { n_{{\mathrm{2}}} }_{ \tau_{{\mathrm{2}}} } }  }  \Rightarrow  \nfhl{ \nfhl{G}_{{\mathrm{2}}} }  $ \\
  \\
  \rfix& $ {\mathcal{N} }  \llbracket \,   \mu \alpha  :  \tau .~ g   \, \rrbracket $     & $=$  & $ \nfhl{ {  {\nfhl{\alpha} }  }_{ \tau } }   \Rightarrow  { \{  { \color{nfcolor}    \nfhl{ {  {\nfhl{\alpha} }  }_{ \tau } }    \rightarrow   \nfhl{N}  }   \mid  { \color{nfcolor}    \nfhl{ { n }_{ \tau } }    \rightarrow   \nfhl{N}  }  \in \nfhl{G} \} }$ 
                                             $  \cup  { \{  { \color{nfcolor}    \nfhl{ { n' }_{ \tau' } }    \rightarrow    \nfhl{ \nfhl{N} ~ \overline{n}' }   }  \mid  { \color{nfcolor}    \nfhl{ { n' }_{ \tau' } }    \rightarrow    \nfhl{  \nfhl{  \nfhl{ {  {\nfhl{\alpha} }  }_{ \tau } }  }  ~ \overline{n}' }   }  \in \nfhl{G} \land   { \color{nfcolor}    \nfhl{ { n }_{ \tau } }    \rightarrow   \nfhl{N}  }  \in \nfhl{G} \} }$ \\
  & & & $  \cup  {{\nfhl{G}}\backslash_{ { \color{nfcolor}    \nfhl{ { n' }_{ \tau' } }    \rightarrow    \nfhl{  \nfhl{  \nfhl{ {  {\nfhl{\alpha} }  }_{ \tau } }  }  ~ \overline{n}' }   } }}
                                             $\\
                   &                       &      & where \enskip  $ {\mathcal{N} }  \llbracket \,  g  \, \rrbracket  =  \nfhl{  \nfhl{ { n }_{ \tau } }  }  \Rightarrow  \nfhl{ \nfhl{G} } $ \\
                   &                       & &  \qquad\quad\, $ \nfhl{G}\backslash_{ { \color{nfcolor}    \nfhl{ { n' }_{ \tau' } }    \rightarrow    \nfhl{  \nfhl{  \nfhl{ {  {\nfhl{\alpha} }  }_{ \tau } }  }  ~ \overline{n}' }   } }$ is $\nfhl{G}$ with all $ { \color{nfcolor}    \nfhl{ { n' }_{ \tau' } }    \rightarrow    \nfhl{  \nfhl{  \nfhl{ {  {\nfhl{\alpha} }  }_{ \tau } }  }  ~ \overline{n}' }   } $ removed for any $n'$, $\tau'$ and $\overline{n}'$
  \\[4pt]
  \rvar & $ {\mathcal{N} }  \llbracket \,   { \alpha }_{ \tau }   \, \rrbracket $           & $=$  & $ \nfhl{ { n }_{ \tau } }   \Rightarrow   \nfhl{\{\,   { \color{nfcolor}    \nfhl{ { n }_{ \tau } }    \rightarrow    \nfhl{  \nfhl{ {  {\nfhl{\alpha} }  }_{ \tau } }  }   }   \,\} } $  \\
\end{tabular}
\end{small}

We also add to typing that
\[
\inferrule{ }{\Gamma  \ottsym{;}  \Delta  \vdash   \nfhl{ { n }_{ \tau } }   \ottsym{:}  \tau}
\]

\noindent
With that, we can type-check any $\nfhl{N}$ according to the typing rules,
by treating $t$ as constants, $ \nfhl{ { n }_{ \tau } } $ as nonterminal of type $\tau$,
and lists as sequences (e.g.~$ \nfhl{ { n_{{\mathrm{1}}} }_{ \tau_{{\mathrm{1}}} } }  \,  \nfhl{ { n_{{\mathrm{2}}} }_{ \tau_{{\mathrm{2}}} } } $ as $ \nfhl{ { n_{{\mathrm{1}}} }_{ \tau_{{\mathrm{1}}} } }  \cdot \nfhl{ { n_{{\mathrm{2}}} }_{ \tau_{{\mathrm{2}}} } } $).

Now we can prove properties about the well-typed normalization.
While those lemmas are proved in the typed normalization,
they naturally hold for the untyped normalization as the two are the same process.

\begin{lemma}
  Given $\Gamma  \ottsym{;}  \Delta  \vdash  g  \ottsym{:}  \tau$,
  and $ {\mathcal{N} }  \llbracket \,  g  \, \rrbracket $ returns $  \nfhl{  \nfhl{ { n }_{ \tau' } }  }  \Rightarrow  \nfhl{ \nfhl{G} } $,
  then $\tau = \tau'$.
\end{lemma}

\proof By a straightforward induction on $\Gamma  \ottsym{;}  \Delta  \vdash  g  \ottsym{:}  \tau$. \qed

\begin{lemma}
  \label{lemma:well-typednf}
  Given $\Gamma  \ottsym{;}  \Delta  \vdash  g  \ottsym{:}  \tau'$,
  and $ {\mathcal{N} }  \llbracket \,  g  \, \rrbracket $ returns $  \nfhl{ \_ }  \Rightarrow  \nfhl{ \nfhl{G} } $,
  then for any $ \nfhl{ { n }_{ \tau } }  \in G$,
  if $\nfhl{N}_{{\mathrm{1}}},...,\nfhl{N}_{\ottmv{i}}$ are all productions of $n$.
  we have
  \begin{itemize}
    \item $\tau = \tau_{{\mathrm{1}}}  \vee  \tau_{{\mathrm{2}}}  \vee  \cdots  \vee  \tau_{\ottmv{i}}$, where
    \item
      ($ { \color{nfcolor}    \nfhl{ { n }_{ \tau } }    \rightarrow   \nfhl{N}_{{\mathrm{1}}}  }  \, \in \, \nfhl{G} \land \Gamma  \ottsym{;}  \Delta  \vdash  \nfhl{N}_{{\mathrm{1}}}  \ottsym{:}  \tau_{{\mathrm{1}}}$)
      and
      ($ { \color{nfcolor}    \nfhl{ { n }_{ \tau } }    \rightarrow   \nfhl{N}_{{\mathrm{2}}}  }  \, \in \, \nfhl{G} \land \Gamma  \ottsym{;}  \Delta  \vdash  \nfhl{N}_{{\mathrm{2}}}  \ottsym{:}  \tau_{{\mathrm{2}}}$)
      and
      $\cdots$
      and
      ($ { \color{nfcolor}    \nfhl{ { n }_{ \tau } }    \rightarrow   \nfhl{N}_{\ottmv{i}}  }  \, \in \, \nfhl{G} \land \Gamma  \ottsym{;}  \Delta  \vdash  \nfhl{N}_{\ottmv{i}}  \ottsym{:}  \tau_{\ottmv{i}}$); and
    \item $ \tau_{{\mathrm{1}}} \;  \#  \;  \tau_{{\mathrm{2}}}  \cdots  \# \; \tau_{\ottmv{i}}$, i.e.~all $\tau_{{\mathrm{1}}}$,
      $\tau_{{\mathrm{2}}}$, $\cdots$, $\tau_{\ottmv{i}}$ are apart from each other.
  \end{itemize}
\end{lemma}

\proof By induction on $\Gamma  \ottsym{;}  \Delta  \vdash  g  \ottsym{:}  \tau$.

\begin{itemize}
  \item The cases for $ \epsilon $, $t$, $ \bot $ and $\alpha$ follow trivially.
  \item The case for $ g_{{\mathrm{1}}}  \cdot  g_{{\mathrm{2}}} $.

    \begin{xtabular}{l}
    $ {\mathcal{N} }  \llbracket \,   g_{{\mathrm{1}}}  \cdot  g_{{\mathrm{2}}}   \, \rrbracket  =  \nfhl{ { n }_{  \tau_{{\mathrm{1}}}  \cdot  \tau_{{\mathrm{2}}}  } }   \Rightarrow  \{  { \color{nfcolor}    \nfhl{ { n }_{  \tau_{{\mathrm{1}}}  \cdot  \tau_{{\mathrm{2}}}  } }    \rightarrow    \nfhl{ \nfhl{N}_{{\mathrm{1}}} ~  \nfhl{ { n_{{\mathrm{2}}} }_{ \tau_{{\mathrm{2}}} } }  }   }  \mid  { \color{nfcolor}    \nfhl{ { n_{{\mathrm{1}}} }_{ \tau_{{\mathrm{1}}} } }    \rightarrow   \nfhl{N}_{{\mathrm{1}}}  }  \in \nfhl{G}_{{\mathrm{1}}}  \}  \cup   \nfhl{ \nfhl{G}_{{\mathrm{1}}}   \cup   \nfhl{G}_{{\mathrm{2}}} }  $ \\
    $ {\mathcal{N} }  \llbracket \,  g_{{\mathrm{1}}}  \, \rrbracket  =  \nfhl{  \nfhl{ { n_{{\mathrm{1}}} }_{ \tau_{{\mathrm{1}}} } }  }  \Rightarrow  \nfhl{ \nfhl{G}_{{\mathrm{1}}} }  \land  {\mathcal{N} }  \llbracket \,  g_{{\mathrm{2}}}  \, \rrbracket  =  \nfhl{  \nfhl{ { n_{{\mathrm{2}}} }_{ \tau_{{\mathrm{2}}} } }  }  \Rightarrow  \nfhl{ \nfhl{G}_{{\mathrm{2}}} }  $ \\
    \end{xtabular}

    By I.H., we know that for each $\nfhl{N}_{{\mathrm{1}}}$,
    we have $\Gamma  \ottsym{;}  \Delta  \vdash  \nfhl{N}_{{\mathrm{1}}}  \ottsym{:}  \tau'_{{\mathrm{1}}}$ for some $\tau'_{{\mathrm{1}}}$, and $\tau_{{\mathrm{1}}}$ is the $ \vee $
    of all $\tau'_{{\mathrm{1}}}$, and all $\tau'_{{\mathrm{1}}}$ is apart ($ \# $) from each other.

    According to well-typedness, we know that $\tau_{{\mathrm{1}}}  \circledast  \tau_{{\mathrm{2}}}$, which says that
    $\tau_{{\mathrm{1}}}.\Followlast \cap \tau_{{\mathrm{2}}}.\First = \emptyset$,
    and $\lnot \tau_{{\mathrm{1}}}.\Null$.

    Since $\tau_{{\mathrm{1}}}$ is the $ \vee $ of all $\tau'_{{\mathrm{1}}}$, we know
    $\tau'_{{\mathrm{1}}}.\Null = \False$, and $\tau'_{{\mathrm{1}}}.\Followlast \cap \tau_{{\mathrm{2}}}.\First =
    \emptyset$,
    and thus $\tau'_{{\mathrm{1}}}  \circledast  \tau_{{\mathrm{2}}}$.

    So $\Gamma  \ottsym{;}  \Delta  \vdash   \nfhl{ \nfhl{N}_{{\mathrm{1}}} ~  \nfhl{ { n_{{\mathrm{2}}} }_{ \tau_{{\mathrm{2}}} } }  }   \ottsym{:}   \tau'_{{\mathrm{1}}}  \cdot  \tau_{{\mathrm{2}}} $

    Moreover,
    since $\tau'_{{\mathrm{1}}}.\Null = \False$,
    we have $ \tau'_{{\mathrm{1}}}  \cdot  \tau_{{\mathrm{2}}} .\First = \tau'_{{\mathrm{1}}}.\First$
    and $ \tau'_{{\mathrm{1}}}  \cdot  \tau_{{\mathrm{2}}} .\Null = \False$. Given that all $\tau'_{{\mathrm{1}}}$ apart
    from each other, we can derive that all $ \tau'_{{\mathrm{1}}}  \cdot  \tau_{{\mathrm{2}}} $ apart from each other.

  \item The case for

    \begin{xtabular}{l}
    $ {\mathcal{N} }  \llbracket \,   g_{{\mathrm{1}}}  \lor  g_{{\mathrm{2}}}   \, \rrbracket =  \nfhl{ { n }_{ \tau_{{\mathrm{1}}}  \vee  \tau_{{\mathrm{2}}} } }   \Rightarrow  \{  { \color{nfcolor}    \nfhl{ { n }_{ \tau_{{\mathrm{1}}}  \vee  \tau_{{\mathrm{2}}} } }    \rightarrow   \nfhl{N}_{{\mathrm{1}}}  }  \mid  { \color{nfcolor}    \nfhl{ { n_{{\mathrm{1}}} }_{ \tau_{{\mathrm{1}}} } }    \rightarrow   \nfhl{N}_{{\mathrm{1}}}  }  \in \nfhl{G}_{{\mathrm{1}}} \}  \cup  \{  { \color{nfcolor}    \nfhl{ { n }_{ \tau_{{\mathrm{1}}}  \vee  \tau_{{\mathrm{2}}} } }    \rightarrow   \nfhl{N}_{{\mathrm{2}}}  }  \mid  { \color{nfcolor}    \nfhl{ { n_{{\mathrm{2}}} }_{ \tau_{{\mathrm{2}}} } }    \rightarrow   \nfhl{N}_{{\mathrm{2}}}  }  \in \nfhl{G}_{{\mathrm{2}}} \} $    \\
    $ \cup  \,\,  \nfhl{ \nfhl{G}_{{\mathrm{1}}}   \cup   \nfhl{G}_{{\mathrm{2}}} }  $ \\
    $ {\mathcal{N} }  \llbracket \,  g_{{\mathrm{1}}}  \, \rrbracket  =  \nfhl{  \nfhl{ { n_{{\mathrm{1}}} }_{ \tau_{{\mathrm{1}}} } }  }  \Rightarrow  \nfhl{ \nfhl{G}_{{\mathrm{1}}} }  \land  {\mathcal{N} }  \llbracket \,  g_{{\mathrm{2}}}  \, \rrbracket  =  \nfhl{  \nfhl{ { n_{{\mathrm{2}}} }_{ \tau_{{\mathrm{2}}} } }  }  \Rightarrow  \nfhl{ \nfhl{G}_{{\mathrm{2}}} }  $ \\
    \end{xtabular}

    By I.H., we know that for each $\nfhl{N}_{{\mathrm{1}}}$,
    we have $\Gamma  \ottsym{;}  \Delta  \vdash  \nfhl{N}_{{\mathrm{1}}}  \ottsym{:}  \tau'_{{\mathrm{1}}}$ for some $\tau'_{{\mathrm{1}}}$, and $\tau_{{\mathrm{1}}}$ is the $ \vee $
    of all $\tau'_{{\mathrm{1}}}$, and all $\tau'_{{\mathrm{1}}}$ is apart ($ \# $) from each other.
    Moreover, for each $\nfhl{N}_{{\mathrm{2}}}$,
    we have $\Gamma  \ottsym{;}  \Delta  \vdash  \nfhl{N}_{{\mathrm{2}}}  \ottsym{:}  \tau'_{{\mathrm{2}}}$ for some $\tau'_{{\mathrm{2}}}$, and $\tau_{{\mathrm{2}}}$ is the $ \vee $
    of all $\tau'_{{\mathrm{2}}}$, and all $\tau'_{{\mathrm{2}}}$ is apart ($ \# $) from each other.

    It's easy to see that $\tau_{{\mathrm{1}}}  \vee  \tau_{{\mathrm{2}}}$ is the $ \vee $ of all $\tau'_{{\mathrm{1}}}$ and $\tau'_{{\mathrm{2}}}$.

    According to well-typedness, we know that $ \tau_{{\mathrm{1}}} \;  \#  \;  \tau_{{\mathrm{2}}} $. That is
    $\tau_{{\mathrm{1}}}.\First \cap \tau_{{\mathrm{2}}}.\First = \emptyset$,
    and $\lnot (\tau_{{\mathrm{1}}}.\Null \land \tau_{{\mathrm{2}}}.\Null) $.
    From the former, we can derive that
    $\tau'_{{\mathrm{1}}}.\First \cap \tau'_{{\mathrm{2}}}.\First = \emptyset$.
    From the latter, we know that at least one of $\tau_{{\mathrm{1}}}$ and $\tau_{{\mathrm{2}}}$ has
    $\Null = \False$, so at least one of $\tau'_{{\mathrm{1}}}$ and $\tau'_{{\mathrm{2}}}$ has $\Null =
    \False$.
    With that, we have $ \tau'_{{\mathrm{1}}} \;  \#  \;  \tau'_{{\mathrm{2}}} $.
    Thus, all $\tau'_{{\mathrm{1}}}$ and $\tau'_{{\mathrm{2}}}$ apart from each other.

  \item The case for $ \mu \alpha  :  \tau .~ g $
 
    \begin{xtabular}{l}
    $ {\mathcal{N} }  \llbracket \,   \mu \alpha  :  \tau .~ g   \, \rrbracket $     $=$   $ \nfhl{ {  {\nfhl{\alpha} }  }_{ \tau } }   \Rightarrow  { \{  { \color{nfcolor}    \nfhl{ {  {\nfhl{\alpha} }  }_{ \tau } }    \rightarrow   \nfhl{N}  }   \mid  { \color{nfcolor}    \nfhl{ { n }_{ \tau } }    \rightarrow   \nfhl{N}  }  \in \nfhl{G} \} }$ 
    $  \cup  { \{  { \color{nfcolor}    \nfhl{ { n' }_{ \tau' } }    \rightarrow    \nfhl{ \nfhl{N} ~ \overline{n}' }   }  \mid  { \color{nfcolor}    \nfhl{ { n' }_{ \tau' } }    \rightarrow    \nfhl{  \nfhl{  \nfhl{ {  {\nfhl{\alpha} }  }_{ \tau } }  }  ~ \overline{n}' }   }  \in \nfhl{G} \land   { \color{nfcolor}    \nfhl{ { n }_{ \tau } }    \rightarrow   \nfhl{N}  }  \in \nfhl{G} \} }$ \\
    \qquad \qquad \quad \qquad \qquad $  \cup  {{\nfhl{G}}\backslash_{ { \color{nfcolor}    \nfhl{ { n' }_{ \tau' } }    \rightarrow    \nfhl{  \nfhl{  \nfhl{ {  {\nfhl{\alpha} }  }_{ \tau } }  }  ~ \overline{n}' }   } }}
    $\\
    $ {\mathcal{N} }  \llbracket \,  g  \, \rrbracket  =  \nfhl{  \nfhl{ { n }_{ \tau } }  }  \Rightarrow  \nfhl{ \nfhl{G} } $ \\
    $ \nfhl{G}\backslash_{ { \color{nfcolor}    \nfhl{ { n' }_{ \tau' } }    \rightarrow    \nfhl{  \nfhl{  \nfhl{ {  {\nfhl{\alpha} }  }_{ \tau } }  }  ~ \overline{n}' }   } }$ is $\nfhl{G}$ with all $ { \color{nfcolor}    \nfhl{ { n' }_{ \tau' } }    \rightarrow    \nfhl{  \nfhl{  \nfhl{ {  {\nfhl{\alpha} }  }_{ \tau } }  }  ~ \overline{n}' }   } $ removed for any $n'$, $\tau'$ and $\overline{n}'$
    \\[4pt]
    \end{xtabular}

    By I.H., we know that for each $\nfhl{N}$,
    we have $\Gamma  \ottsym{;}  \Delta  \vdash  \nfhl{N}  \ottsym{:}  \tau''$ for some $\tau''$, and $\tau$ is the $ \vee $
    of all $\tau''$, and all $\tau''$ is apart ($ \# $) from each other.

    The goal for $ \nfhl{ {  {\nfhl{\alpha} }  }_{ \tau } } $ follows from $ \nfhl{ { n }_{ \tau } } $. The remaining is to
    show that the goal holds for each $ \nfhl{ { n' }_{ \tau' } } $ that has a production that starts
    with $ \nfhl{ {  {\nfhl{\alpha} }  }_{ \tau } } $. Essentially what happens is that one production
    $ { \color{nfcolor}    \nfhl{ { n' }_{ \tau' } }    \rightarrow    \nfhl{  \nfhl{  \nfhl{ {  {\nfhl{\alpha} }  }_{ \tau } }  }  ~ \overline{n}' }   } $ is replaced by multiple productions
    $ { \color{nfcolor}    \nfhl{ { n' }_{ \tau' } }    \rightarrow    \nfhl{ \nfhl{N} ~ \overline{n}' }   } $ for each $ { \color{nfcolor}    \nfhl{ { n }_{ \tau } }    \rightarrow   \nfhl{N}  }  \, \in \, \nfhl{G}$ where $\Gamma  \ottsym{;}  \Delta  \vdash  \nfhl{N}  \ottsym{:}  \tau''$.

    First, we need to show that $ \nfhl{ \nfhl{N} ~ \overline{n}' } $ is well-typed. We already
    know each individual terminal or nonterminal in $ \nfhl{ \nfhl{N} ~ \overline{n}' } $ is well-typed,
    so the only requirement is the $ \circledast $ condition during type-checking. 
    Given that $ \nfhl{ {  {\nfhl{\alpha} }  }_{ \tau } }  \, \overline{n}'$ is well-typed, we know that $\tau.\Null =
    \False$, so  $\tau''.\Null = \False$. Moreover, $\tau''.\Followlast
    \subseteq \tau.\Followlast$. With that, and the fact that $ \nfhl{ {  {\nfhl{\alpha} }  }_{ \tau } }  \, \overline{n}'$
    is well-typed, we can derive that the $ \circledast $
    condition is always satisfied when type-checking $ \nfhl{ \nfhl{N} ~ \overline{n}' } $.
    Therefore, $ \nfhl{ \nfhl{N} ~ \overline{n}' } $ is well-typed.

    Because $\tau$ is the $ \vee $ of all $\tau''$, it's easy to show that
    the type of $ \nfhl{ {  {\nfhl{\alpha} }  }_{ \tau } }  \, \overline{n}'$ is the $ \vee $ of the types of all $ \nfhl{ \nfhl{N} ~ \overline{n}' } $.
    Therefore,
    the type of $n'$ is the same as before.
    Also, 
    all types of the productions
    of $n'$ are still apart with each other.

\end{itemize}

\qed

\section{Normalization is well-defined (proof)}
\label{section:well-definedness}

\lemmatotalnullfalse*

\proof

  \textbf{Left to right}
  According to \Cref{lemma:well-typednf}, we must have one $ { \color{nfcolor}    \nfhl{ { n }_{ \tau } }    \rightarrow   \nfhl{N}  }  \, \in \, \nfhl{G}$,
  where $\Gamma  \ottsym{;}  \Delta  \vdash  \nfhl{N}  \ottsym{:}  \tau$, and $\tau.\Null = \True$. We case analyze the
  shape of $\nfhl{N}$:

  \begin{itemize}
  \item If $\nfhl{N} =  \epsilon $, then we have proved (1).
  \item If $\nfhl{N} =  \nfhl{  \nfhl{ t }  ~ \overline{n} } $, then it's impossible that $\tau.\Null = \True$.
  \item If $\nfhl{N} =  {\nfhl{\alpha} }  \, \overline{n}$. Since $ {\nfhl{\alpha} }  \, \overline{n}$ is well-typed, if $\overline{n}$ is
    not empty, then the type must have $\Null = \False$. Therefore $\overline{n}$
    must be empty, and $\alpha$ has its type $\Null = \True$. So we have proved (2).
  \end{itemize}

  \textbf{Right to left}
    Following \Cref{lemma:well-typednf}, the type $\tau$ is
    the $ \vee $ of all types. If either $ { \color{nfcolor}   n   \rightarrow   \epsilon  } $
    of $\alpha$ has type $\Null = \True$, we know that $\tau.\Null = \True$.

\qed

\lemmatotality*

\proof

By induction on $g$. Most cases are straightforward. The only interesting
cases are when $g =  g_{{\mathrm{1}}}  \cdot  g_{{\mathrm{2}}} $ or $g  \ottsym{=}   \mu \alpha .~ g' $.

\begin{itemize}
\item $g  \ottsym{=}   g_{{\mathrm{1}}}  \cdot  g_{{\mathrm{2}}} $. We have:

  \begin{xtabular}{l}
  $ {\mathcal{N} }  \llbracket \,   g_{{\mathrm{1}}}  \cdot  g_{{\mathrm{2}}}   \, \rrbracket $ $=$  $ n  \Rightarrow  \{  { \color{nfcolor}   n   \rightarrow    \nfhl{ \nfhl{N}_{{\mathrm{1}}} ~ n_{{\mathrm{2}}} }   }  \mid  { \color{nfcolor}   n_{{\mathrm{1}}}   \rightarrow   \nfhl{N}_{{\mathrm{1}}}  }  \in \nfhl{G}_{{\mathrm{1}}}  \}  \cup   \nfhl{ \nfhl{G}_{{\mathrm{1}}}   \cup   \nfhl{G}_{{\mathrm{2}}} }  $ \\
  $ {\mathcal{N} }  \llbracket \,  g_{{\mathrm{1}}}  \, \rrbracket  =  \nfhl{ n_{{\mathrm{1}}} }  \Rightarrow  \nfhl{ \nfhl{G}_{{\mathrm{1}}} }  \land  {\mathcal{N} }  \llbracket \,  g_{{\mathrm{2}}}  \, \rrbracket  =  \nfhl{ n_{{\mathrm{2}}} }  \Rightarrow  \nfhl{ \nfhl{G}_{{\mathrm{2}}} }  $ \\
  \end{xtabular}

  As $ g_{{\mathrm{1}}}  \cdot  g_{{\mathrm{2}}} $ is well-typed, we know
  that the type of $g_{{\mathrm{1}}}$ has $\Null =  \mathsf{false} $.
  By \Cref{lemma:total:null:false}, $\nfhl{N}_{{\mathrm{1}}}$ is not $ \epsilon $, ensuring that
  $ \nfhl{ \nfhl{N}_{{\mathrm{1}}} ~ n_{{\mathrm{2}}} } $ is a valid form.
\item $g  \ottsym{=}   \mu \alpha .~ g' $. We have:

  \begin{xtabular}{l}
    $\Gamma  \ottsym{;}  \Delta  \ottsym{,}  \alpha  \ottsym{:}  \tau  \vdash  g  \ottsym{:}  \tau$ \\
    $ {\mathcal{N} }  \llbracket \,   \mu \alpha .~ g   \, \rrbracket $      $=$   $ {\nfhl{\alpha} }   \Rightarrow   \{  { \color{nfcolor}    {\nfhl{\alpha} }    \rightarrow   \nfhl{N}  }   \mid  { \color{nfcolor}   n   \rightarrow   \nfhl{N}  }  \in \nfhl{G} \} $
    $  \cup   \{  { \color{nfcolor}   n'   \rightarrow    \nfhl{ \nfhl{N} ~ \overline{n}' }   }  \mid  { \color{nfcolor}   n'   \rightarrow    \nfhl{  \nfhl{  {\nfhl{\alpha} }  }  ~ \overline{n}' }   }  \in
      \nfhl{G} \land   { \color{nfcolor}   n   \rightarrow   \nfhl{N}  }  \in \nfhl{G} \}
     \cup  {\nfhl{G}}\backslash_{ { \color{nfcolor}   n'   \rightarrow    \nfhl{  \nfhl{  {\nfhl{\alpha} }  }  ~ \overline{n}' }   } } $\\
    $ {\mathcal{N} }  \llbracket \,  g  \, \rrbracket  =  \nfhl{ n }  \Rightarrow  \nfhl{ \nfhl{G} } $ \\
    $ \nfhl{G}\backslash_{ { \color{nfcolor}   n'   \rightarrow    \nfhl{  \nfhl{  {\nfhl{\alpha} }  }  ~ \overline{n}' }   } }$ is $\nfhl{G}$ with all $ { \color{nfcolor}   n'   \rightarrow    \nfhl{  \nfhl{  {\nfhl{\alpha} }  }  ~ \overline{n}' }   } $ removed for any $n'$ and $\overline{n}'$
  \end{xtabular}%

  We need to show that $ \nfhl{ \nfhl{N} ~ \overline{n}' } $ is
  valid, requiring either $\nfhl{N}$ to not be $ \epsilon $,
  or $\overline{n}'$ to be empty.
  Since $ {\nfhl{\alpha} }  \, \overline{n}'$ is well-typed (\Cref{lemma:well-typednf}), we know that
  either $\overline{n}'$ is empty, or $ {\nfhl{\alpha} } $ must have $\Null = \False$.
  In the first case we are done.
  In the second case, following \Cref{lemma:total:null:false},
  we know $\nfhl{N}$ cannot be $ \epsilon $.
\end{itemize}

\qed

\section{Normalization returns  DGNF grammars (proof)}

\subsection{Normalizing closed expressions produces no $ {\nfhl{\alpha} }  \, \overline{n}$ form}

\lemmanafv*

\proof

\textbf{Part 1}
By induction on $\Gamma  \ottsym{;}  \Delta  \vdash  g  \ottsym{:}  \tau$, most cases are straightforward.
We discuss the following three cases:

\begin{itemize}
\item $g  \ottsym{=}  \alpha$. As $g$ is well-typed, it must be $\alpha \, \in \, \mathsf{dom} \, \ottsym{(}  \Gamma  \ottsym{)}$. The goal
  follows directly.
\item $g  \ottsym{=}   g_{{\mathrm{1}}}  \cdot  g_{{\mathrm{2}}} $. The goal follows by the I.H. on $g_{{\mathrm{1}}}$.
\item $g  \ottsym{=}   \mu \alpha .~ g' $. 
  As the well-typedness of $g'$ adds $\alpha$ to $\Delta$,
  the goal follows directly by the I.H. on $g'$.
\end{itemize}

\textbf{Part 2}
By induction on $\Gamma  \ottsym{;}  \Delta  \vdash  g  \ottsym{:}  \tau$.
The only interesting case is when $g =  \mu \alpha .~ g' $.
We have

\begin{xtabular}{l}
  $\Gamma  \ottsym{;}  \Delta  \ottsym{,}  \alpha  \ottsym{:}  \tau  \vdash  g  \ottsym{:}  \tau$ \\
  $ {\mathcal{N} }  \llbracket \,   \mu \alpha .~ g   \, \rrbracket $      $=$   $ {\nfhl{\alpha} }   \Rightarrow  \underset{\sshl{(1)}}{ \{  { \color{nfcolor}    {\nfhl{\alpha} }    \rightarrow   \nfhl{N}  }   \mid  { \color{nfcolor}   n   \rightarrow   \nfhl{N}  }  \in \nfhl{G} \} }$
  $  \cup  \underset{\sshl{(2)}}{ \{  { \color{nfcolor}   n'   \rightarrow    \nfhl{ \nfhl{N} ~ \overline{n}' }   }  \mid  { \color{nfcolor}   n'   \rightarrow    \nfhl{  \nfhl{  {\nfhl{\alpha} }  }  ~ \overline{n}' }   }  \in
    \nfhl{G} \land   { \color{nfcolor}   n   \rightarrow   \nfhl{N}  }  \in \nfhl{G} \} }
   \cup  \underset{\sshl{(3)}}{{\nfhl{G}}\backslash_{ { \color{nfcolor}   n'   \rightarrow    \nfhl{  \nfhl{  {\nfhl{\alpha} }  }  ~ \overline{n}' }   } }} $\\
  $ {\mathcal{N} }  \llbracket \,  g  \, \rrbracket  =  \nfhl{ n }  \Rightarrow  \nfhl{ \nfhl{G} } $ \\
  $ \nfhl{G}\backslash_{ { \color{nfcolor}   n'   \rightarrow    \nfhl{  \nfhl{  {\nfhl{\alpha} }  }  ~ \overline{n}' }   } }$ is $\nfhl{G}$ with all $ { \color{nfcolor}   n'   \rightarrow    \nfhl{  \nfhl{  {\nfhl{\alpha} }  }  ~ \overline{n}' }   } $ removed for any $n'$ and $\overline{n}'$
\end{xtabular}%

By I.H., we know that for all $\ottsym{(}   \nfhl{  { \color{nfcolor}   n''   \rightarrow    \nfhl{  \nfhl{  {\nfhl{\beta} }  }  ~ \overline{n} }   }  }   \ottsym{)}  \in  \nfhl{G}\backslash_{ { \color{nfcolor}   n'   \rightarrow    \nfhl{  \nfhl{  {\nfhl{\alpha} }  }  ~ \overline{n}' }   } }$, $\beta \, \in \, \mathsf{fv} \, \ottsym{(}  g  \ottsym{)}$.

For $\nfhl{N}$, if it is $ {\nfhl{\beta} }  \, \overline{n}$, then either $\beta \, \in \, \mathsf{fv} \, \ottsym{(}   \mu \alpha .~ g   \ottsym{)}$, or $\beta = \alpha$. By Part 1, we know that $\beta \, \in \, \mathsf{dom} \, \ottsym{(}  \Gamma  \ottsym{)}$, so $\beta \neq \alpha$. So it can only be $\beta \, \in \, \mathsf{fv} \, \ottsym{(}   \mu \alpha .~ g   \ottsym{)}$. And the goal follows.

\qed

\lemmaformcorrect*

\proof Follows directly from \Cref{lemma:na:fv}.
\qed

\subsection{A nonterminal's non-$\epsilon$ productions start with distinct terminals.}

\lemmaterminalfirst*

\proof

\textbf{Left to right}
According to \Cref{lemma:well-typednf}, we must have one $ { \color{nfcolor}    \nfhl{ { n }_{ \tau } }    \rightarrow   \nfhl{N}  }  \, \in \, \nfhl{G}$,
where $\Gamma  \ottsym{;}  \Delta  \vdash  \nfhl{N}  \ottsym{:}  \tau$, and $t \in \tau.\First$. We case analyze the
shape of $\nfhl{N}$:

\begin{itemize}
\item If $\nfhl{N} =  \epsilon $, then it's impossible.
\item If $\nfhl{N} =  \nfhl{  \nfhl{ t }  ~ \overline{n} } $, then we have proved (1).
\item If $\nfhl{N} =  {\nfhl{\alpha} }  \, \overline{n}$. Since $ {\nfhl{\alpha} }  \, \overline{n}$ is well-typed,
  the $\First$ of the type of $ {\nfhl{\alpha} }  \, \overline{n}$ is equivalent to
  the $\First$ of the type of $\alpha$. 
   So we have proved (2).
\end{itemize}

\textbf{Right to left}
Following \Cref{lemma:well-typednf}, the type $\tau$ is
the $ \vee $ of all types. If either $ { \color{nfcolor}   n   \rightarrow    \nfhl{  \nfhl{ t }  ~ \overline{n} }   } $
of $\alpha$ has type $t \in \First$, we know that $t \in \tau.\First$.

\qed

\begin{restatable}[Productions with distinct terminals]{lemma}{lemmanormalformt}
    \label{lemma:normal:form:t}
    If $\Gamma  \ottsym{;}  \Delta  \vdash  g  \ottsym{:}  \tau$,
    and $ {\mathcal{N} }  \llbracket \,  g  \, \rrbracket  $ returns $  \nfhl{ \_ }  \Rightarrow  \nfhl{ \nfhl{G} } $,
    then for any two productions $\ottsym{(}   { \color{nfcolor}   n   \rightarrow    \nfhl{  \nfhl{ t_{{\mathrm{1}}} }  ~ \overline{n}_{{\mathrm{1}}} }   }   \ottsym{)} \, \in \, \nfhl{G}$
    and $\ottsym{(}   { \color{nfcolor}   n   \rightarrow    \nfhl{  \nfhl{ t_{{\mathrm{2}}} }  ~ \overline{n}_{{\mathrm{2}}} }   }   \ottsym{)} \, \in \, \nfhl{G}$,
    we have $t_{{\mathrm{1}}} \neq t_{{\mathrm{2}}}$.
\end{restatable}

\proof

Suppose there are $ { \color{nfcolor}   n   \rightarrow    \nfhl{  \nfhl{ t }  ~ \overline{n}_{{\mathrm{1}}} }   } $ and $ { \color{nfcolor}   n   \rightarrow    \nfhl{  \nfhl{ t }  ~ \overline{n}_{{\mathrm{2}}} }   } $.

By \Cref{lemma:well-typednf}, we know that the types of $ \nfhl{  \nfhl{ t }  ~ \overline{n}_{{\mathrm{1}}} } $
and $ \nfhl{  \nfhl{ t }  ~ \overline{n}_{{\mathrm{2}}} } $ must be apart. Therefore they have disjoint $\First$.

By \Cref{lemma:terminal:first}, we know that both $ \nfhl{  \nfhl{ t }  ~ \overline{n}_{{\mathrm{1}}} } $ and $ \nfhl{  \nfhl{ t }  ~ \overline{n}_{{\mathrm{2}}} } $
have $t \in \First$.
However, since their types have disjoint $\First$, this is impossible. So contradiction.

\qed

\subsection{The $ \epsilon $-production may only be used when
  other productions do not apply.}

We defined the notion of containment of types as follows. The key
of the definition is \rref{st-base}, which says that a grammar $g_{{\mathrm{1}}}$
is a subtype grammar of $g_{{\mathrm{2}}}$, if
$g_{{\mathrm{1}}}$ is of type $\tau_{{\mathrm{1}}}$,
$g_{{\mathrm{2}}}$ is of type $\tau_{{\mathrm{2}}}$, and $\tau_{{\mathrm{1}}}  \ottsym{=}  \tau_{{\mathrm{2}}}  \vee  \tau$ for some $\tau$.
Notably, we have $ \Gamma  ;  \Delta   \vdash   g  \lesssim  g $ for any well-typed grammar $\Gamma  \ottsym{;}  \Delta  \vdash  g  \ottsym{:}  \tau$,
as we have $\tau = \tau  \vee  \mktype{\False}{\emptyset}{\emptyset}$.

\drules[st]{$ \Gamma  ;  \Delta   \vdash   g_{{\mathrm{1}}}  \lesssim  g_{{\mathrm{2}}} $}{containment of types}{base,trans,con,union}

\begin{lemma}
  \label{lemma:subtype:grammar}
  If $\Gamma  \ottsym{;}  \Delta  \vdash  g_{{\mathrm{1}}}  \ottsym{:}  \tau_{{\mathrm{1}}}$,
  and $ \Gamma  ;  \Delta   \vdash   g_{{\mathrm{1}}}  \lesssim  g_{{\mathrm{2}}} $,
  then
  $\Gamma  \ottsym{;}  \Delta  \vdash  g_{{\mathrm{2}}}  \ottsym{:}  \tau_{{\mathrm{2}}}$,
  and $\tau_{{\mathrm{1}}}  \ottsym{=}  \tau_{{\mathrm{2}}}  \vee  \tau$ for some $\tau$.
\end{lemma}

\proof  By induction on $ \Gamma  ;  \Delta   \vdash   g_{{\mathrm{1}}}  \lesssim  g_{{\mathrm{2}}} $.

\begin{itemize}
  \item \rref{st-base} follows trivially.
  \item Case

    \drule{st-trans}

    We have $g_{{\mathrm{1}}}$ of type $\tau_{{\mathrm{1}}}$.

    By I.H., $g_{{\mathrm{2}}}$ of type $\tau_{{\mathrm{2}}}$, and $\tau_{{\mathrm{1}}}  \ottsym{=}  \tau_{{\mathrm{2}}}  \vee  \tau$.

    By the second I.H., $g_{{\mathrm{3}}}$ of type $\tau_{{\mathrm{3}}}$, and $\tau_{{\mathrm{2}}}  \ottsym{=}  \tau_{{\mathrm{3}}}  \vee  \tau'$.

    Therefore, $\tau_{{\mathrm{1}}}  \ottsym{=}  \tau_{{\mathrm{3}}}  \vee  \ottsym{(}  \tau  \vee  \tau'  \ottsym{)}$.

  \item Case

    \drule{st-con}

    We have $ g_{{\mathrm{1}}}  \cdot  g_{{\mathrm{2}}} $ of type $ \tau_{{\mathrm{1}}}  \cdot  \tau_{{\mathrm{2}}} $ with
    $g_{{\mathrm{1}}}$ of type $\tau_{{\mathrm{1}}}$ and $g_{{\mathrm{2}}}$ of type $\tau_{{\mathrm{2}}}$
    and $\tau_{{\mathrm{1}}}  \circledast  \tau_{{\mathrm{2}}}$.

    By I.H., $g'_{{\mathrm{1}}}$ of type $\tau'_{{\mathrm{1}}}$, and $\tau_{{\mathrm{1}}}  \ottsym{=}  \tau'_{{\mathrm{1}}}  \vee  \tau$.

    By the second I.H., $g'_{{\mathrm{2}}}$ of type $\tau'_{{\mathrm{2}}}$, and $\tau_{{\mathrm{2}}}  \ottsym{=}  \tau'_{{\mathrm{2}}}  \vee  \tau'$.

    Now we want to show $ g'_{{\mathrm{1}}}  \cdot  g'_{{\mathrm{2}}} $ is of type $ \tau'_{{\mathrm{1}}}  \cdot  \tau'_{{\mathrm{2}}} $. For that,
    we need to prove $\tau'_{{\mathrm{1}}}  \circledast  \tau'_{{\mathrm{2}}}$.

    That means we need to prove
    $\tau'_1.\Followlast \cap \tau'_2.\First = \emptyset \land \lnot \tau'_1.\Null$

    We already know $\tau_{{\mathrm{1}}}  \circledast  \tau_{{\mathrm{2}}}$, which means
    $\tau_1.\Followlast \cap \tau_2.\First = \emptyset \land \lnot \tau_1.\Null$

    Since $\tau_{{\mathrm{1}}}  \ottsym{=}  \tau'_{{\mathrm{1}}}  \vee  \tau$ and $\tau_{{\mathrm{2}}}  \ottsym{=}  \tau'_{{\mathrm{2}}}  \vee  \tau'$, we can derive
    $\tau'_1.\Followlast \cap \tau'_2.\First = \emptyset \land \lnot \tau'_1.\Null$

    Therefore, $\tau'_{{\mathrm{1}}}  \circledast  \tau'_{{\mathrm{2}}}$,
    and $ g'_{{\mathrm{1}}}  \cdot  g'_{{\mathrm{2}}} $ is of type $ \tau'_{{\mathrm{1}}}  \cdot  \tau'_{{\mathrm{2}}} $.

    Now the goal is to relate $ \tau_{{\mathrm{1}}}  \cdot  \tau_{{\mathrm{2}}} $ with $ \tau'_{{\mathrm{1}}}  \cdot  \tau'_{{\mathrm{2}}} $.

    $ \tau_{{\mathrm{1}}}  \cdot  \tau_{{\mathrm{2}}}  = \mktype[long]{\tau_1.\Null \land \tau_2.\Null}
    {\tau_1.\First \;\cup \tau_1.\Null \Implies \tau_2.\First }
    {\tau_2.\Followlast \;\cup \tau_2.\Null \Implies (\tau_2.\First \cup \tau_1.\Followlast)} $

    given $\lnot \tau_1.\Null$

    $ \tau_{{\mathrm{1}}}  \cdot  \tau_{{\mathrm{2}}}  = \mktype[long]{\False}
    {\tau_1.\First }
    {\tau_2.\Followlast \;\cup \tau_2.\Null \Implies (\tau_2.\First \cup \tau_1.\Followlast)} $

    Similarly,

     $ \tau'_{{\mathrm{1}}}  \cdot  \tau'_{{\mathrm{2}}}  =
    \mktype[long]{\False}
    {\tau'_1.\First }
    {\tau'_2.\Followlast \;\cup \tau'_2.\Null \Implies (\tau'_2.\First \cup \tau'_1.\Followlast)} $

    We have $\tau_{{\mathrm{1}}}  \ottsym{=}  \tau'_{{\mathrm{1}}}  \vee  \tau$ and $\tau_{{\mathrm{2}}}  \ottsym{=}  \tau'_{{\mathrm{2}}}  \vee  \tau'$.
    Therefore, with $\lnot \tau_2.\Null$ implying $\lnot \tau'_2.\Null$,

    $ \tau_{{\mathrm{1}}}  \cdot  \tau_{{\mathrm{2}}}  = \ottsym{(}   \tau'_{{\mathrm{1}}}  \cdot  \tau'_{{\mathrm{2}}}   \ottsym{)}  \vee 
    \mktype[long]{\False}
    {\tau.\First }
    {\tau'.\Followlast \;\cup \tau_2.\Null \Implies (\tau_2.\First \cup \tau_1.\Followlast)} $

  \item Case

    \drule{st-union}

    We have $ g_{{\mathrm{1}}}  \lor  g_{{\mathrm{2}}} $ of type $\tau_{{\mathrm{1}}}  \vee  \tau_{{\mathrm{2}}}$ with
    $g_{{\mathrm{1}}}$ of type $\tau_{{\mathrm{1}}}$ and $g_{{\mathrm{2}}}$ of type $\tau_{{\mathrm{2}}}$
    and $ \tau_{{\mathrm{1}}} \;  \#  \;  \tau_{{\mathrm{2}}} $.

    By I.H., $g'_{{\mathrm{1}}}$ of type $\tau'_{{\mathrm{1}}}$, and $\tau_{{\mathrm{1}}}  \ottsym{=}  \tau'_{{\mathrm{1}}}  \vee  \tau$.

    By the second I.H., $g'_{{\mathrm{2}}}$ of type $\tau'_{{\mathrm{2}}}$, and $\tau_{{\mathrm{2}}}  \ottsym{=}  \tau'_{{\mathrm{2}}}  \vee  \tau'$.

    Now we want to show $ g'_{{\mathrm{1}}}  \lor  g'_{{\mathrm{2}}} $ is of type $\tau'_{{\mathrm{1}}}  \vee  \tau'_{{\mathrm{2}}}$. For that,
    we need to prove $ \tau'_{{\mathrm{1}}} \;  \#  \;  \tau'_{{\mathrm{2}}} $.

    That means we need to prove
    $(\tau'_1.\First \cap \tau'_2.\First = \emptyset) \land \lnot (\tau'_1.\Null \land \tau'_2.\Null)$

    We already know $ \tau_{{\mathrm{1}}} \;  \#  \;  \tau_{{\mathrm{2}}} $, which means
    $(\tau_1.\First \cap \tau_2.\First = \emptyset) \land \lnot (\tau_1.\Null \land \tau_2.\Null)$

    Since $\tau_{{\mathrm{1}}}  \ottsym{=}  \tau'_{{\mathrm{1}}}  \vee  \tau$ and $\tau_{{\mathrm{2}}}  \ottsym{=}  \tau'_{{\mathrm{2}}}  \vee  \tau'$, we can derive
    $(\tau'_1.\First \cap \tau'_2.\First = \emptyset) \land \lnot (\tau'_1.\Null \land \tau'_2.\Null)$

    Therefore, $ \tau'_{{\mathrm{1}}} \;  \#  \;  \tau'_{{\mathrm{2}}} $,
    and $ g'_{{\mathrm{1}}}  \lor  g'_{{\mathrm{2}}} $ is of type $\tau'_{{\mathrm{1}}}  \vee  \tau'_{{\mathrm{2}}}$.

    Finally, we have $\tau_{{\mathrm{1}}}  \vee  \tau_{{\mathrm{2}}} = \ottsym{(}  \tau'_{{\mathrm{1}}}  \vee  \tau'_{{\mathrm{2}}}  \ottsym{)}  \vee  \ottsym{(}  \tau  \vee  \tau'  \ottsym{)}$.

\end{itemize}

\qed

\begin{lemma}[Expansion preserves typing]
  \label{lemma:extension:typing}
  Given $\Gamma  \ottsym{;}  \Delta  \vdash  g  \ottsym{:}  \tau$,
  $ {\mathcal{N} }  \llbracket \,  g  \, \rrbracket  $ returns $  \nfhl{ \_ }  \Rightarrow  \nfhl{ \nfhl{G} } $,
  if $ \nfhl{G}   \vdash  \nfhl{  \nfhl{ { n }_{ \tau } }  }  \leadsto    \nfhl{ \overline{t} }  \,  \nfhl{ n' }  \,  \nfhl{ \overline{n} }  $,
  then $\Gamma;\Delta   \vdash   \nfhl{ \overline{t} }  \,  \nfhl{ n' }  \,  \nfhl{ \overline{n} }  : \tau_{{\mathrm{1}}}$,
  and $\tau  \ottsym{=}  \tau_{{\mathrm{1}}}  \vee  \tau'$ for some $\tau'$.
\end{lemma}

\proof

By induction on $ \nfhl{G}   \vdash  \nfhl{  \nfhl{ { n }_{ \tau } }  }  \leadsto    \nfhl{ \overline{t} }  \,  \nfhl{ n' }  \,  \nfhl{ \overline{n} }  $.

\begin{itemize}

\item In the base case,  $ \nfhl{G}   \vdash  \nfhl{  \nfhl{ { n }_{ \tau } }  }  \leadsto    \nfhl{  \nfhl{ { n }_{ \tau } }  }  $. The goal follows
  trivially.
\item In the inductive case, we have
  $ \nfhl{G}   \vdash  \nfhl{  \nfhl{ { n }_{ \tau } }  }  \leadsto    \nfhl{ \overline{t} }  \,  \nfhl{ n' }  \,  \nfhl{ \overline{n} }  $,
  $ { \color{nfcolor}   n'   \rightarrow   \nfhl{N}  }  \, \in \, \nfhl{G}$
  and so $ \nfhl{G}   \vdash  \nfhl{ n }  \leadsto    \nfhl{ \overline{t} }  \,  \nfhl{ \nfhl{N} }  \,  \nfhl{ \overline{n} }  $,

By I.H., we have
$\Gamma;\Delta   \vdash   \nfhl{ \overline{t} }  \,  \nfhl{ n' }  \,  \nfhl{ \overline{n} }  : \tau_{{\mathrm{1}}}$, and $\tau  \ottsym{=}  \tau_{{\mathrm{1}}}  \vee  \tau'$.

According to \Cref{lemma:well-typednf}, we know that
$\Gamma; \Delta  \vdash  n'  \lesssim  \nfhl{N}$ by \rref{st-base}.

Therefore,
$\Gamma; \Delta  \vdash   \nfhl{ \overline{t} }  \,  \nfhl{ n' }  \,  \nfhl{ \overline{n} }   \lesssim   \nfhl{ \overline{t} }  \,  \nfhl{ \nfhl{N} }  \,  \nfhl{ \overline{n} } $ by \rref{st-con}.

By \Cref{lemma:subtype:grammar},
$\Gamma;\Delta   \vdash   \nfhl{ \overline{t} }  \,  \nfhl{ \nfhl{N} }  \,  \nfhl{ \overline{n} }  : \tau_{{\mathrm{2}}}$, and $\tau_{{\mathrm{1}}}  \ottsym{=}  \tau_{{\mathrm{2}}}  \vee  \tau''$.

Therefore,
$\tau  \ottsym{=}  \tau_{{\mathrm{2}}}  \vee  \ottsym{(}  \tau'  \vee  \tau''  \ottsym{)}$.

\end{itemize}

\qed

\begin{restatable}[Guarded $ \epsilon $-production]{lemma}{lemmanormalformepsilon}
    \label{lemma:epsilon:use}
    Given $\Gamma  \ottsym{;}  \Delta  \vdash  g  \ottsym{:}  \tau$,
    $ {\mathcal{N} }  \llbracket \,  g  \, \rrbracket  $ returns $  \nfhl{ n }  \Rightarrow  \nfhl{ \nfhl{G} } $,
    and $\nfhl{G}  \vdash n  \leadsto^{*} 
    \cdots n_{{\mathrm{1}}} n_{{\mathrm{2}}} \cdots$,
    if $\ottsym{(}   { \color{nfcolor}   n_{{\mathrm{1}}}   \rightarrow   \epsilon  }   \ottsym{)} \, \in \, \nfhl{G}$,
    then either
    $\ottsym{(}   { \color{nfcolor}   n_{{\mathrm{1}}}   \rightarrow    \nfhl{  \nfhl{ t }  ~ \overline{n}_{{\mathrm{1}}} }   }   \ottsym{)} \, \notin \, \nfhl{G}$
    or $\ottsym{(}   { \color{nfcolor}   n_{{\mathrm{2}}}   \rightarrow    \nfhl{  \nfhl{ t }  ~ \overline{n}_{{\mathrm{2}}} }   }   \ottsym{)} \, \notin \, \nfhl{G}$ for any $t$, $\overline{n}_{{\mathrm{1}}}$, $\overline{n}_{{\mathrm{2}}}$.
  \end{restatable}

\proof We have:

\begin{xtabular}{ll|l}
  & $\cdots n_{{\mathrm{1}}} n_{{\mathrm{2}}} \cdots$ is well-typed & By \Cref{lemma:extension:typing} \\
  & The type of $\cdots n_{{\mathrm{1}}}$ is $ \tau  \cdot  \tau_{{\mathrm{1}}} $, the type of $n_{{\mathrm{1}}}$ is $\tau_{{\mathrm{1}}}$,
  and the type of $n_{{\mathrm{2}}}$ is $\tau_{{\mathrm{2}}}$ & Suppose \\
  & $ \tau  \cdot  \tau_{{\mathrm{1}}}   \circledast  \tau_{{\mathrm{2}}}$ & By typing \\
  & $ \tau  \cdot  \tau_{{\mathrm{1}}} .\Followlast \cap \tau_{{\mathrm{2}}}.\First = \emptyset$ & By $ \circledast $ \\
  & $ \tau  \cdot  \tau_{{\mathrm{1}}} .\Followlast = \tau_{{\mathrm{1}}}.\Followlast \cup \tau_{{\mathrm{1}}}.\Null \Implies
  (\tau_{{\mathrm{1}}}.\First \cup \tau.\Followlast) $
  & By definition \\
  & $ { \color{nfcolor}   n_{{\mathrm{1}}}   \rightarrow   \epsilon  }  \, \in \, \nfhl{G}$ & Given \\
  & $n_{{\mathrm{1}}}.\Null = \True$ & \Cref{lemma:total:null:false} \\
  & $ \tau  \cdot  \tau_{{\mathrm{1}}} .\Followlast = \tau_{{\mathrm{1}}}.\Followlast \cup (\tau_{{\mathrm{1}}}.\First \cup \tau.\Followlast) $
  & Follows \\
  & $\tau_{{\mathrm{1}}}.\First \cap \tau_{{\mathrm{2}}}.\First = \emptyset $
  & Follows \\
  & $ { \color{nfcolor}   n_{{\mathrm{1}}}   \rightarrow    \nfhl{  \nfhl{ t }  ~ \overline{n}_{{\mathrm{1}}} }   }  \, \in \, \nfhl{G} \land  { \color{nfcolor}   n_{{\mathrm{2}}}   \rightarrow    \nfhl{  \nfhl{ t }  ~ \overline{n}_{{\mathrm{2}}} }   }  \, \in \, \nfhl{G}$
  & Assume \\
  & $t \in \tau_{{\mathrm{1}}}.\First  \land t \in \tau_{{\mathrm{2}}}.\First $
  & \Cref{lemma:terminal:first} \\
  & Contradiction with $\tau_{{\mathrm{1}}}.\First \cap \tau_{{\mathrm{2}}}.\First = \emptyset $
  \\
\end{xtabular}%

\qed

\subsection{Final result}

\theoremgnf*

\proof Follows from \Cref{lemma:form:correct},
\Cref{lemma:normal:form:t},
and \Cref{lemma:epsilon:use}. \qed

\section{Soundness (proof)}

\subsection{An alternative normalization}

To make proofs easier, we consider the definition $ {\mathcal{N}\!\!\!\!\!\mathcal{N} } $, which has the
same definition as $ {\mathcal{N} } $ except for the case of $ \mu \alpha .~ g $, where we do
not substitute $ {\nfhl{\alpha} } $:

\begin{center}
  \begin{tabular}{llll}\toprule
    $ {\mathcal{N}\!\!\!\!\!\mathcal{N} }  (   \mu \alpha  :  \tau .~ g   ) $     & $=$  & $ {\nfhl{\alpha} }   \Rightarrow  { \{  { \color{nfcolor}    {\nfhl{\alpha} }    \rightarrow   \nfhl{N}  }   \mid  { \color{nfcolor}   n   \rightarrow   \nfhl{N}  }  \in \nfhl{G} \} }  \cup  \nfhl{G} $ \\
                                   &                         & where \enskip  $ {\mathcal{N}\!\!\!\!\!\mathcal{N} }  (  g  )  =  \nfhl{ n }  \Rightarrow  \nfhl{ \nfhl{G} } $ \\
    \bottomrule
  \end{tabular}
\end{center}

While $ {\mathcal{N}\!\!\!\!\!\mathcal{N} } $ does not return a DGNF grammar,
it is easy to see that $ {\mathcal{N} } $ and $ {\mathcal{N}\!\!\!\!\!\mathcal{N} } $ defines the same language:

\begin{lemma}
  \label{lemma:norm:nnorm}
  If $ {\mathcal{N} }  \llbracket \,  g  \, \rrbracket $ return $   \nfhl{ n_{{\mathrm{1}}} } \Rightarrow \nfhl{ \nfhl{G}_{{\mathrm{1}}} } $,
  and $ {\mathcal{N}\!\!\!\!\!\mathcal{N} }  (  g  )  $ return $  \nfhl{ n_{{\mathrm{2}}} } \Rightarrow \nfhl{ \nfhl{G}_{{\mathrm{2}}} } $,
  then for all $w$,
  $\nfhl{G}_{{\mathrm{1}}}  \vdash  n_{{\mathrm{1}}}  \leadsto^{*}  w$ if and only if $\nfhl{G}_{{\mathrm{2}}}  \vdash  n_{{\mathrm{2}}}  \leadsto^{*}  w$.
\end{lemma}

\proof By straightforward induction on $g$. \qed

\subsection{Subexpression}

The subexpression relation essentially defines a subset relation between the
grammars denoted by context-free expressions.

\drules[sg]{$ g_{{\mathrm{1}}}  \sqsubseteq  g_{{\mathrm{2}}} $}{Subexpression}{refl,trans,con-l,con-r,union-l,union-r,mu}

We can show that what subexpression means in terms of the alternative normalization.

\begin{lemma}
  \label{lemma:subgrammar}
  If $ g_{{\mathrm{1}}}  \sqsubseteq  g_{{\mathrm{2}}} $,
  and $ {\mathcal{N}\!\!\!\!\!\mathcal{N} }  (  g_{{\mathrm{1}}}  )  $ returns $   \nfhl{ n_{{\mathrm{1}}} } \Rightarrow \nfhl{ \nfhl{G}_{{\mathrm{1}}} } $,
  and $ {\mathcal{N}\!\!\!\!\!\mathcal{N} }  (  g_{{\mathrm{2}}}  )  $ returns $   \nfhl{ n_{{\mathrm{2}}} } \Rightarrow \nfhl{ \nfhl{G}_{{\mathrm{2}}} } $,
  then for all $n \, \in \, \mathsf{dom} \, \ottsym{(}  \nfhl{G}_{{\mathrm{1}}}  \ottsym{)}$,
  $\ottsym{(}   { \color{nfcolor}   n   \rightarrow   \nfhl{N}  }   \ottsym{)} \, \in \, \nfhl{G}_{{\mathrm{1}}}$ if and only if $\ottsym{(}   { \color{nfcolor}   n   \rightarrow   \nfhl{N}  }   \ottsym{)} \, \in \, \nfhl{G}_{{\mathrm{2}}}$.
\end{lemma}

\proof By straightforward induction on $ g_{{\mathrm{1}}}  \sqsubseteq  g_{{\mathrm{2}}} $. \qed

\subsection{Proof of soundness}

\newcommand{\num}{\mathbb{n}}

In the following lemma statement, we denote a natural number as $\num$,
and the length of a word $w$ as $\ottsym{\mbox{$\mid$}}  w  \ottsym{\mbox{$\mid$}}$. The relations $\gamma  \vDash  \Gamma$
and $\delta  \vDash  \Delta$ mean that
$\gamma$ and $\delta$ give interpretations (i.e.~languages $\mathtt{L}$) of variables in $\Gamma$ and $\Delta$
respectively.

\begin{mathpar}
  \inferrule{ }{\bullet  \vDash  \bullet}
  \and
  \inferrule{ \delta  \vDash  \Delta \quad \mathtt{L}  \vDash  \tau} {\delta  \ottsym{,}   \mathtt{L}  /  \alpha   \vDash  \Delta  \ottsym{,}  \alpha  \ottsym{:}  \tau} \\
  \mathtt{L}  \vDash  \tau \triangleq \Null(LL) \Rightarrow \tau.\Null \land
  \First(\mathtt{L}) \subseteq \tau.\First \land \Followlast(\mathtt{L}) \subseteq \tau.\Followlast
\end{mathpar}

\begin{lemma}
  \label{lemma:correct}
  Given $\Gamma  \ottsym{;}  \Delta  \vdash  g  \ottsym{:}  \tau$,
  and $\gamma  \vDash  \Gamma$,
  and $\delta  \vDash  \Delta$,
  and $ {\mathcal{N}\!\!\!\!\!\mathcal{N} }  (  g  )  $ returns $   \nfhl{ n } \Rightarrow \nfhl{ \nfhl{G} } $,
  if
  \begin{enumerate}
    \item
   $ g  \sqsubseteq  g' $,
   where
  $\bullet  \ottsym{;}  \bullet  \vdash  g'  \ottsym{:}  \tau'$
  and $ {\mathcal{N}\!\!\!\!\!\mathcal{N} }  (  g'  ) $ returns $  \nfhl{ n' } \Rightarrow \nfhl{ \nfhl{G}' } $; and
  \item
  $\forall \alpha \, \in \, \mathsf{dom} \, \ottsym{(}  \gamma  \ottsym{)}$,
  $\forall \ottsym{\mbox{$\mid$}}  w_{{\mathrm{1}}}  \ottsym{\mbox{$\mid$}} \leq \num$, $w_{{\mathrm{1}}} \, \in \, \gamma  \ottsym{(}  \alpha  \ottsym{)}$ if and only if $\nfhl{G}'  \vdash   {\nfhl{\alpha} }   \leadsto^{*}  w_{{\mathrm{1}}}$; and
  \item
  $\forall \alpha \, \in \, \mathsf{dom} \, \ottsym{(}  \delta  \ottsym{)}$,
  $\forall \ottsym{\mbox{$\mid$}}  w_{{\mathrm{2}}}  \ottsym{\mbox{$\mid$}} < \num $,  $w_{{\mathrm{2}}} \, \in \, \delta  \ottsym{(}  \alpha  \ottsym{)}$ if and only if $\nfhl{G}'  \vdash   {\nfhl{\alpha} }   \leadsto^{*}  w_{{\mathrm{2}}}$,
  \end{enumerate}
  then
  $\forall w \leq \num$,
  $w \, \in \,  { {\llbracket  g  \rrbracket}_{ \ottsym{(}  \gamma  \ottsym{,}  \delta  \ottsym{)} } } $ if and only if
  and $\nfhl{G}'  \vdash  n  \leadsto^{*}  w$.
\end{lemma}

\proof

By first induction on $\num$. The base case of $0$ is trivial. In the inductive
case, we have that the lemma holds for $ \ottsym{\mbox{$\mid$}}  w  \ottsym{\mbox{$\mid$}} < \num$, and we want to prove it for
$\ottsym{\mbox{$\mid$}}  w  \ottsym{\mbox{$\mid$}} \leq \num$.

Now we perform induction on $g$.

\begin{itemize}
\item  The cases for $g  \ottsym{=}   t $, $g  \ottsym{=}  \epsilon$, and $g  \ottsym{=}  \bot$ are straightforward.
\item  $g  \ottsym{=}  \alpha$. Then $ {\mathcal{N}\!\!\!\!\!\mathcal{N} }  (  \alpha  )  =  \nfhl{ n } \Rightarrow \nfhl{  \nfhl{  { \color{nfcolor}   n   \rightarrow    \nfhl{  {\nfhl{\alpha} }  }   }  }  } $.
  By \Cref{lemma:subgrammar}, we know $\ottsym{(}   { \color{nfcolor}   n   \rightarrow    \nfhl{  {\nfhl{\alpha} }  }   }   \ottsym{)} \, \in \, \nfhl{G}'$,
  and there is no other production for $n$ in $\nfhl{G}'$.

  Since $g$ is well-typed, it must be $\alpha \, \in \, \mathsf{dom} \, \ottsym{(}  \Gamma  \ottsym{)}$,
  and thus $\alpha \, \in \, \mathsf{dom} \, \ottsym{(}  \gamma  \ottsym{)}$. Then $ { {\llbracket  g  \rrbracket}_{ \ottsym{(}  \gamma  \ottsym{,}  \delta  \ottsym{)} } }  = \gamma  \ottsym{(}  \alpha  \ottsym{)}$.

  As given, we know that $\forall \ottsym{\mbox{$\mid$}}  w  \ottsym{\mbox{$\mid$}} \leq \num $,  $w \, \in \, \gamma  \ottsym{(}  \alpha  \ottsym{)}$ if and
  only if $\nfhl{G}'  \vdash   {\nfhl{\alpha} }   \leadsto^{*}  w$.

  Since we know $\ottsym{(}   { \color{nfcolor}   n   \rightarrow    \nfhl{  {\nfhl{\alpha} }  }   }   \ottsym{)} \, \in \, \nfhl{G}'$,
  we have
  $\forall \ottsym{\mbox{$\mid$}}  w  \ottsym{\mbox{$\mid$}} \leq \num $,  $w \, \in \, \gamma  \ottsym{(}  \alpha  \ottsym{)}$ if and only if $\nfhl{G}'  \vdash  n  \leadsto^{*}  w$.

\item  $g  \ottsym{=}   g_{{\mathrm{1}}}  \lor  g_{{\mathrm{2}}} $. Then $ { {\llbracket   g_{{\mathrm{1}}}  \lor  g_{{\mathrm{2}}}   \rrbracket}_{ \ottsym{(}  \gamma  \ottsym{,}  \delta  \ottsym{)} } }  =  { {\llbracket  g_{{\mathrm{1}}}  \rrbracket}_{ \ottsym{(}  \gamma  \ottsym{,}  \delta  \ottsym{)} } }  \, \cup \,  { {\llbracket  g_{{\mathrm{2}}}  \rrbracket}_{ \ottsym{(}  \gamma  \ottsym{,}  \delta  \ottsym{)} } } $

  We have

  \begin{xtabular}{ll}
    $\{  { \color{nfcolor}   n   \rightarrow   \nfhl{N}_{{\mathrm{1}}}  }  \mid  { \color{nfcolor}   n_{{\mathrm{1}}}   \rightarrow   \nfhl{N}_{{\mathrm{1}}}  }  \in \nfhl{G}_{{\mathrm{1}}} \}  \cup  \{  { \color{nfcolor}   n   \rightarrow   \nfhl{N}_{{\mathrm{2}}}  }  \mid  { \color{nfcolor}   n_{{\mathrm{2}}}   \rightarrow   \nfhl{N}_{{\mathrm{2}}}  }  \in \nfhl{G}_{{\mathrm{2}}} \}  \cup   \nfhl{ \nfhl{G}_{{\mathrm{1}}}   \cup   \nfhl{G}_{{\mathrm{2}}} }  $    \\
    $ {\mathcal{N}\!\!\!\!\!\mathcal{N} }  (  g_{{\mathrm{1}}}  )  =  \nfhl{ n_{{\mathrm{1}}} } \Rightarrow \nfhl{ \nfhl{G}_{{\mathrm{1}}} } $ \\
    $ {\mathcal{N}\!\!\!\!\!\mathcal{N} }  (  g_{{\mathrm{2}}}  )  =  \nfhl{ n_{{\mathrm{2}}} } \Rightarrow \nfhl{ \nfhl{G}_{{\mathrm{2}}} } $ \\
  \end{xtabular}

  The goal follows from I.H. on $g_{{\mathrm{1}}}$ and $g_{{\mathrm{2}}}$.

\item  $g  \ottsym{=}   g_{{\mathrm{1}}}  \cdot  g_{{\mathrm{2}}} $. Then $ { {\llbracket   g_{{\mathrm{1}}}  \cdot  g_{{\mathrm{2}}}   \rrbracket}_{ \ottsym{(}  \gamma  \ottsym{,}  \delta  \ottsym{)} } }  = \{  w_{{\mathrm{1}}}  \cdot  w_{{\mathrm{2}}}  \mid
  w_{{\mathrm{1}}} \, \in \,  { {\llbracket  g_{{\mathrm{1}}}  \rrbracket}_{ \ottsym{(}  \gamma  \ottsym{,}  \delta  \ottsym{)} } }  \land w_{{\mathrm{2}}} \, \in \,  { {\llbracket  g_{{\mathrm{2}}}  \rrbracket}_{ \ottsym{(}  \gamma  \ottsym{,}  \delta  \ottsym{)} } }  \}$.

  According to $ {\mathcal{N}\!\!\!\!\!\mathcal{N} } $, we have

  \begin{xtabular}{ll}
    $\{  { \color{nfcolor}   n   \rightarrow    \nfhl{ \nfhl{N}_{{\mathrm{1}}} ~ n_{{\mathrm{2}}} }   }  \mid  { \color{nfcolor}   n_{{\mathrm{1}}}   \rightarrow   \nfhl{N}_{{\mathrm{1}}}  }  \in \nfhl{G}_{{\mathrm{1}}}  \}  \cup   \nfhl{ \nfhl{G}_{{\mathrm{1}}}   \cup   \nfhl{G}_{{\mathrm{2}}} }  $ \\
    $ {\mathcal{N}\!\!\!\!\!\mathcal{N} }  (  g_{{\mathrm{1}}}  )  =  \nfhl{ n_{{\mathrm{1}}} } \Rightarrow \nfhl{ \nfhl{G}_{{\mathrm{1}}} } $  \\
    $ {\mathcal{N}\!\!\!\!\!\mathcal{N} }  (  g_{{\mathrm{2}}}  )  =  \nfhl{ n_{{\mathrm{2}}} } \Rightarrow \nfhl{ \nfhl{G}_{{\mathrm{2}}} } $ \\
  \end{xtabular}

  According to typing, we have

  \begin{xtabular}{ll}
    $\Gamma  \ottsym{;}  \Delta  \vdash  g_{{\mathrm{1}}}  \ottsym{:}  \tau_{{\mathrm{1}}}$ \\
    $\Gamma  \ottsym{,}  \Delta  \ottsym{;}  \bullet  \vdash  g_{{\mathrm{2}}}  \ottsym{:}  \tau_{{\mathrm{2}}} $ \\
  \end{xtabular}
  
  By I.H. on $g_{{\mathrm{1}}}$, we have

  $\forall \ottsym{\mbox{$\mid$}}  w_{{\mathrm{1}}}  \ottsym{\mbox{$\mid$}} \leq \num $,  $w_{{\mathrm{1}}} \, \in \,  { {\llbracket  g_{{\mathrm{1}}}  \rrbracket}_{ \ottsym{(}  \gamma  \ottsym{,}  \delta  \ottsym{)} } } $ if and only if $\nfhl{G}'  \vdash  n_{{\mathrm{1}}}  \leadsto^{*}  w_{{\mathrm{1}}}$.

  By I.H. on $g_{{\mathrm{2}}}$, we have the following. Here we use $<$ instead of $\leq$
  as its typing context $\Gamma  \ottsym{,}  \Delta$ includes $\Delta$ that only has
  interpretations for $\ottsym{\mbox{$\mid$}}  w_{{\mathrm{2}}}  \ottsym{\mbox{$\mid$}} < \num$.

  $\forall \ottsym{\mbox{$\mid$}}  w_{{\mathrm{2}}}  \ottsym{\mbox{$\mid$}} < \num $,  $w_{{\mathrm{2}}} \, \in \,  { {\llbracket  g_{{\mathrm{2}}}  \rrbracket}_{ \ottsym{(}  \gamma  \ottsym{,}  \delta  \ottsym{)} } } $ if and only if $\nfhl{G}'  \vdash  n_{{\mathrm{2}}}  \leadsto^{*}  w_{{\mathrm{2}}}$.

  We first prove the conclusion from left to right.
  Given $w \leq \num$, and $w \, \in \,  { {\llbracket   g_{{\mathrm{1}}}  \cdot  g_{{\mathrm{2}}}   \rrbracket}_{ \ottsym{(}  \gamma  \ottsym{,}  \delta  \ottsym{)} } } $, it must be
  $w =  w_{{\mathrm{1}}}  \cdot  w_{{\mathrm{2}}} $ and $w_{{\mathrm{1}}} \, \in \,  { {\llbracket  g_{{\mathrm{1}}}  \rrbracket}_{ \ottsym{(}  \gamma  \ottsym{,}  \delta  \ottsym{)} } } $ and $w_{{\mathrm{2}}} \, \in \,  { {\llbracket  g_{{\mathrm{2}}}  \rrbracket}_{ \ottsym{(}  \gamma  \ottsym{,}  \delta  \ottsym{)} } } $. As $ g_{{\mathrm{1}}}  \cdot  g_{{\mathrm{2}}} $ is well-typed, we know $\tau_{{\mathrm{1}}}.{\Null} =  \mathsf{false} $, so
  $w_{{\mathrm{1}}}$ cannot be empty, and thus $w_{{\mathrm{2}}}$ must have length $< \num$.
  So following I.H., and that $n$ represents the same language as $n_{{\mathrm{1}}} \, n_{{\mathrm{2}}}$, we have $\nfhl{G}'  \vdash  n  \leadsto^{*}   w_{{\mathrm{1}}}  \cdot  w_{{\mathrm{2}}} $.

  Now we move to the conclusion from right to left.
  Given $\nfhl{G}'  \vdash  n  \leadsto^{*}  w$, it must be $w =  w_{{\mathrm{1}}}  \cdot  w_{{\mathrm{2}}} $,
  and $\nfhl{G}'  \vdash  n_{{\mathrm{1}}}  \leadsto^{*}  w_{{\mathrm{1}}}$, and $\nfhl{G}'  \vdash  n_{{\mathrm{2}}}  \leadsto^{*}  w_{{\mathrm{2}}}$.
  As $ g_{{\mathrm{1}}}  \cdot  g_{{\mathrm{2}}} $ is well-typed, we know that $\tau_{{\mathrm{1}}}.{\Null} =  \mathsf{false} $,
  so by \Cref{lemma:total:null:false}, $w_{{\mathrm{1}}}$ cannot be empty,
  and thus $w_{{\mathrm{2}}}$ must have length  $< \num$.
  So following I.H., we have $w_{{\mathrm{1}}} \, \in \,  { {\llbracket  g_{{\mathrm{1}}}  \rrbracket}_{ \ottsym{(}  \gamma  \ottsym{,}  \delta  \ottsym{)} } } $,
  and $w_{{\mathrm{2}}} \, \in \,  { {\llbracket  g_{{\mathrm{2}}}  \rrbracket}_{ \ottsym{(}  \gamma  \ottsym{,}  \delta  \ottsym{)} } } $,
  and thus $w \, \in \,  { {\llbracket   g_{{\mathrm{1}}}  \cdot  g_{{\mathrm{2}}}   \rrbracket}_{ \ottsym{(}  \gamma  \ottsym{,}  \delta  \ottsym{)} } } $.

\item  $g  \ottsym{=}   \mu \alpha .~ g_{{\mathrm{1}}} $. Then $ { {\llbracket   \mu \alpha .~ g_{{\mathrm{1}}}   \rrbracket}_{ \ottsym{(}  \gamma  \ottsym{,}  \delta  \ottsym{)} } } 
  =  { {\llbracket  g_{{\mathrm{1}}}  \rrbracket}_{ \ottsym{(}  \gamma  \ottsym{,}  \delta  \ottsym{,}    { {\llbracket   \mu \alpha .~ g_{{\mathrm{1}}}   \rrbracket}_{ \ottsym{(}  \gamma  \ottsym{,}  \delta  \ottsym{)} } }   /  \alpha   \ottsym{)} } }   $.

  We have

  \begin{xtabular}{ll}
    $ {\mathcal{N}\!\!\!\!\!\mathcal{N} }  (   \mu \alpha  :  \tau .~ g_{{\mathrm{1}}}   ) $     $=$  $ {\nfhl{\alpha} }   \Rightarrow  { \{  { \color{nfcolor}    {\nfhl{\alpha} }    \rightarrow   \nfhl{N}  }   \mid  { \color{nfcolor}   n   \rightarrow   \nfhl{N}  }  \in \nfhl{G} \} }  \cup  \nfhl{G} $ \\
    $ {\mathcal{N}\!\!\!\!\!\mathcal{N} }  (  g_{{\mathrm{1}}}  )  =  \nfhl{ n }  \Rightarrow  \nfhl{ \nfhl{G} } $ \\
  \end{xtabular}

  According to typing, we have $\Gamma  \ottsym{;}  \Delta  \ottsym{,}  \alpha  \ottsym{:}  \tau  \vdash  g_{{\mathrm{1}}}  \ottsym{:}  \tau$.

  According to the I.H. on $\num$, we have

  $\forall w' < \num$,
  $w' \, \in \,  { {\llbracket   \mu \alpha .~ g_{{\mathrm{1}}}   \rrbracket}_{ \ottsym{(}  \gamma  \ottsym{,}  \delta  \ottsym{)} } } $ if and only if and $\nfhl{G}'  \vdash   {\nfhl{\alpha} }   \leadsto^{*}  w'$.

  We have $\ottsym{(}  \gamma  \ottsym{,}  \delta  \ottsym{,}    { {\llbracket   \mu \alpha .~ g_{{\mathrm{1}}}   \rrbracket}_{ \ottsym{(}  \gamma  \ottsym{,}  \delta  \ottsym{)} } }   /  \alpha   \ottsym{)}  \ottsym{(}  \alpha  \ottsym{)} =  { {\llbracket   \mu \alpha .~ g_{{\mathrm{1}}}   \rrbracket}_{ \ottsym{(}  \gamma  \ottsym{,}  \delta  \ottsym{)} } } $.

  That means we have

  $\forall \beta \, \in \, \mathsf{dom} \, \ottsym{(}  \delta  \ottsym{,}    { {\llbracket   \mu \alpha .~ g_{{\mathrm{1}}}   \rrbracket}_{ \ottsym{(}  \gamma  \ottsym{,}  \delta  \ottsym{)} } }   /  \alpha   \ottsym{)}$,

  $\forall \ottsym{\mbox{$\mid$}}  w'  \ottsym{\mbox{$\mid$}} < \num $,  $w' \, \in \, \ottsym{(}  \delta  \ottsym{,}    { {\llbracket   \mu \alpha .~ g_{{\mathrm{1}}}   \rrbracket}_{ \ottsym{(}  \gamma  \ottsym{,}  \delta  \ottsym{)} } }   /  \alpha   \ottsym{)}  \ottsym{(}  \beta  \ottsym{)}$
  if and only if $\nfhl{G}'  \vdash   {\nfhl{\beta} }   \leadsto^{*}  w'$.

  Now by I.H. on $g_{{\mathrm{1}}}$,

  $\forall w \leq \num$,
  $w \, \in \,  { {\llbracket  g_{{\mathrm{1}}}  \rrbracket}_{ \ottsym{(}  \gamma  \ottsym{,}  \delta  \ottsym{,}    { {\llbracket   \mu \alpha .~ g_{{\mathrm{1}}}   \rrbracket}_{ \ottsym{(}  \gamma  \ottsym{,}  \delta  \ottsym{)} } }   /  \alpha   \ottsym{)} } } $ if and only if
  $\nfhl{G}'  \vdash  n  \leadsto^{*}  w$

  equivalent to

  $\forall w \leq \num$,
  $w \, \in \,  { {\llbracket   \mu \alpha .~ g_{{\mathrm{1}}}   \rrbracket}_{ \ottsym{(}  \gamma  \ottsym{,}  \delta  \ottsym{)} } } $ if and only if
  $\nfhl{G}'  \vdash  n  \leadsto^{*}  w$.

  We have $ { \color{nfcolor}    {\nfhl{\alpha} }    \rightarrow   \nfhl{N}  }  \in  {\mathcal{N}\!\!\!\!\!\mathcal{N} }  (   \mu \alpha .~ g_{{\mathrm{1}}}   ) $,
  where $ { \color{nfcolor}   n   \rightarrow   \nfhl{N}  }  \in  {\mathcal{N}\!\!\!\!\!\mathcal{N} }  (   \mu \alpha .~ g_{{\mathrm{1}}}   ) $.
  By \Cref{lemma:subgrammar},
  we have $ { \color{nfcolor}    {\nfhl{\alpha} }    \rightarrow   \nfhl{N}  }  \, \in \, \nfhl{G}'$ and there is no other productions for $ {\nfhl{\alpha} } $.

  Therefore,

  $\forall w \leq \num$,
  $w \, \in \,  { {\llbracket   \mu \alpha .~ g_{{\mathrm{1}}}   \rrbracket}_{ \ottsym{(}  \gamma  \ottsym{,}  \delta  \ottsym{)} } } $ if and only if
  $ \nfhl{G}'   \vdash  \nfhl{  {\nfhl{\alpha} }  }  \leadsto   w $.

\end{itemize}

\qed

\theoremcorrectness*

\proof Follows by \Cref{lemma:correct}, making use of \Cref{lemma:norm:nnorm}. \qed

 }

\end{document}